# Attosecond Molecular Spectroscopy and Dynamics

Denitsa Baykusheva[1] and Hans Jakob Wörner[1*]

[1]*Laboratorium für Physikalische Chemie, ETH Zürich, Zürich, Switzerland*

## 1. Introduction

Time-resolved spectroscopy, or the investigation of fundamental natural processes in real time, is a scientific field that lies at the intersection of physics, chemistry, and biology. Historically, the application of Schlieren photography by Alfred Toepler to study the instantaneous density profiles of fluids and propagating shockwaves by means of two subsequent light flashes (Krehl and Engemann, 1995) can be considered as the first time-resolved study of a microscopic process on a sub-second time-scale. These experiments laid out the methodology of the pump-probe technique, whereas further progress has remained coupled to the state of technological development setting the temporal limit to the cross-correlation of available pump/probe pulses. The nanosecond time scale was reached already in the 19th century, when Abraham and Lemoine (Abraham and Lemoine, 1899) demonstrated that the Kerr cell shutter can be operated with a time resolution of less than $10^{-8}$ s. In the first half of the 20th-century, the development of the relaxation methods and the flash photolysis to study chemical reactions occurring on the micro- to nanosecond timescales allowed for the identification of transient reaction intermediates, a work whose importance was acknowledged by the Nobel prize in chemistry in 1967 awarded to M. Eigen, N. Porter and W. Norrish (Eigen, 1954; Norrish and Porter, 1949). Subsequently, the development of ultrafast laser technology, starting with the invention of the laser in 1960 and the subsequent introduction of pulsed-laser operation through Q-switching, modelocking, and frequency up-conversion techniques based on non-linear optical processes led to the improvement of the achievable time resolution by nearly six orders of magnitude. By the mid-1980s, pulses of few-femtosecond (fs) duration containing only few cycles of the fundamental carrier light wave became accessible using dye lasers (Fork et al., 1987). Stimulated by the work of A. Zewail (Nobel Prize in Chemistry 1999) (Zewail, 2000), these technologies have enabled the real-time tracking of detailed chemical reactions, including the monitoring of rotational and vibrational motion, dissociation processes occurring on bound and repulsive potential energy surfaces, and dynamics at conical intersections.

Since pulsed lasers operate mainly in the visible (VIS) and the (near-)infrared (NIR) part of the spectrum, the associated duration of a single optical cycle (e.g. 2.67 fs for 800-nm-carrier wave) imposes a fundamental lower limit to the pulse duration. Further extension of the achievable time resolution to the attosecond (as, $10^{-18}$ s) frontier hence requires a different approach, namely the transfer of the ultrafast technology to shorter wavelengths via non-linear frequency up-conversion.

* hwoerner@ethz.ch

Historically, this requirement entailed up-scaling of the energy of the femtosecond pulses by three orders of magnitude from the nJ- to the mJ-energy scale in order to reach the necessary laser field intensities. A multitude of techniques were developed toward meeting this goal, among those are the introduction of titanium-sapphire (Ti:Sa) as a broadband solid-state laser medium (Moulton, 1986), the Kerr-lens mode-locking (KLM) (Keller et al., 1991; Spence et al., 1991), the invention of semiconductor saturable absorbers (Sutter et al., 1999), and the chirped-pulse amplification (CPA) scheme (Maine et al., 1988; Strickland and Mourou, 1985). The latter development had far-reaching implications for fields not bearing direct relationships to fundamental research (i.e. industry/medicine) and was honored with the Nobel prize in physics in 2018. At the same time, the discovery of high-harmonic generation (HHG) (McPherson et al., 1987), to be discussed in detail in one of the subsequent sections, a highly non-linear non-perturbative process involving the up-conversion of the frequency of VIS or NIR laser pulses to a multiple of their frequency, provided a fully coherent, bright, broadband source of radiation in the vacuum- and extreme-ultraviolet (VUV-XUV) to soft-X-ray (SXR) ranges. Further developments, which will be treated in detail in Section 3, consolidated the role of HHG-based methods for the generation of attosecond pulse trains or even isolated attosecond pulses and the measurement and the control of the sub-cycle electric field evolution. These developments represent the tools necessary for the real-time observation and steering of electron dynamics on a sub-fs time scale.

The present review addresses the technical advances and the theoretical developments to realize and rationalize attosecond-science experiments that reveal a new dynamical time scale ($10^{-15}$-$10^{-18}$ s), with a particular emphasis on molecular systems and the implications of attosecond processes for chemical dynamics. After a brief outline of the theoretical framework for treating non-perturbative phenomena in Section 2, we introduce the physical mechanisms underlying high-harmonic generation and attosecond technology. The relevant technological developments and experimental schemes are covered in Section 3. Throughout the remainder of the chapter, we report on selected applications in molecular attosecond physics, thereby addressing specific phenomena mediated by purely electronic dynamics: charge localization in a simple molecule ($H_2$), charge migration and delays in molecular photoionization. A particular focus will be placed on the description of the corresponding experimental methodology. An in-depth discussion of the state-of-the-art theoretical developments can be found in a recent review (Nisoli et al., 2017) as well as in a recent monograph (Vrakking and Lepine, 2019). Further reviews on attosecond science and technology can be found in (Gallmann et al., 2012; Kling and Vrakking, 2008; Kraus and Wörner, 2018a, 2018b; Krausz, 2016; Krausz and Ivanov, 2009; Wörner and Corkum, 2011).

At this point, it is important to outline the new insights that attosecond metrology adds to the field of molecular physics. First and foremost, the majority of the experimental schemes enabling access to sub-fs dynamics rely on strong-field processes, i.e. highly non-linear light-matter interactions. The resulting novel phenomena pose a fundamental challenge to the well-established perturbative framework of quantum mechanics and make advanced theoretical methods indispensable for the rationalization of the experimental outcomes. Second, recalling that (within the simplest version of Bohr's model) the orbiting period of an electron in the hydrogen atom is about 152 as, attosecond techniques allow one to directly address the electronic degrees of freedom. One direction of research follows closely the extension of the operating principle of femtochemistry to the attosecond regime. Generally, femtochemistry studies nuclear rearrangements taking place on a timescale of tens to hundreds of femtoseconds (a vibration associated with the H-H bond has a fundamental period of 7.6 fs) and has further inspired the notion of coherent control, or the steering of reactions into pre-defined pathways. The attosecond analogue of this scheme (termed 'attochemistry' (Lépine et al., 2014; Remacle and Levine, 2006)) exploits the fact that (as far as the concept of an electronic state retains its validity) the potential energy surface (PES) dictating the nuclear motion is formed by the fast-moving electrons. Thus, perturbing the PES in a very specific manner can potentially give rise to a rearrangement of the nuclear framework, providing a steering mechanism solely based on electron dynamics. Third, attosecond technology allows one to address fundamental questions in chemical dynamics. One of them is the possibility to measure the finite duration of a photoionization event: Delays in photoemission ranging from several to several tens of attoseconds have successfully been resolved in atoms (Guénot et al., 2014; Heuser et al., 2016; Klünder et al., 2011; Sabbar et al., 2015; Schultze et al., 2010), surfaces (Cavalieri et al., 2007; Neppl et al., 2012; Okell et al., 2015) and, more recently, in molecules (Huppert et al., 2016). Another exciting aspect concerns the role of many-body interactions, in particular the influence of electron correlation phenomena beyond the mean-field approximation. A particular example that has been widely explored theoretically and experimentally is the charge migration process (Cederbaum et al., 1986; Wörner et al., 2017), or the ultrafast electronic reorganization along the nuclear framework following a prompt ionization event. In general, the study of coherent electron dynamics and its implications for chemical reactivity have become possible only with the advancement of attosecond technology, specifically the generation of (isolated) pulses of attosecond duration. The significant bandwidths of the latter enable the coherent excitation of several electronic states; simultaneously, the short pulse durations ensure that the interaction time is confined within a temporal window below the typical lifetime of excited electronic states.

## 2. Theoretical description of strong-field phenomena

### *2.1 Overview of the basic terminology*

The term "strong-field physics" is broadly used to classify the regime in which the treatment of laser-matter interactions in terms of perturbative non-linear optics is no longer applicable. Within the perturbative picture, the induced non-linear response is quantified in terms of a multipole expansion with respect to the driving field amplitude. For the case of transverse electromagnetic (TEM) fields, this treatment is applicable when the condition $U_\mathrm{p} < \omega_0$ holds, whereby $\omega_0$ denotes the angular frequency of the oscillating field and $U_\mathrm{p}$ stands for the ponderomotive energy. The latter is defined as the cycle-averaged energy associated with the periodic motion of a free electron in a plane-wave field in the frame of reference in which it is minimal. In the commonly encountered case of a linearly polarized laser field with a peak electric-field amplitude $F_0$, $U_\mathrm{p}$ is given by:

$$U_\mathrm{p} = \left(\frac{F_0}{2\omega_0}\right)^2 . \tag{2.1}$$

The latter implies that the ponderomotive energy grows quadratically with increasing wavelength of the driving field. Here and in what follows, unless explicitly stated otherwise, we use atomic units for convenience. In the system of atomic units, the unit of mass is the electron mass, the unit of charge is the elementary charge, the unit of length is the bohr, symbol $a_0$ ($1\ a_0 \approx 53$ pm), and the unit of energy is the hartree, symbol $E_\mathrm{h}$ ($1\ E_\mathrm{h} \approx 4.6\ 10^{-18}$ J). When using atomic units, one usually reports relative quantities, i.e. a length $r$ is given as $r/a_0$, an energy $E$ is given as $E/E_\mathrm{h}$ (see also (Cohen et al., 2008)). The Keldysh parameter $\gamma_K$, which relates the electron's binding energy in its ground state, $E_\mathrm{b}$, with the ponderomotive energy accumulated in the field via the expression:

$$\gamma_\mathrm{K} = \sqrt{\frac{|E_\mathrm{b}|}{2U_\mathrm{p}}} = \frac{\omega_0}{F_0}\sqrt{2|E_\mathrm{b}|}, \tag{2.2}$$

is commonly employed in order to distinguish between perturbative and non-perturbative regimes in the context of strong-field ionization. As such, it has found a wide acceptance, although its universality has been subject to debate (Reiss, 2008a, 2008b). Strong-field phenomena occur in the regime of $\gamma_\mathrm{K} \ll 1$, also known as the "tunneling" or quasi-static regime and characterized by low laser frequencies ($\omega_0$) and/or strong electric field amplitudes $F_0$. The opposite regime is known as the multi-photon ionization (MPI) limit. The reasoning behind this convention can be inferred from the schematic depiction of the two processes presented in Fig. 2.1. Tunneling ionization (TI) is typically modelled on the basis of a static Coulomb-like potential well modified by the strong, quasi-static field. A strong enough field can suppress/lower one side of the Coulomb potential, leading to the formation of a barrier of finite width, s. Fig. 2.1 A. A prerequisite for this situation to arise is that the time scale of the electric-field oscillation has to be slow enough compared to the time required for the electron to cross the tunneling barrier (in a naïve and incorrect classical picture) to permit adiabatic tunneling. Conversely, in weak and/or rapidly oscillating fields, the MPI picture (Fig. 2.1 B) may be more appropriate. For the case of Ti:Sa wavelengths and the typical field strengths encountered under realistic conditions, $\gamma_\mathrm{K} \sim 1$, implying that most experiments reported to date take place in a regime that is best described as intermediate between the two limiting cases.

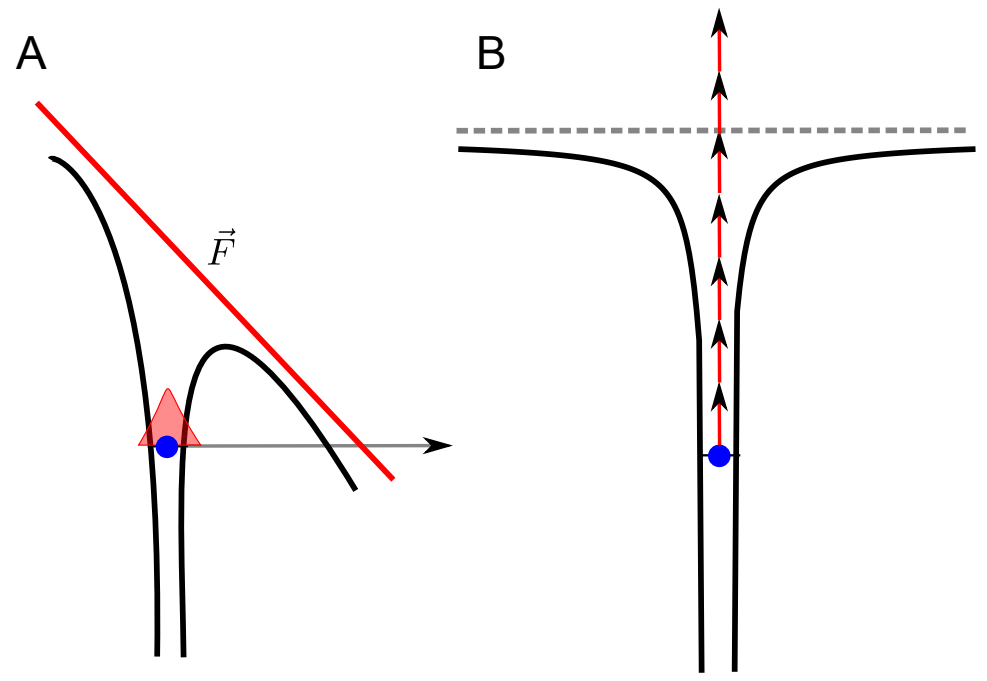

*Figure 2.1: Schematic illustration of the strong-field ionization process in the tunneling (A) and the multiphoton (B) regimes.*

The nomenclature "Keldysh parameter" derives its origin from the 1965 publication of L. Keldysh (Keldysh, 1965) on the topic of the quasi-static limit of strong-field ionization of a one-electron atom. This analysis was subsequently generalized to cover both tunnel and multi-photon limits in the works of Perelomov, Popov, and Terent'ev (also known as the PPT model (Perelomov et al., 1966)). Another popular approach for the quantitative treatment of the limiting case of tunnel ionization rates is the Ammosov, Delone, and Krainov model (Ammosov et al., 1986) (ADK), initially formulated for atoms and subsequently extended to molecules (Tong et al., 2002), or the work of G. Yudin and M. Ivanov (Yudin and Ivanov, 2001) addressing the aspect of non-adiabatic effects in few-cycle fields. Since the subject has been extensively covered in various reviews (Ivanov et al., 2005; Popruzhenko, 2014; Smirnova et al., 2007), we refrain from a more detailed mathematical description and only emphasize the following central result. With exponential accuracy, the total tunnel ionization rate is given by:

$$w_{\mathrm{TI}} \propto \exp\left(-\frac{2(2|E_{\mathrm{b}}|)^{3/2}}{3|E_0|}\right). \tag{2.3}$$

The exponential dependence of the TI rate on the strength of the applied field implied by the above equation has far-reaching consequences for the intensity requirements for technological applications of attosecond physics (cp. Section 3).

*2.2 Electric-dipole approximation and gauge invariance*

The theoretical descriptions of strong-field phenomena often employ the electric-dipole approximation (EDA, also simply referred to as "dipole approximation"). The latter amounts to neglecting the spatial variation of the propagating TEM field, which for a plane-wave treatment is governed by the phase factor:

$$\omega_0 t - \overrightarrow{k_{\mathrm{p}}} \cdot \vec{r} \approx \omega_0 t, \tag{2.4}$$

where $\overrightarrow{k_{\mathrm{p}}}$ denotes the wave vector of the field. This assumption is justified in the case where the wavelength of the TEM field significantly exceeds the size of the system it interacts with. The direct consequence of Eq. (2.4) is the vanishing of the spatial dependence of the vector potential $\vec{A}(\vec{r},t) \approx \vec{A}(\vec{0},t) \equiv \vec{A}(t)$ associated with the TEM field as well as its magnetic-field component: $\vec{B}(\vec{r},t) = \vec{\nabla} \times \vec{A}(\vec{r},t) \approx 0$. These approximations are well-justified for most realistic experimental conditions, generally, the breakdown of the EDA is reached at sufficiently high intensities ($I \sim 10^{15}$ W/cm$^2$ for Ti:Sa wavelengths) or extremely long driving wavelengths (Ludwig et al., 2014).

At this point, we wish to address one specific aspect of the theoretical description of non-perturbative phenomena, namely the gauge invariance (Bandrauk et al., 2013; Reiss, 1980). Whereas the choice of the gauge is of no particular importance when describing the perturbative response, it can have far-reaching consequences for the theoretical or numerical results when dealing with strong-field phenomena. There are two commonly employed theoretical frameworks, referred to as length or velocity gauge, both of which represent special cases of the EM Coulomb gauge ($\vec{\nabla} \cdot \vec{A}(\vec{r},t) = 0$). Below we briefly discuss these two approaches in the context of the Schrödinger equation for a free particle in an electromagnetic field. The latter problem poses the foundation of the strong-field approximation (Lewenstein et al., 1994) (SFA), which forms the starting point for the interpretation of a plethora of attosecond phenomena, including strong-field ionization, high-harmonic generation and above-threshold ionization. The SFA will be covered in detail for the case of HHG in Section 2.4.

In its essence, the velocity gauge is the Coulomb gauge from electromagnetism, whereby the EDA has been imposed: $\vec{A}(\vec{r},t) \rightarrow \vec{A}(t)$. The term "velocity" originates from the fact that the associated Hamiltonian contains the term $\vec{p} + \vec{A}(t)$, which is proportional to the kinematic momentum $\vec{k}(t)$ and thus to the velocity. Here and in what follows, the symbol $\vec{p}$ will be used to denote the canonical momentum, which is a conserved quantity. The corresponding Hamiltonian for a free electron in an EM field reads in the velocity gauge:

$$i\,\partial_t|\Psi^{\mathrm{V}}\rangle = \frac{1}{2}\left[-i\vec{\nabla} + \vec{A}(t)\right]^2 |\Psi^{\mathrm{V}}\rangle. \tag{2.5}$$

The associated solution can be obtained in a straightforward manner by switching to the momentum-space representation:

$$\Psi^{\mathrm{V}}(\vec{r},t) \equiv \langle \vec{r}|\Psi^{\mathrm{V}}\rangle = \frac{1}{(2\pi)^{3/2}} \exp\left\{ i\vec{p}\cdot\vec{r} - \frac{i}{2}\int_{-\infty}^{t} \left[\vec{p} + \vec{A}(t')\right]^2 dt' \right\} \tag{2.6}$$

and is also known as the Volkov solution.

Within the length gauge, which is defined within the EDA, a transverse field is expressed in terms of the scalar potential:

$$\varphi(\vec{r},t) = -\vec{r}\cdot\vec{F}(t). \tag{2.7}$$

The above equation implies that the (transverse) plane-wave laser field is treated as a quasi-static, longitudinal field. The length gauge gives rise to many intuitive interpretations of strong-field phenomena. For instance, the fact that the laser field and the scalar potential reduce to additive scalars in a one-dimensional treatment gives rise to the tunneling picture illustrated in Fig. 2.1 A. The free-electron Schrödinger equation in this gauge becomes:

$$i\,\partial_t|\Psi^{\mathrm{L}}\rangle = \left[\frac{1}{2}\left(-i\vec{\nabla}\right)^2 - \vec{r}\cdot\vec{F}\right]|\Psi^{\mathrm{L}}\rangle. \tag{2.8}$$

Since there is only a scalar potential in the length gauge, one cannot establish a direct relationship between the above equation and the Volkov solutions defined by Eq. (2.6), which are defined for a transverse field. In order to circumvent this problem, one resorts to a gauge transformation of the velocity-gauge solution $\Psi^{\mathrm{V}}(\vec{r},t)$. For this purpose, the velocity-gauge solution in Eq. (2.6) is multiplied by the transformation factor

$$\exp\left(i\vec{r}\cdot\vec{A}(t)\right), \tag{2.9}$$

which yields the length-gauge Volkov wavefunctions:

$$\Psi^{\mathrm{L}}(\vec{r},t) = \frac{1}{(2\pi)^{3/2}} \exp\left\{ i\left(\vec{p} + \vec{A}(t)\right)\cdot\vec{r} - \frac{i}{2}\int_{-\infty}^{t} \left[\vec{p} + \vec{A}(t')\right]^2 dt' \right\}. \tag{2.10}$$

### *2.3 The three-step model of high-harmonic generation*

Having defined the basic theoretical framework, we proceed with the description of the high-harmonic generation process, first in semi-classical terms (Section 2.3), and then within the quantum-mechanical framework of the strong-field approximation. The emphasis on HHG in this short overview is dictated by the central importance of this process for the development of attosecond science and technology, mainly through its application as a table-top source of bright, coherent radiation in the XUV and the soft-X-ray ranges, and in the generation of attosecond pulses. HHG is a parametric process which occurs in the course of the interaction of a strong ultrashort (< ps) laser field (of frequency $\omega_0$) with matter and leads to the emission of high-frequency radiation. The emitted intensity spectrum features three characteristic regions: 1) an abrupt decrease spanning 2-3 orders of magnitude for the first few harmonics, followed by 2) a "plateau" region, where the intensity of subsequent harmonic orders is nearly constant, and 3) a rapid falloff ("cut-off-region"). In centrosymmetric media, the harmonic comb consists of odd harmonic multiples ($n\omega_0, n \in \mathcal{N}_+$) of the driving field. Most of the underlying physics can be explained with the aid of a simple, non-perturbative approach introduced by P. Corkum, K. C. Kulander and K. Schafer in 1993 (Corkum, 1993; Kulander et al., 1993; Schafer et al., 1993), which became known as the "three-step model" (TSM).

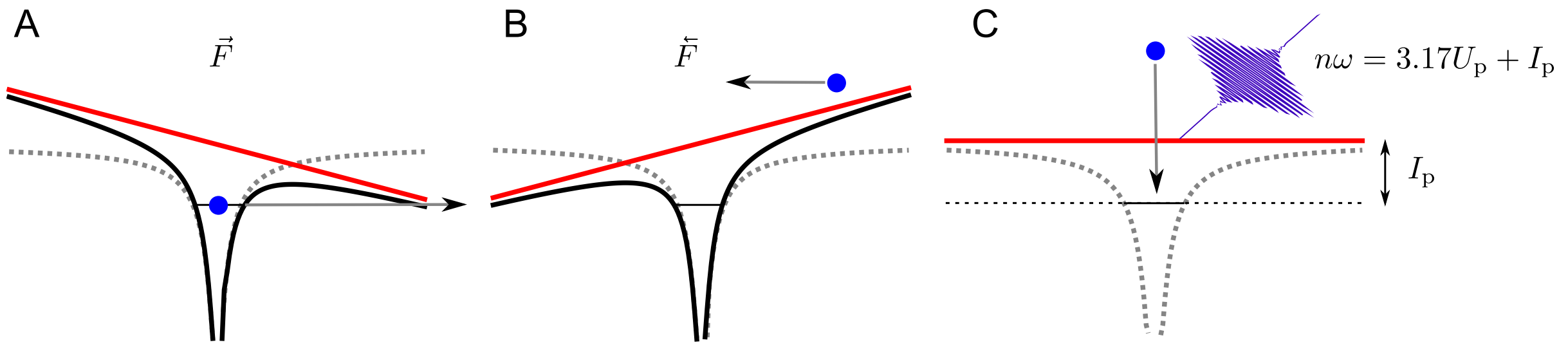

*Figure 2.2: Schematic illustration of each of the three processes constituting the three-step model: (A) ionization, (B) propagation, and (C) recombination. The approximate location of the cutoff is indicated in pane (C).*

The TSM model is summarized graphically in Fig. 2.2. First, the strong field, implicitly treated in the quasi-static limit, distorts ("bends") the Coulomb barrier, leading to the formation of a potential barrier of a finite width. The subsequent tunneling process can release an initially bound electron into a state belonging to the continuum. In the second step, the dynamics of the liberated electron is essentially governed by the laser field, while the Coulomb potential is assumed to represent a minor perturbation. The electron is accelerated and driven away from the core, after which the field switches its sign (for a linearly polarized field, this occurs after $1/4^{\text{th}}$ of an optical cycle), it decelerates and is driven back to the origin. In the ensuing recombination, a pulse of high-energy radiation is emitted, which forms a part of the harmonic comb. The generation of a highly energetic photon is only one (and by far not the most probable one) of several strong-field processes that take place in parallel, among which laser-induced electron diffraction, re-scattering, non-sequential double ionization etc. Depending on the exact sub-cycle timing of the ionization event, some electrons are driven away by the field instead of being accelerated back.

Despite its simplicity, the intuitive picture implied by the TSM enables one to explain some characteristic experimental observations associated with HHG from a purely classical perspective. By treating the electron motion in the second step as a classical laser-driven acceleration of an initially stationary electron with zero velocity, the trajectories of the accelerated electrons can be obtained from integration of Newton's equation of motion by applying the constraint that the electron returns to its original position ($\vec{r}(t') = 0$) at a certain recombination time $t'$. For a monochromatic field linearly polarized along the $\hat{x}$-direction, i.e. $\vec{F}(t) = F_0 \cos(\omega_0 t)\,\hat{x}$, the classical trajectory $\vec{r}(t) \equiv x(t)$ is given by:

$$\ddot{x}(t) = -F_0 \cos(\omega_0 t) \tag{2.11a}$$

$$x(t) = \frac{F_0}{\omega_0^2}(\cos(\omega_0 t) - \cos(\omega_0 t')) + \frac{F_0}{\omega_0}\sin(\omega_0 t')(t - t')\,. \tag{2.11b}$$

The above classical interpretation accounts for the observed position of the harmonic cutoff at ≈ $3.2U_\mathrm{p} + I_\mathrm{p}$. Further, it allows one to map the kinetic energy at the return time, $\frac{\dot{x}^2(t)}{2}$ (and thus the photon energy of the emitted radiation) to a given electron trajectory and thus to a given travel time of the electron in the continuum. For the classical case, restricted to recombination events within one cycle, this mapping is not unique, as there is a pair of trajectories that map to the same photon energy, as evident from Fig. 2.3. These two "branches" of solutions are associated with families of "long" and "short" trajectories, which refers to the length of their transit time in the continuum. The short trajectories originate from ionization after $t'/T_0 \sim 0.05$, whereas the long ones begin prior to this instant. The harmonic emission associated with these two trajectories is characterized by distinctly different properties. In particular, it implies that the emitted energy changes with time on a sub-cycle scale, giving rise to an intrinsic chirp of the high-harmonic emission, termed "atto-chirp". As evident from Fig. 2.3, the long and the short branches have opposite chirps (negative and positive, respectively). The two contributions can be differentiated in an experiment by exploiting macroscopic phase matching considerations owing to their different divergence properties. The latter enables a unique mapping of the harmonic energy to a given transit time of the electron in the continuum. As a result, ultrafast dynamics taking place between the two instants will be mapped onto the properties of the high-harmonic emission, i.e. intensity, polarization and ellipticity. This constitutes the self-probing aspect of high-harmonic spectroscopy (HHS). Moreover, recombining trajectories are characterized by durations of a fraction of a laser cycle, or several hundreds of attoseconds for a Ti:Sa-wavelength driver, thus endowing HHG with a sub-cycle temporal resolution in spite of the fact that the process is driven by multicycle (femtosecond) pulses.

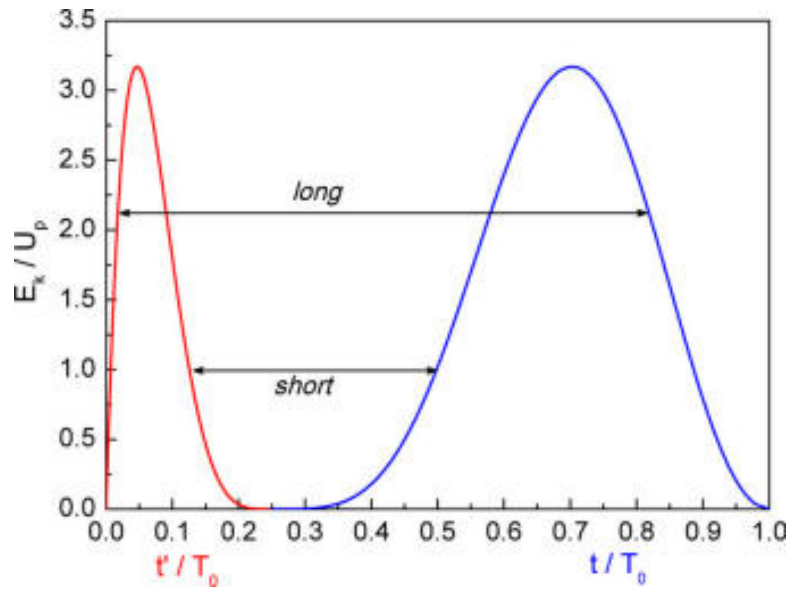

*Figure 2.3: Kinetic energy (in units of $U_p$) of the re-colliding electron at the instant of recombination, as a function of the ionization time t' (red curve, in units of the laser period $T_0$) and the recombination time t (blue curve). The ionization/recombination times of the two trajectory branches are indicated with black arrows. Figure adapted from Ref. (Nisoli et al., 2017).*

### *2.4 High-harmonic generation within the strong-field approximation*

Despite its remarkable success, the TSM does not allow for a more quantitative analysis of the HHG spectra in terms of its specific features, e.g. the intensity distribution of individual harmonics. We therefore conclude this section by outlining the quantum-mechanical treatment of HHG within the strong-field approximation. The starting point of the derivation of the HHG amplitude is the semi-classical length-gauge Schrödinger equation within the single-active electron approximation (SAE):

$$i\,\partial_t|\Psi(t)\rangle = \hat{H}(t)|\Psi(t)\rangle = \left[\frac{1}{2}\left(-i\vec{\nabla}\right)^2 + \hat{V}(\vec{r},t)\right]|\Psi(t)\rangle. \tag{2.12}$$

It will prove useful to separate the Hamiltonian into a field-free part $\widehat{H_0} = \frac{\overrightarrow{p^2}}{2} + U(\vec{r})$ and the perturbation induced by the time-dependent classical field: $V_{\mathrm{L}}(t) = -\vec{r}\cdot\vec{F}(t)$:

$$i\,\partial_t|\Psi(t)\rangle = \left[\frac{\overrightarrow{p^2}}{2} + U(\vec{r}) + V_{\mathrm{L}}(t)\right]|\Psi(t)\rangle = \left[\widehat{H_0} + V_{\mathrm{L}}(t)\right]|\Psi(t)\rangle. \tag{2.13}$$

where $U(\vec{r})$ denotes the effective SAE potential. The time evolution of the eigenstates of the system in the presence of the field ($|\Psi(t)\rangle$) from $t'$ to $t$ can be expressed with the aid of the propagation operator $U(t,t')$:

$$|\Psi(t)\rangle = \hat{U}(t,t')|\Psi(t')\rangle. \tag{2.14}$$

Employing the Dyson series expansion, one arrives at the following formal expression for the propagator:

$$\hat{U}(t,t') = \widehat{U_0}(t,t') - i\int_{t'}^{t}\hat{U}(t,t'')V_L(t'')\widehat{U_0}(t'',t')\mathrm{d}t'', \tag{2.15}$$

where $\widehat{U_0}(t,t')$ denotes the propagator associated with the system in the absence of the external field. The formal solution for the perturbed wavefunction $|\Psi(t)\rangle$ then becomes:

$$|\Psi(t)\rangle = \widehat{U_0}(t,t_0)|\Psi(t_0)\rangle - i\int_{t_0}^{t}\hat{U}(t,t')V_L(t')\widehat{U_0}(t',t_0)|\Psi(t_0)\rangle dt', \tag{2.16}$$

which is exact in terms of the single-active-electron treatment. The first assumption that constituted the SFA is to truncate the full set of bound-state wavefunctions to a single component only, i.e. $|\Psi(t_0)\rangle \equiv |\Psi_0(t_0)\rangle = e^{-iE_{\mathrm{b}}t_0}|\Psi_0\rangle$. Next, the core ansatz that constitutes the essence of the SFA is the substitution of the full propagator $\hat{U}(t,t')$ in Eq. (2.16) by the propagator pertaining to a free particle in an EM field (cp. Eq. (2.6)), the so-called "Volkov propagator" $\widehat{U_{\mathrm{V}}}(t,t')$:

$$\widehat{U_{\mathrm{V}}}(t,t') = \int \mathrm{d}^3\vec{p}\,|\Psi_{\mathrm{V}}(\vec{p},t)\rangle\langle\Psi_{\mathrm{V}}(\vec{p},t')|, \tag{2.17}$$

where $|\Psi_{\mathrm{V}}(\vec{p},t)\rangle$ is the length-gauge Volkov function. The spatial representation of the latter can also be cast in the form:

$$\Psi_{\mathrm{V}}^{\vec{p}}(\vec{r},t) = \langle\vec{r}|\Psi_{\mathrm{V}}(\vec{p},t)\rangle = e^{i[\vec{p}+\vec{A}(t)]\cdot\vec{r}-i\vec{S}(t)} \equiv \langle\vec{r}|\vec{p}+\vec{A}(t)\rangle e^{-iS(t)} \tag{2.18}$$

with the dependence on the canonical momentum denoted explicitly. $S(t) = \frac{1}{2}\int^{t}\left[\vec{p}+\vec{A}(t')\right]^2\mathrm{d}\,t'$ is the equivalent of the classical action. After inserting the Volkov resolvent into the expression for the time-dependent induced HHG dipole:

$$\vec{d}(t) = \langle\Psi(t)|\vec{r}|\Psi(t)\rangle \tag{2.19}$$

and neglecting the continuum-continuum couplings, one arrives at:

$$\vec{d}(t) \propto -i\int_{-\infty}^{t}\mathrm{d}\,t'\int \mathrm{r}d^3\vec{p}\,e^{-iS(\vec{p},t,t')}\langle\Psi_0|\vec{r}|\vec{p}+\vec{A}(t)\rangle\langle\vec{p}+\vec{A}(t')|\vec{r}\cdot\vec{F}(t')|\Psi_0\rangle, \tag{2.20}$$

where the phase factor originating from the field-free propagation of the ground state has been incorporated into the semiclassical action (by means of the relation $I_{\mathrm{p}} = -E_{\mathrm{b}}$):

$$S(\vec{p},t,t') = \frac{1}{2}\int_{t'}^{t}\left[\vec{p}+\vec{A}(t'')\right]^2 \mathrm{d}t'' + I_\mathrm{p}(t-t'). \tag{2.21}$$

In the above, $I_\mathrm{p}$ stands for the ionization potential. According to Eq. (2.20), the dipole response can be reduced to an integral involving transition matrix elements between the ground state and the plane-wave continuum. Indeed, the terms in the bra-kets can be related to the dipole couplings associated with the ionization resp. recombination steps of the three-step model:

$$d_\mathrm{ion}\left(\vec{p}+\vec{A}(t)\right) \equiv \langle \vec{p}+\vec{A}(t)|\vec{r}\cdot\vec{F}(t)|\Psi_0\rangle \tag{2.22a}$$

and

$$\vec{\mathrm{d}}_\mathrm{rec}\left(\vec{p}+\vec{A}(t)\right) \equiv \langle \Psi_0|\vec{r}|\vec{p}+\vec{A}(t)\rangle. \tag{2.22b}$$

Despite this significant reduction in complexity, Eq. (2.20) is still not (easily) amenable to a numerical treatment, due to the presence of the rapidly oscillating factor associated with the semi-classical action. As an alternative route, one can deform the integration path in the complex plane and utilize the steepest-descent method. This approach results in the saddle-point treatment of the SFA, whereby the integral in Eq. (2.20) reduces to a coherent summation over the contributions of a finite set of electron trajectories defined in terms of the saddle points $\left(\overrightarrow{p_\mathrm{st}}^{(j)}, t^{(j)}, t'^{(j)}\right)$ of the semiclassical action (cp. Eq. (2.21)). In the language of the Feynman's quantum-path formalism (Salières et al., 2001), these trajectories are often referred to as "quantum orbits". The HHG amplitude, expressed as the Fourier transform of the time-dependent dipole in Eq. (2.20), then reads:

$$\tilde{\vec{d}}(\omega) = \sum_j \left(\frac{2\pi}{t^{(j)} - t'^{(j)} + i\epsilon}\right)^{3/2} \left[\det\left(\frac{\partial^2 S}{\partial t'^{(j)2}}\right)\right]^{-1/2} \left[\det\left(\frac{\partial^2 S}{\partial t^{(j)2}}\right)\right]^{-1/2} \times$$
$$\vec{d}_\mathrm{rec}\left(\vec{p}_\mathrm{st}^{(j)} + \vec{A}(t^{(j)})\right) \vec{d}_\mathrm{ion}\left(\vec{p}_\mathrm{st}^{(j)} + \vec{A}\left(t'^{(j)}\right)\right) e^{-iS\left(\vec{p}_\mathrm{st}^{(j)}, t^{(j)}, t'^{(j)}\right) + i\omega t^{(j)}}, \tag{2.23}$$

where the additional multiplicative factors result from the contributions of the Hessian determinant. With the above approximation one can easily recover the physical picture contained in the three-step model. It implies that the dipole emission can be decomposed in terms of the contributions from electrons emitted at the ionization instant $t'$ and recombining with the parent system at time $t$, whereby the weight of each trajectory is proportional to the dipole-matrix element terms. One essential difference, however, is the complex character of the QM saddle points, which originate from the contribution of the tunneling process in the first step of the process. These quantum corrections have been confirmed and quantified experimentally using high-harmonic generation in a two-color field (Shafir et al., 2012) (s. Fig. 2.4).

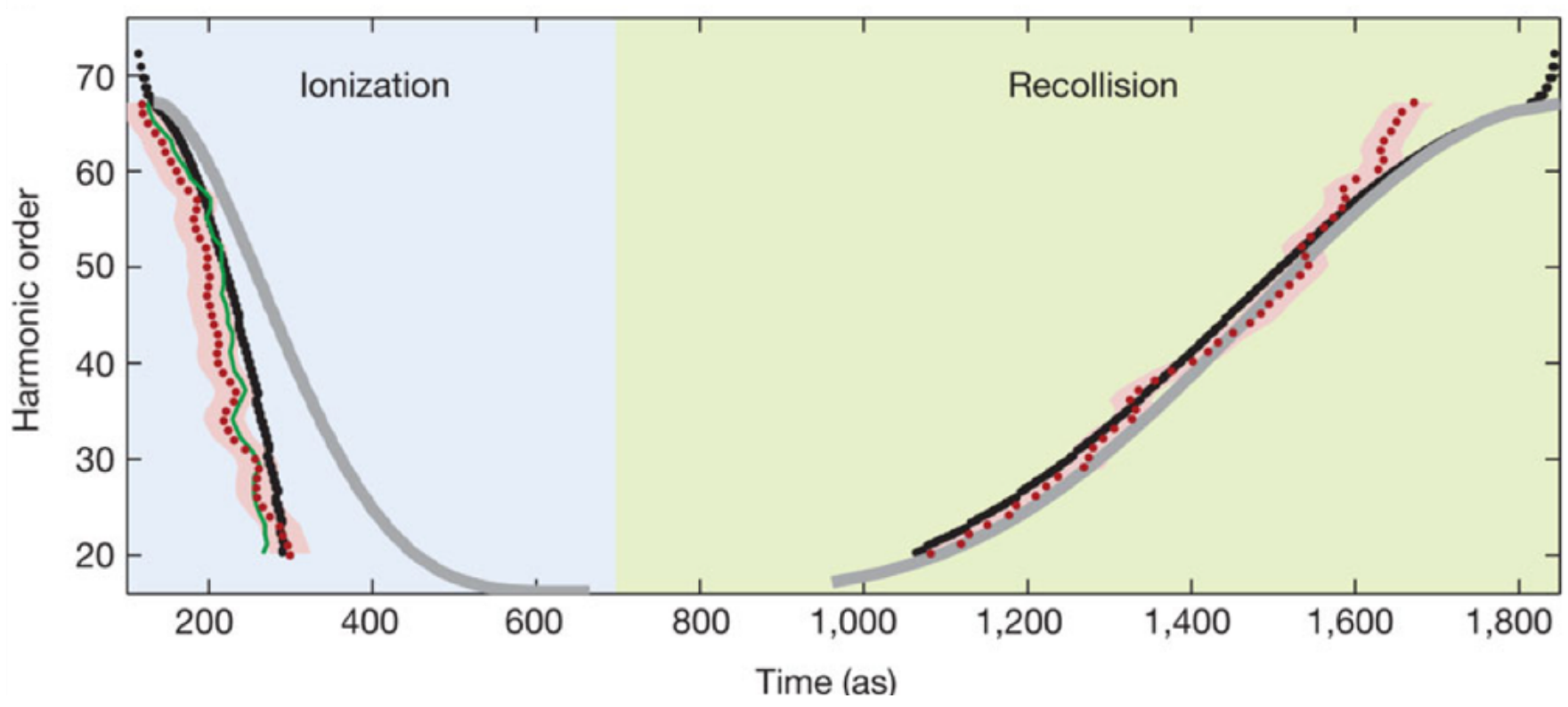

*Figure 2.4: Comparison of the experimentally extracted ionization and recombination times (red dots) with the predictions of the semiclassical model (grey) and the quantum-trajectory solutions (black). Figure adapted from (Shafir et al., 2012) with permission.*

## 3. Attosecond technology

In this section we provide a concise overview of the recent experimental progress in attosecond source development. First, we review the principal methods for generating few-cycle laser pulses with high peak power and then proceed with the description of the various schemes for attosecond pulse generation. Afterwards, we present the operating principle of several dedicated techniques for attosecond spectroscopy experiments. The latter will be covered in a greater detail in the subsequent sections, where we focus on selected applications.

### *3.1 Chirped-pulse amplification*

The invention of the chirped pulse amplification (CPA) technique in 1985 (Maine et al., 1988; Strickland and Mourou, 1985) has had a profound impact on the development of ultrafast laser physics. Within the CPA scheme, the low-energy sub-1ps pulses delivered by a mode-locked oscillator are first chirped and stretched in the time domain to a considerably longer duration (e.g. several hundreds of ps or even ns) by means of a dispersive element such as a grating pair or a long fiber. In the subsequent amplification stage, the energy after passing the gain medium is increased to a mJ- or even to a J-level. At the same time, the peak power is kept below the damage threshold of the optical components and the non-linear pulse distortion limit. Afterwards, a dispersive element (e.g. a second grating pair) is used to compensate the chirp and to re-compress the pulse to a value close to the original pulse duration, thereby enabling peak powers reaching the TW-range and beyond. The CPA technique, in conjunction with the Ti:Sa technology, has undergone multiple improvements throughout the following decades. To date, peak powers up to the PW level are available at large-scale facilities, whereas table-top high-power systems delivering pulses with energies reaching 200 J and 10 fs in duration have been developed (see, e.g. (Chu et al., 2015; Gan et al., 2017), or Ref. (Danson et al., 2015) for a recent review).

*3.2 Carrier-envelope phase stabilization*

A further challenge associated with the realization of spectroscopic techniques relying on isolated attosecond pulses is the control over the carrier-envelope phase of the laser pulse. For a few-cycle pulse with an electric field amplitude profile $E_0(t)$, the instantaneous electric field assumes the form:

$$E(t) = E_0(t)\cos(\omega_0 t + \phi)\,. \tag{3.1}$$

Thereby, the phase offset $\phi$ is referred to as carrier-envelope phase (CEP) and is formally defined as $\phi = \Delta t\ \omega_0$, i.e. the offset between the maximum of the envelope and the position of the nearest field antinode. For pulse durations approaching three cycles or less, the slowly-varying-envelope approximation, commonly employed to simplify the description of ultrashort pulse propagation, breaks down as the CEP starts to influence the pulse propagation and interaction with other media. In a typical Ti:Sa oscillator or amplifier, the CEP strongly fluctuates from shot to shot: whereas the envelope travels with the group velocity through the active medium, the carrier wave is propagated at the phase velocity. The resulting velocity mismatch gives rise to a phase slippage $\Delta\phi$ between carrier and envelope between consecutive pulses separated by the cavity round-trip time. This circumstance is additionally aggravated by energy fluctuations (e.g. in the pump laser, or by Kerr non-linearities in the active material), which cause a variation of $\Delta\phi$ as a function of time. A variety of methods for *(i)* measuring and *(ii)* controlling the pulse-to-pulse CE phase-slip have been developed to address this problem. The earliest approaches for quantifying the CEP slippage occurring in an oscillator were based on interferometric cross-correlation between two consecutive pulses in a pulse train (Xu et al., 1996) but had only limited precision. The adaptation of radio-frequency heterodyne detection schemes (Telle et al., 1999) enabled significant progress in this respect. The most common implementations of the heterodyne scheme include the *f-2f* –interferometry, which is based on heterodyning the fundamental and the second harmonic, or difference-frequency generation (DFG, "*0-f*-interferometry") between different parts of the same frequency comb (see, e.g. Ref. (Fuji et al., 2005)). Thereby, the frequency comb corresponds to the mode-locked pulse train. In case the oscillator output is not octave-spanning, external spectral broadening is necessary, which can be conveniently realized in a photonic crystal fiber medium (PCF) (Dudley et al., 2006). These methods are not directly applicable to amplified pulses, as the repetition rates of the latter lie in the kHz-range or below. In this case, CEP characterization is performed using spectral interferometry (Kakehata et al., 2001; Lepetit et al., 1995; Mehendale et al., 2000) by analyzing the spectra of two different harmonics that overlap temporally. A different method, termed "stereo-above-threshold-ionization" (stereo-ATI) was pioneered by G. Paulus *et al.* (Kreß et al., 2006; Paulus et al., 2001; Sayler et al., 2011).

The control of the CEP, on the other hand, is achieved using either *active* (i.e. based on electronic feedback loops) or *passive* (based on all-optical approaches such as DFG) methods. In its most widespread implementation, the active scheme consists of two dedicated loops, a "fast" and a "slow" one, designed to stabilize the CEP at the output of the oscillator and after the amplification medium. The "fast" loop is based on measuring and stabilizing the phase slippage $\Delta\phi$ using an electronic feedback on an acousto-optic modulator (AOM), thereby adjusting the oscillator pump power and selecting only pulses sharing the same CEP (cp. Ref. (Calegari et al., 2015)). A different method utilizes an acousto-optic frequency-shifter (AOFS) for splitting the oscillator output by diffraction, whereby the zeroth order is used for measuring the CEP, whereas the first order (60-70%) is shifted to compensate the offset and subsequently used for amplification (Koke et al., 2010). The conventional implementation of the "slow loop" is based on the *f-2f*-characterization scheme. Thereby, a portion of the amplifier output is focused into a sapphire plate to generate an octave-spanning supercontinuum and subsequently frequency-doubled in a β-barium-borate (BBO) crystal. The extracted CEP variation $\Delta\phi$ is then used to control the AOM in the oscillator or compensated with the aid of wedges in front of the amplifier. Another scheme using acousto-optic programmable dispersive filter (AOPDF) has been described in Ref. (Canova et al., 2009).
The passive scheme, introduced by Baltuška *et al.* (Baltuška et al., 2002), is based on the all-optical process of DFG between pulses sharing the same CEP variation. Upon frequency-mixing in a non-linear crystal, the CEP of the generated wave cancels out (up to a constant phase factor). Most commonly, the passive scheme is exploited for CEP stabilization of the idler pulse produced in the process of optical parametric amplification (see Section 3.4).

### *3.3 Pulse post-compression techniques*

On the basis of the technologies described in the preceding paragraphs, present-day commercial Ti:Sa-based systems routinely deliver CEP-stale pulses of 20 fs duration, with a central wavelength around 800 nm and at an energy in the mJ range. Generation of isolated pulses for HHG with high peak intensities requires further pulse compression to the few-cycle regime (i. e. < 8 fs) in a post-compression stage. The latter is most commonly realized by means of spectral broadening via self-phase modulation (SPM) in a gas-filled hollow-core fiber (HCF), followed by re-compression in a chirped-mirror assembly for ultrabroadband dispersion compensation. Owing to the long interaction lengths employed (~ 1 m) and the efficient coupling (achieved by using HCF inner diameters on the order of 200-300 μm), significant SPM levels can be maintained even at moderate gas pressures of 200-300 mbar. First demonstrated in 1996 (Nisoli et al., 1996), when it was used to compress 140 fs pulses of 0.66 mJ energy down to 10 fs (0.24 mJ), HCF-based pulse compression is now routinely employed to yield sub-5 fs pulse durations. Thereby, the optical cycle of the carrying wave (2.66 fs) represents a fundamental lower bound on the realizable pulse compression. In the initial version, the available energies were limited to the mJ-level. A further increase requires the implementation of a differential pumping scheme and a pressure gradient (Suda et al., 2005) to mitigate the detrimental effects of ionization. Other schemes rely on the usage of circularly-polarized drivers to suppress the onset of ionization. Using a combination of a pressure gradient and chirped pulses, the generation of 5 fs pulses of 5 mJ energy was reported in 2010 (Bohman et al., 2010). High-energy pulses of up to 13.7 mJ and with a pulse duration of 11.4 fs were obtained using a HCF filled with helium at a low gas pressure (Dutin et al., 2010). Spectral broadening in an HCF has also been extensively used for post-compressing pulses with a central energy red-shifted compared to the typical Ti:Sa output, such as the ones generated using the OPA scheme. In the spectral range around 1.8 μm and above, one can conveniently exploit the negative dispersion properties of bulk materials (e.g. fused silica) in order to shorten the pulse duration down to 2-3 cycles (Schmidt et al., 2011, 2010), thus circumventing the necessity for employing chirped mirrors.
Another related post-compression technique relies on filamentation (Hauri et al., 2004) by loose focusing of an intense laser pulse in a gas-filled cell, whereby argon is conventionally used as the generation medium. As a consequence of the dynamical equilibrium between self-focusing (due to the Kerr effect) and defocusing (due to ionization and plasma generation), the generated filament (typically 100 μm in diameter) can propagate over distances up to several tens of centimeters, thereby undergoing self-compression.

A conceptually different technique for generating optical pulses (i.e. covering the VIS region) with sub-fs duration was presented by Wirth *et al.* in 2011 (Wirth et al., 2011). In this scheme, termed "light-wave synthesizer", a broad supercontinuum (260-1100 nm) generated in a hollow-core-fiber is first decomposed into three spectral regions (UV: 350-500 nm, VIS: 500-700 nm, and NIR:700-1100 nm) with the aid of dichroic beam-splitters. After compressing each spectral region with the aid of dedicated chirped mirrors, the individual components are re-combined. By using active stabilization in combination with dispersive elements controlling the CEP in each arm, the electric field of the light wave can be modulated on a sub-cycle time scale. In a later version of the synthesizer (Hassan et al., 2016), a fourth arm covering the deep UV-region (DUV: 260-350 nm) was added, which allowed for the synthesis of optical attosecond pulses with a duration of 415 as, as confirmed by the streaking measurements (s. Section 3.6.1).

### *3.4 Attosecond sources in the mid-infrared*

Ti:Sa-based femtosecond sources are limited around 800 nm. Under typical experimental conditions, HHG driven by pulses in this wavelength range (NIR) yields access to photon energies of only up to 100 eV, thus precluding the investigation of electron dynamics of core-excited electrons. The demand for attosecond pulses with higher photon energies is mainly dictated by the high biological relevance of the spectral region between 280 eV and 543 eV, where the absorption edges of the main constituents of organic matter, carbon (K-edge, 284 eV), oxygen (K-edge, 543 eV) and nitrogen (K-edge, 397 eV) are located, whereas water (i.e. the cell environment) is transparent. On a different note, photon energies in the keV range are also required for the investigation of ultrafast processes in solid-state systems with X-ray absorption spectroscopy employing elements such as Ti, Cu, Al, Si (Seres et al., 2006).

One way to up-scale the photon energy involves increasing the laser intensity, thereby reaching the conditions of strong ionization. Using helium gas as a medium, soft-X-ray harmonics of energies up to 460 eV were reported in 1997 (Chang et al., 1997). By exploiting phase-matching techniques, bright high-harmonic generation reaching the keV range has been demonstrated (Popmintchev et al., 2009). Nevertheless, the photon flux achievable via these methods still does not meet the level required for practical applications, as a consequence of the phase mismatch caused by the generated electron plasma.

The quadratic dependence of the harmonic cutoff on the driving field wavelength (cp. ponderomotive term in Eq. 2.1) provides an alternative route towards increasing the photon energy of the generated harmonics. This fact has therefore spawned the development of mid-IR-sources (MIR) for high-harmonic generation (Heyl et al., 2016; Ishii et al., 2014; Wolter et al., 2015). Conventionally, the optical parametric amplification (OPA), covered in detail in Ref. (Cerullo and De Silvestri, 2003), a wavelength down-conversion method in which a seed pulse is amplified by an ultrafast pump in a non-linear crystal, provides a straightforward route towards ultrashort pulses in the 1.1 – 3.0 μm range. Relying on the OPA technology, extension of high-harmonic generation to the water window was first demonstrated in 2008 (Takahashi et al., 2008), whereby photon energies of up to 300 eV from Ne and 450 eV from He were generated using a 1.5 μm driver. However, the limited pulse energies (up to mJ) resp. peak powers (up to few GW), as well as the considerable output pulse durations (50 fs) pose fundamental limitations for the power-scalability of the OPA method.

Some of the inherent limitations of the OPA process can be overcome in a dual-chirp OPA scheme (DC-OPA), whereby properly optimized, chirped seed and pump pulses are utilized. First theoretically proposed in 2011 (Zhang et al., 2011), this scheme has been implemented to generate TW, sub-two-cycle IR pulses in the mid-IR region (Fu et al., 2015). Further, the generation of multi-TW, fs pulses with an energy exceeding 100 mJ was reported (Fu et al., 2018). In this aspect, optical parametric chirped-pulse amplification (OPCPA), first proposed in 1992 (Dubietis et al., 1992), a technology which unifies the OPA and the CPA concepts, has been increasingly gaining in popularity. OPCPA utilizes the main characteristics of the CPA scheme, i.e. stretching in the temporal domain, amplification, and recompression, however, the amplification stage is based on an optical parametric process involving energy transfer from a pump to a seed beam, instead of stimulated emission (as in Ti:Sa amplifiers). This method possesses an octave-spanning amplification bandwidth and is capable of generating sub-10-fs spectrum in the VIS range. Further, the detrimental influence of thermal distortions is considerably mitigated as the instantaneous nature of the OPA process does not require storage of the energy in the amplification medium. Utilizing this scheme, multiple groups have reported the generation of few-cycle pulses at varying central wavelengths: at 0.88 μm (Rothhardt et al., 2012), 2.1 μm (Deng et al., 2012), 3.2 μm (Chalus et al., 2010) or 3.9 μm (Andriukaitis et al., 2011). A comprehensive review of the state-of-the art performance level of OPCPA systems including few-cycle table-top sources has been given by Vaupel *et al.* in Ref. (Vaupel et al., 2013). The OPCPA architecture itself has been reviewed in Refs. (Butkus et al., 2004; Dubietis et al., 2006; Witte and Eikema, 2012). Another recent review on the topic of IR sources can be found in Ref. (Ciriolo et al., 2017). We conclude the discussion of MIR sources with the Fourier-domain optical parametric amplification (FOPA) method, a novel pulse amplification scheme that possesses the potential for overcoming the fundamental limitations of traditional methods (such as the gain bandwidth, or material damage threshold). Within FOPA, the dispersed spectral components of a transform-limited pulse are amplified independently of each other utilizing different non-linear crystals in the Fourier plane. In this manner, simultaneous up-scaling of energy and amplified bandwidth is achieved. The experimental realization of the FOPA concept utilizes a symmetric *4f*-scheme and has been first demonstrated via the generation of CEP-stable, 1.43 mJ pulses at 1.8 μm (Schmidt et al., 2014). A later work reported on the generation of two-cycle (~ 11 fs), 1.8 μm pulses with peak powers reaching 2.5 TW (Gruson et al., 2017).

### *3.5 Generation of isolated attosecond pulses*

As evident from the theoretical discussion in the preceding section, high-harmonic generation driven by sufficiently intense multi-cycle pulses involves the emission of attosecond pulse trains (APTs) consisting of XUV light bursts separated by half a period of the driving field. The first experimental observation of attosecond pulse trains was reported by Paul *et al.* in 2001 (Paul et al., 2001), whereas experiments by Mairesse *et al.* 2003 (Mairesse et al., 2003) revealed the chirped nature of the high-harmonic emission, in agreement with the semi-classical prediction of the three-step model. The generation of isolated attosecond pulses (IAPs) thus requires confining the harmonic generation process to a single emission event within the driving pulse. This possibility was first theoretically investigated by Farkas and Tóth in 1992 (Farkas and Tóth, 1992), who considered the generation of pulses with durations of 30-70 as based on HHG. To date, the most popular methods for generating isolated attosecond pulses can be broadly grouped into *amplitude gating* and *temporal gating* techniques. Both gating schemes necessitate an electric field form which is reproducible from shot to shot, making CEP stabilization a prerequisite. The amplitude-gating approach resides on the spectral filtering of the high-harmonic emission: as the cut-off portion of the harmonic spectrum is generated in the most-intense half-cycle of the driving field, spectral isolation of this region corresponds to an attosecond pulse in the temporal domain. In order to ensure that the high-energy portion of the XUV spectrum is generated only within one half cycle of the driving pulse, strict requirements on the driver pulse duration (< 5-fs) and intensity apply. IAPs with durations as short as 80 as have been produced using the amplitude-gating method (Goulielmakis et al., 2008).

The temporal-gating method, on the other hand, requires appropriate tailoring of the properties of the driving electric field in time (e.g. polarization) in order to confine the HHG event within a single half-cycle. This scheme is much more versatile, and here we limit the discussion to its most popular variants: polarization gating (PG), double-optical gating (DOG), and ionization gating (IG). The polarization-gating method originally proposed by Corkum *et al.* (Corkum et al., 1994) exploits the recollision nature of the HHG event (on a microscopic scale) to confine the emission to a single cycle by manipulating the polarization of the driving pulse. HHG driven by a monochromatic pulse exhibits a pronounced ellipticity dependence: the efficiency of the process drops by a factor of two for ellipticities above 10 %. By temporally modulating the polarization of the driving field from circular to linear and then to circular again, one can thus restrict the XUV emission to the central part of the pulse, where the field is linearly polarized. An all-optical scheme based on the usage of two birefringent elements (a multiple-order quarter wave-plate (QWP) followed by a second zero-order QWP) was proposed by Tcherbakoff *et al.* (Tcherbakoff et al., 2003) and experimentally realized in 2006 with the generation of IAPs of durations reaching 130 as in the 25-50 eV spectral range (Sansone et al., 2006; Sola et al., 2006). The duration of two consecutive XUV bursts imposes an upper boundary for the duration of the linearly polarized segment of the pulse, consequently, driving pulses with durations of sub-2-3 optical periods are required (or ~ 7 fs for 800 nm driving fields). This requirement can be relaxed by breaking the symmetry of the driving field, e.g. by adding a second color with the appropriate pulse energy and phase delay. This is the principle behind the DOG scheme, first demonstrated in Ref. (Chang, 2007). An additional advantage of the DOG technique is the superior generation efficiency related to the lower depletion of the generation medium at the leading pulse edge, as well as the possibility to employ higher field intensities. Generation of XUV-supercontinuum with a FWHM of 200 eV using the DOG scheme in combination with a sub-8-fs 800 nm driver was reported by Mashiko *et al.* in 2009. Other variants of the polarization gating techniques, designed to mitigate the effects of ionization depletion, include generalized DOG (GDOG), whereby the leading and the trailing edges of the pulse have a lower ellipticity (0.5 instead of 1.0) compared to the DOG scheme. This technique allowed the generation of IAPs with durations of nearly 150 as by using 28 fs drivers at 800 nm (Feng et al., 2009). Interferometric polarization gating is another, more sophisticated approach based on the same principle, whereby up to 4 electric field components with properly optimized ellipticities, relative energies and phases are combined. In this manner, XUV-supercontinua supporting 260-as-pulses were generated using driving pulses longer than 50 fs. The high relevance of this method is the ability to employ a loose-focusing geometry, which translates into higher XUV flux (e.g. 20 nJ in the 20-70 eV spectral range (Skantzakis et al., 2009; Tzallas et al., 2007)).

As an alternative to the temporal gate provided by polarization gating, the ionization gate relies on the complete ionization depletion during the leading edge of the pulse, which suppresses HHG emission throughout the rest of the pulse due to phase mismatch. Typically, the temporal gate realized in this manner is not narrow enough to restrict the emission to a single-cycle event, therefore, additional bandpass-filtering is required (Bouhal et al., 1998; Jullien et al., 2008; Pfeifer et al., 2007). Alternatively, on can spatially filter the XUV beam in order to tailor its temporal characteristics (Ferrari et al., 2010). PASSAGE (Timmers et al., 2016), or polarization-assisted amplitude gating, is a technique which allows the generation of IAPs with a central frequency tunable in the range from 50 to 130 eV, and combines amplitude and polarization gates, with the additional advantage of relaxed requirements with regard to the pulse duration of the driver.
In the light of the discussion presented in Section 3.4, recent efforts center on transferring the technology described in the preceding for the case of Ti:Sa pulses to MIR drivers, thus targeting the generation of IAPs in the water window spectral range. IAPs with central energies near the carbon K-edge (248 eV), a pulse duration of 400 as and a spectral bandwidth supporting 30 as pulses were recently reported, whereby a spatiotemporal gating scheme based on wave-front rotation was employed (Silva et al., 2015). The current world record for the shortest pulse (43 as) was generated using a 1.8 μm driver in a xenon target (Gaumnitz et al., 2017). This result superseded the previously achieved pulse durations of 53 as for IAPs in the water-window range (Li et al., 2017).
As a concluding remark, we briefly address the recent development of novel schemes for generating IAPs that do not rely on gating. The lighthouse method first proposed in 2012 (Vincenti and Quéré, 2012) and experimentally demonstrated shortly thereafter (Kim et al., 2013; Wheeler et al., 2012) employs a driving field with a rotating wave front in order to generate high harmonics with a non-negligible pulse-front tilt. As each component of the attosecond pulse train is emitted in a different instantaneous direction, a far-field spatial separation of the individual attosecond pulses composing the train is achieved. The non-collinear optical gating (NOG) scheme, also called angular streaking, introduced in Ref. (Louisy et al., 2015), is based on a similar concept. Thereby, an intensity grating is created in the focus by superimposing two pulses at a small angle. As a result, the attosecond bursts are emitted along the bisector angle of the driving fields, and the emission direction corresponds to the orientation of the driving wavefront. By adjusting the delay between the two drivers, an effective wavefront rotation on a sub-cycle time scale can be achieved, which enables the spatial separation of the emitted XUV bursts in the far-field.

### *3.6 Attosecond spectroscopic techniques*

The final paragraphs of this section are devoted to the presentation of several of the most commonly employed experimental configurations for conducting attosecond time-resolved pump-probe experiments. Currently, the low conversion efficiencies of the HHG process (lying in the $10^{-9}$-$10^{-5}$ range) preclude the application of high-harmonic sources to direct XUV-pump XUV-probe experiments. Therefore, the established approach for studying events occurring on the sub-cycle scale involves the combination of an XUV pulse with an ultrashort optical pulse. The most common realizations of this principle are the RABBITT (reconstruction of attosecond beating by interference of two-photon transitions) and the attosecond streaking schemes, which are reviewed in the following. In both cases, the XUV pump serves to initiate the electron dynamics, which are subsequently probed by the infrared pulse as a function of the mutual delay between both pulses.

#### *3.6.1 Reconstruction of attosecond beating by interference of two-photon transitions*

The RABBIT technique utilizes an APT in combination with a less intense and longer, multi-cycle NIR field, and was initially developed and utilized to characterize the duration of the XUV bursts emitted in the pulse train (Paul et al., 2001). As discussed in Section 2, the frequency-domain manifestation of an APT is given by a harmonic comb consisting of a finite number of peaks separated by twice the frequency of the driving field. In a RABBITT measurement, the XUV photons are detected indirectly through photoemission from a target using the photoelectric effect. Adding a NIR perturbing field (typically derived from the same field used for generating the APT) leads to the emergence of additional maxima in the photoelectron spectrum, shifted by the NIR photon energy with respect to the main harmonic peaks. These peaks, referred to as sidebands, result from the emission or absorption of an NIR photon under the influence of the dressing field in addition to the ionizing XUV photon. The non-distinguishability of the two pathways contributing to a given sideband will result in a modulation of the sideband intensity at a frequency given by twice the frequency of the NIR field. Denoting the (two-photon) amplitudes of the absorption resp. the emission pathway contributing to a given sideband by $M_{\mathrm{a}}$ resp. $M_{\mathrm{e}}$, the proability for the emission of a photoelectron at an energy $E_k = \omega_{\mathrm{XUV}} + \omega_{\mathrm{IR}}$ can be expressed mathematically as a sum of the squared amplitudes of the individual pathways:

$$P_k \approx |M_{\mathrm{a}} + M_{\mathrm{e}}|^2 = |M_{\mathrm{a}}|^2 + |M_{\mathrm{e}}|^2 + 2\mathcal{R}e\{M_{\mathrm{a}}^* M_{\mathrm{e}}\}. \quad (3.2)$$

The total probability thus depends on the relative phase $\phi$ between the two contributing quantum paths through the last term in Eq. (3.2). The latter can be controlled by changing the relative delay $\tau$ between the XUV and the IR, i.e. $\phi \equiv \omega_0 \tau$. The dependence of the amplitudes of the individual pathways on the dressing field is given by $M_{\mathrm{a}} \propto |E_{\mathrm{IR}}| e^{i\omega_0\tau + \phi_{\mathrm{a}}}$ resp. $M_{\mathrm{e}} \propto |E_{\mathrm{IR}}| e^{-i\omega_0\tau + \phi_{\mathrm{e}}}$, implying that the last term in Eq. (3.2) will vary as: $\propto \cos(2\omega_0\tau)$, i.e. at twice the frequency of the NIR pulse. If one considers the two steps in the NIR-assisted XUV photoionization as decoupled from each other, the relative delay can be further decomposed into its individual contributions as follows:

$$\tau = \tau_{\mathrm{GD}} + \tau_{\mathrm{XUV}} + \tau_{\mathrm{cc}}. \quad (3.3)$$

The first term, $\tau_{\mathrm{GD}}$, corresponds to the relative group delay of the XUV relative to the NIR probe. The second and the third terms originate from the intrinsic phases of the individual matrix elements characterizing each pathway. Eq. (3.3) illustrates the utility of the RABBIT technique in characterization measurements. In case the intrinsic phase difference can be calculated from *ab-initio* methods, which holds true for simple atomic systems, the relative phases between consecutive bursts in an APT can be inferred from a RABITT spectrogram, thus providing a direct characterization of the HHG spectrum. Inversely, in case the group delay can be measured by other means or eliminated by adopting a proper reference scheme, the sideband oscillation phases provide a direct access to the phases of the two-photon matrix elements. The latter implementation of the RABBIT scheme is most commonly utilized to infer photoionization delays associated with the absorption of a XUV photon.

### *3.6.2 Attosecond streaking*

The attosecond streak camera is closely related to the RABBITT scheme, with several major differences. First, the photoionization event is initiated with an isolated attosecond pulse instead of a pulse train, which imposes severe requirements on the NIR field in terms of pulse duration and CEP stability. Further, the dressing (or streaking) field, typically derived from the one used for the IAP generation in order to ensure optimal stabilization, is orders of magnitudes stronger than the typical field strengths employed in RABBITT. Consequently, the interaction of the system with the NIR is no longer perturbative as multiple photons of the IR are absorbed. From this reason, one more conveniently resorts to a semiclassical treatment of the dynamics of the liberated electron in the NIR field, which results in the following expression for the electron velocity at time instant $t$:

$$v(t) = -\int^{t} E(t')dt' + [v_0 + A(t_i)] = -A(t) + [v_0 + A(t_i)]. \quad (3.4)$$

In the above, $t_i$ denotes the instant of ionization, $v_0$ is the initial velocity, and $E(t)$ resp. $A(t)$ denote the electric field resp. the associated vector potential. The first term of the right-most side represents the electron's quiver energy which goes to zero when the pulse is over. The motivation behind the term "streak camera" used for this technique can be inferred from the fact that the above equation represents a mapping of a given instant of ionization, $t_i$, or "birth in the continuum", onto a velocity displacement $v(t)$ that is a function of the vector potential of the NIR laser field (Kling and Vrakking, 2008). When recorded over a large set of delays between the IAP and the NIR, one recovers a cross-correlation between the EWP and the NIR field. The dependence on the electric field (through the vector potential $A(t)$) rather than the intensity yields sub-fs temporal resolution, despite the few-fs duration of the streaking field. The ability to reproduce the sub-cycle oscillations of the electric field amplitude renders the streaking technique the most accurate method for characterizing attosecond pulses. Initially introduced as a characterization method ((Hentschel et al., 2001; Itatani et al., 2002)), it was later implemented to infer the underlying attosecond dynamics of the ejected electron. Since an absolute reference for the exact emission time is lacking, one typically records streaking traces originating from electrons emitted from multiple (two or more) electronic shells of the system in a single measurement. The latter is possible given the large bandwidths of the employed APTs. Subsequently one analyzes the small phase shifts between the individual streaking traces, which are interpreted as temporal delays between the photoionization events associated with the underlying electronic shells. Although the spectral overlap resulting from the large IAP bandwidths has prevented widespread application of the streak camera for the photoionization dynamics in molecules, we include this brief description in the current review as this technique has been instrumental in the time-resolving of fundamental events in atomic systems (Drescher et al., 2002; Ossiander et al., 2018, 2016; Schultze et al., 2010; Uiberacker et al., 2007).

### *3.6.3 Photoelectron and photoion spectroscopy*

Prior to concluding the current section, we address another unique aspect of attosecond spectroscopy: besides the unprecedented temporal resolution, the utilization of large-bandwidth pulses in the XUV/SXR ranges offers the possibility for spatial imaging with Ångstrom resolution. The photon energies associated with the attosecond pulses derived from HHG exceed by far the ionization energies of valence electrons, leading to the emission of photoelectrons with very short de-Broglie wavelengths. Recording the angular distribution of the ejected electrons thus gives access to scattering dynamics in the molecular frame (for a recent perspective on the topic, see Ref. (Vrakking, 2014)). In general, in experimental configurations based on the detection of charged particles, i.e. photoelectron or photoion spectroscopy, the dynamics of the investigated process are encoded in the kinetic energy distributions of the detected species. The most commonly employed detection schemes are based on time-of-flight (ToF) electron or ion-mass spectrometry. The usage of magnetic-bottle spectrometers allows for superior detection efficiencies. Velocity-map imaging (VMI) techniques in combination with Abel inversion algorithms are employed in order to detect and retrieve the photoelectron angular distribution in the laboratory frame. In order to achieve a complete reconstruction of the scattering event in the molecular frame, a simultaneous detection of the emitted photoelectrons and photoions in coincidence is required. In this respect, the Cold Target Recoil Ion Momentum Spectroscopy (COLTRIMS) technique provides a way towards kinematically complete scattering experiments (Dörner et al., 2000; Ullrich et al., 2003, 1997).

## 4. Attosecond electron/ion imaging spectroscopy

Given the high photon energy of attosecond pulses, the interaction of attosecond pulses with molecules results in ionization, leaving behind charged particles in potentially highly excited states, which can undergo further relaxation pathways specific to the target. The associated dynamics particularly lends itself to investigation using the techniques of attosecond electron/ion spectroscopy introduced in Section 3.

The examples covered in this section are seminal in attosecond spectroscopy as they constitute the first extension of attosecond pump-probe techniques to molecular systems, thereby addressing the fundamental role of vibronic interactions on the sub-fs localization of the electronic charge distributions in small molecules. The majority of the studies address the paradigmatic case of molecular hydrogen ($H_2$) or deuterium ($D_2$).

In 2010, Sansone *et al.* (Sansone et al., 2010) used the combination of isolated attosecond pulses and waveform-controlled, few-cycle intense NIR fields to track the charge localization dynamics in $H_2$ and $D_2$ following dissociative ionization (DI). Intense-field DI results in the production of $H^+/D^+$ fragments, and their velocity and angular distributions are detected in the experiment. Pronounced angular asymmetries in the fragment momentum distributions induced by the presence of the NIR field encode the spatial localization of the charge distribution and its dynamics.

In the experiment, ionization is induced using intense isolated attosecond EUV pulses with a bandwidth extending from 20 eV to 40 eV. Given the large spectral width of the IAP, the kinetic energy distribution (KED) spectra contain contributions from several ionization channels. The potential energy surfaces of the two lowest electronic states of the cation ( ${}^2\Sigma_g^+(1s\sigma_g)$ and ${}^2\Sigma_u^+(2p\sigma_u)$) that are of relevance for this study are reproduced in Fig. 4.1 A, together with the autoionizing (AI) states ($Q_1$ and $Q_2$) located in this energy region. Direct ionization is the dominant process at photon energies up to 25 eV and leads to the population of the ${}^2\Sigma_g^+(1s\sigma_g)$ -state of the ion, releasing a small fraction of low-KE fragments (s. spectrum in Fig. 4.1 B). Excitation to the doubly-excited $Q_1$ ${}^1\Sigma_u^+$ -states embedded in the continuum prevails at photon energies in the range between 25 and 36 eV, whereby the symmetry of the transition dipole moment (parallel) preferentially selects molecules aligned parallel to the laser polarization axis. These transitions are characterized by large oscillator strengths; subsequent AI leads to the population of the ${}^2\Sigma_g^+(1s\sigma_g)$ state and releases ionic fragments at characteristic kinetic energies between 2 and 7 eV (Fig. 4.1 B). However, due to the large cone opening angle of the spectrometer employed in the experiment, the measured spectra above 31 eV contain contributions from the perpendicular transition, which involves the $Q_2$ ${}^1\Pi_u$ -doubly-excited states. The ionization channel resulting in the population of the ${}^2\Sigma_u^+(2p\sigma_u)$ state of the ion opens at 30 eV. At these photon energies, the ionization process can thus leave the ion in either the ${}^2\Sigma_g^+(1s\sigma_g)$- or the ${}^2\Sigma_u^+(2p\sigma_u)$-state, leading to different fragment energies.

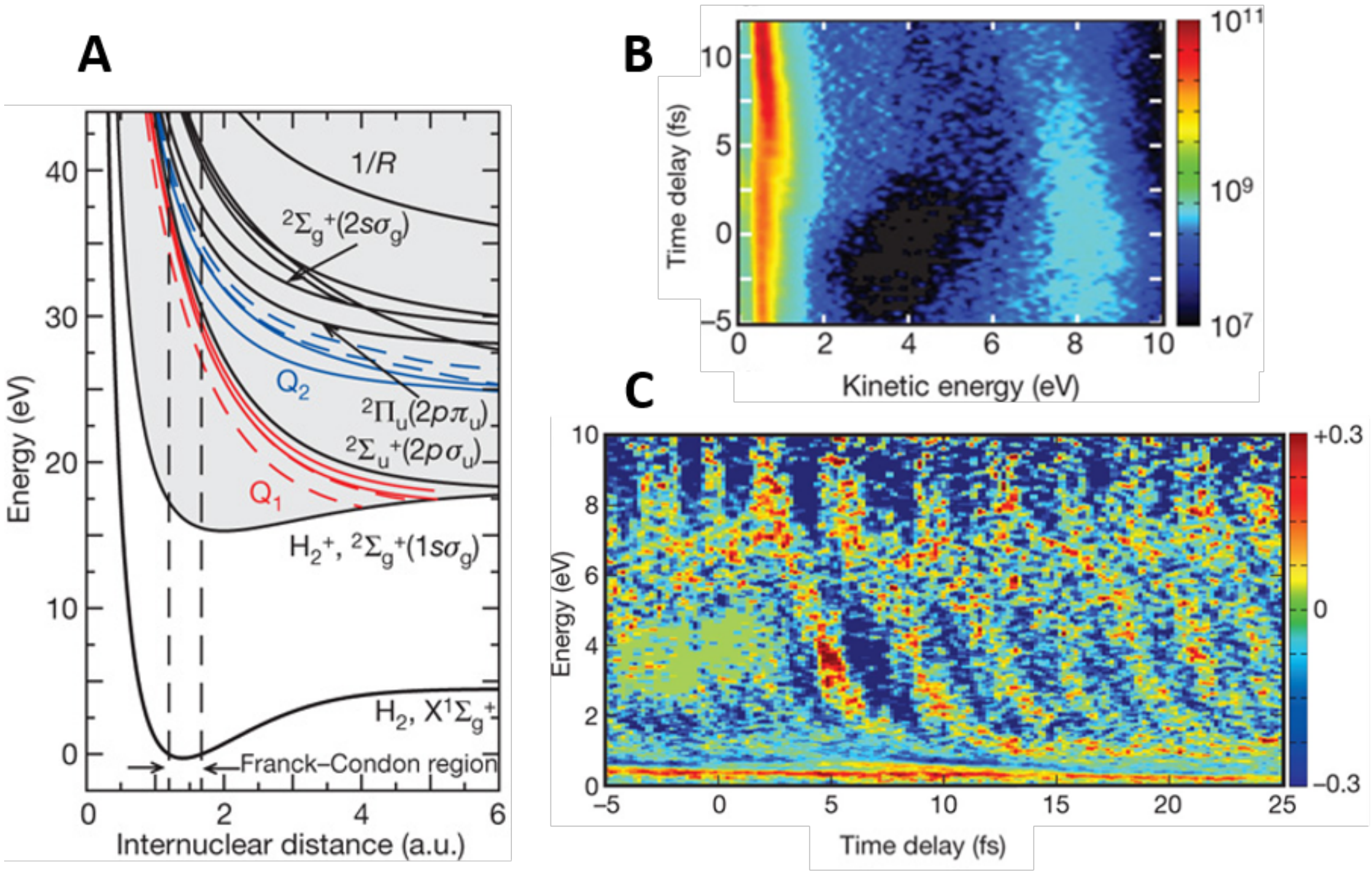

*Figure 4.1: (A) Potential energy curves of the states participating in the dissociative ionization of neutral hydrogen (bottom curve). Photoexcitation results in excitation of the $Q_1$ (red) and $Q_2$ (blue) doubly excited states and ionization to the $1s\sigma_g$ and $2p\sigma_u$ states. R denotes the internuclear distance, reported in atomic units (a.u.). (B) Experimental kinetic energy distributions of the $D^+$-fragments as a function of the delay between the XUV and the NIR pulses. The color scale represents the fragment yield (arb. u.). (C) Experimentally extracted asymmetry parameter associated with the production of $D^+$-fragments in the dissociative ionization of $D_2$ as a function of the kinetic energy and the temporal delay between the two pulses. The asymmetry parameter exhibits pronounced oscillations as a function of time that vary strongly with kinetic energy. Figure adapted from (Sansone et al., 2010) with permission.*

A time-delayed, intense 6-fs NIR pulse subsequently redistributes the charge within the molecular ion, leaving its imprint in the form of an asymmetric angular distribution of the ejected fragments with respect to the laser polarization axis. This asymmetry is quantified via the laboratory-frame asymmetry parameter:

$$A(E_k,\tau) = \frac{N_\uparrow(E_k,\tau) - N_\downarrow(E_k,\tau)}{N_\uparrow(E_k,\tau) + N_\downarrow(E_k,\tau)}, \quad (4.1)$$

where $N_\uparrow(E_k,\tau)$ and $N_\downarrow(E_k,\tau)$ denote the fragments emitted in opposite directions (within 45°) relative to the polarization axis. This parameter bears a direct relationship to the localization of the charge after the ionization event. As evident from Fig. 4.1 C, the asymmetry parameter displays distinct oscillations as a function of the delay between the XUV and the NIR: a shift between the two pulses by half the period of the IR results in a reversal of the charge localization site.

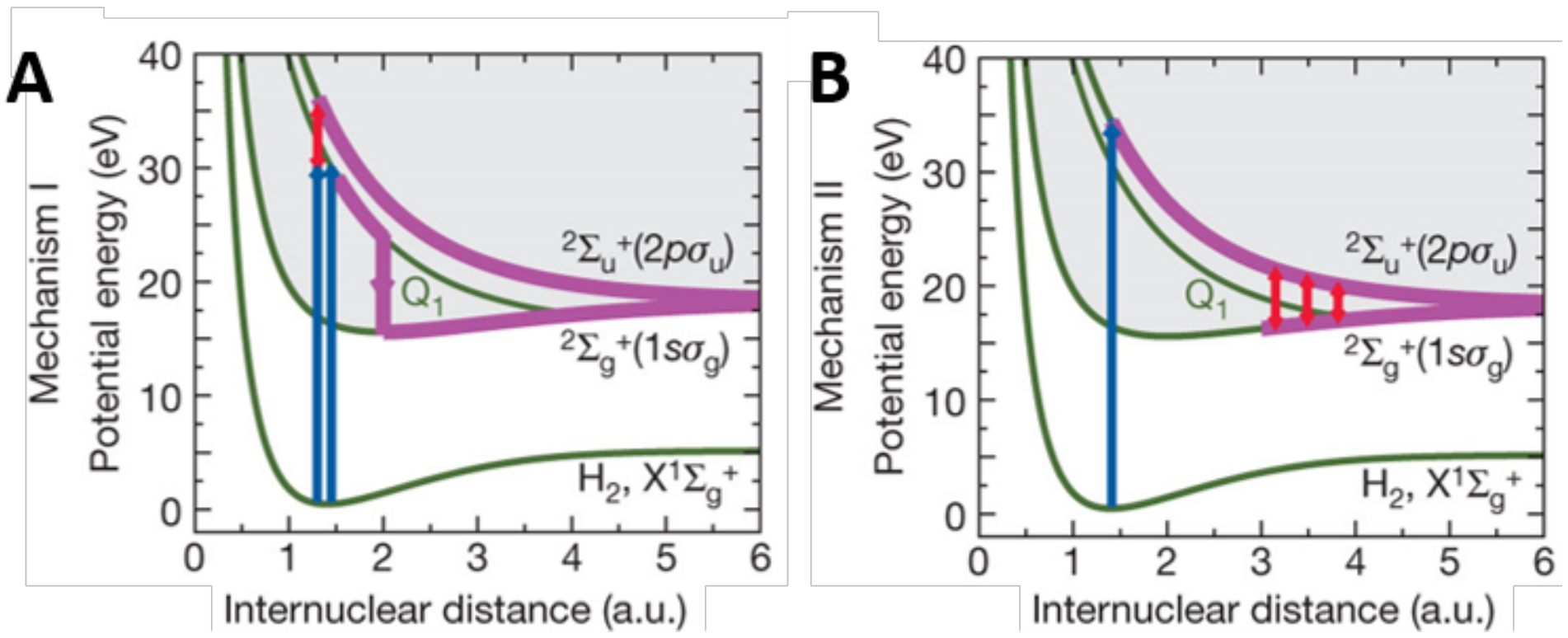

*Figure 4.2: Mechanisms underlying the observed asymmetry in XUV-NIR dissociative ionization. (A) Asymmetry caused by the interference of a wave packet initially launched in the 2p\sigma_u state by direct XUV ionization (blue arrows) and by rapid ionization of the $Q_1\ {}^1\Sigma_u^+$ doubly excited states by the IR pulse with a wavepacket in the $1s\sigma_g$-state resulting from autoionization of the $Q_1\ {}^1\Sigma_u^+$ states. Purple lines signify intrinsic molecular dynamics. (B) Asymmetry caused by the interference of a wave packet prepare directly in the $2p\sigma_u$ state by direct XUV ionization with a wave packet in the $1s\sigma_g$-state resulting from the stimulated emission during dissociation. Figure adapted from (Sansone et al., 2010) with permission.*

In absence of the IR, the electronic states $1s\sigma_g$ and $2p\sigma_u$ that can be accessed via a single-photon absorption of a XUV photon possess opposite, but well-defined parities, which results in symmetric distribution of electrons emitted "up" ($N_\uparrow(E_k,\tau)$) or "down" ($N_\downarrow(E_k,\tau)$). A coherent superposition of the $1s\sigma_g$ and $2p\sigma_u$ states is therefore a prerequisite for the emergence of charge localization. The dipole selection rules governing the single-photon excitation determine the angular momentum symmetry ($l_{g,u}$) of the outgoing photoelectron (*gerade* ($l_g$) in the case of $2p\sigma_u$ and *ungerade* ($l_u$) for $1s\sigma_g$). The two photoelectrons are thus distinguishable, erasing the coherence between the states. This situation changes under the influence of the NIR pulse, which can coherently couple the two states and prepare a coherent superposition. Two coupling mechanisms, illustrated schematically in Fig. 4.2 A and B), that can lead to this scenario are invoked. The first one involves an interference between the autoionization into the $1s\sigma_g$ state proceeding through the $Q_1\ {}^1\Sigma_u^+$ doubly-excited states that is accompanied by the emission of a photoelectron of *p*-symmetry, and the direct ionization of the WP launched on the $2p\sigma_u$-potential energy surface leading to the emission of an *s*-electron. In this scenario, the role of the NIR absorption consists in changing the angular momentum character of the outgoing electron, thereby erasing the link between the parity of the cationic wavefunction and the angular momentum character of the photoelectron. Interaction with the NIR thus leaves the molecular ion and the electron in an entangled state. In the second mechanism, the NIR acts upon the molecular ion in the course of its dissociation on the $2p\sigma_u$-potential, thereby coupling it to the $1s\sigma_g$-surface, forming a coherent superposition between the two states. The asymmetric dissociation is then determined by the relative phase of the two molecular wave functions of opposite parity. This mechanism is particularly efficient at large internuclear distances as the energy separation between the relevant curves approaches one NIR photon.

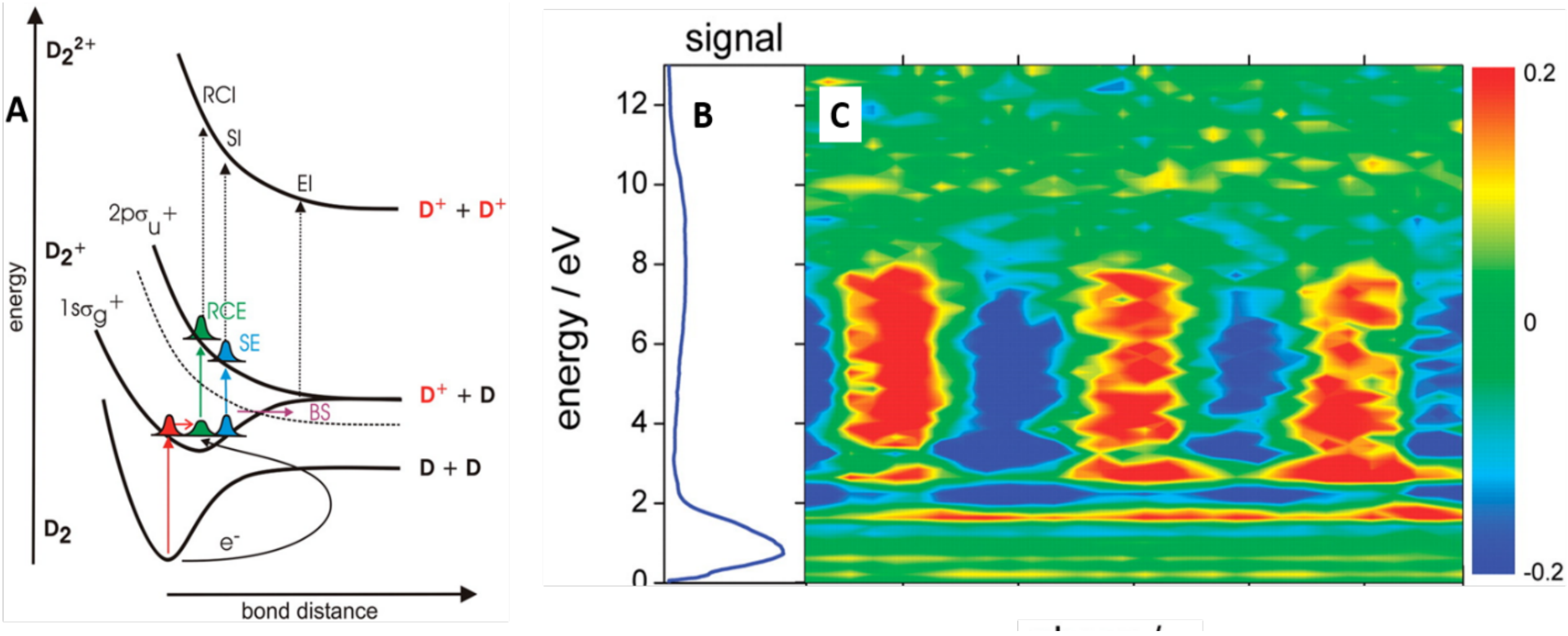

*Figure 4.3: Possible pathways for the production of $D^+$-fragments from $D_2$ wither by photodissociation of the molecular ion (through bond softening (BS), sequential excitation (SE) or recollision excitation (RCE) or by Coulomb explosion (through RCI, SI, EI). (B) Kinetic energy distribution of the $D^+$-fragments originating from the dissociation of $D_2$ induced by a 5-fs , $10^{14}$ W/cm$^2$ laser pulses without CEP stabilization. (C) Asymmetry parameter as a function of the $D^+$ kinetic energy and the CEP phase offset. Figure adapted from (Kling et al., 2006) with permission.*

Prior to this study, electron dynamics evolving on the attosecond timescale have been observed by Kling *et al.*(Kling et al., 2006) in the dissociative ionization of $D_2$ using intense, CEP-controlled NIR (800 nm) few-cycle pulses of duration 5 fs and intensities up to $10^{14}$ W/cm$^2$. As in the preceding study, electron localization is inferred by measuring the $D^+$-fragment kinetic energy and angular distributions in a VMI apparatus, whereby in the current case the asymmetry of the fragments is monitored as a function of the CEP of the pulse. At the field intensities employed in the experiment, ionization of $D_2$ leaves the molecular ion in the $1s\sigma_g$-state, whereas the dynamics of the outgoing photoelectron is (almost completely) governed by the oscillatory laser field. Depending on the instant of the ionization within the NIR laser cycle, the electron can revisit the core in a manner reminiscent of the three-step-model for HHG introduced in Section 2.3. One of the possible outcomes of the recollision event is the excitation of the molecular ion to the higher-lying $2p\sigma_u$-state (cp. Fig. 4.3 A). Due to the repulsive nature of this state, this scenario leads to the ejection of high-kinetic-energy $D^+$-fragments (up to 12 eV), which are observed as a broad peak (8-10 eV) in the KER (cp. Fig. 4.3 B). The recollision nature of the mechanism leading to the production of these fragments is further confirmed by conducting the experiment using a CPL driver, in which case the suppressed recombination probability leads to the disappearance of the features in the energy range above 8 eV. The central outcome of the experiment is the observation of pronounced (a modulation depth of ≈50 %, s. Fig. 4.3 C) periodic oscillations of the asymmetry parameter associated with these recollision fragments as a function of the CEP of the ionizing pulse. In a similar fashion to the case study presented above, the mechanistic origin of the observed modulation resides in the formation of a coherent wave packet with no well-defined parity, leading to an asymmetry of the fragment ejection that depends on the exact phase of the electric field. In this case, the $D_2^+$-molecular ion is formed in its ground state $1s\sigma_g$ around the maximum of the electric field. After an excursion time of up to 1.7 fs, the electron re-collides with the core, promoting the ion to the strongly repulsive $2p\sigma_u$- state, which promptly dissociates, resulting in D/$D^+$ fragments with kinetic energies up to 10 eV. During the dissociation event, the NIR field can induce a partial population transfer back to the ground state, thereby forming a dissociative wavepacket with a large excess kinetic energy. The coherent superposition of these two pathways underlies the observed time-dependent localization of the electron density within the molecule that can be controlled by switching the polarity of the field on a sub-cycle scale by varying the CEP.
In the experiments discussed in the preceding, the observed molecular attosecond dynamics are associated with the relaxation pathways of the ionized particles generated by the action of the attosecond pulse. A similar experimental principle has been transferred by Neidel *et al.* (Neidel et al., 2013) to a series of small-to-medium sized molecules in order to realize a "molecular attosecond Stark spectroscopy". The experimental approach is based on using a two-color pump-probe sequence, whereby a 1 kHz, moderately intense ($10^{13}$ W/cm$^2$) 800-nm-NIR field of 30 fs duration serves as the pump, whereas the probe pulses are provided by a XUV attosecond pulse train derived from harmonic generation in argon. Both pulses are focused with a toroidal mirror in the active region of an VMI apparatus, in a molecular beam of $N_2$, $CO_2$, or $C_2H_4$. The collected TOF spectra of the molecular ion yields reveal pronounced periodic modulations as a function of the pump-probe delay. These oscillations are interpreted in terms of a time-dependent dipole moment induced during the interaction of the molecules with the NIR field in the pump step, which gives rise to a time-dependent modification of the electronic density and thus to a modulation of the XUV photoionization yield of the molecule. The amplitude of the XUV yield oscillations was found to correlate with the magnitude of the component of the polarizability tensor along the internuclear axis $\alpha_z$.
The possibility of attosecond control in the molecular photoionization of H2 using an APT in the place of an isolated pulse has been further explored in a work by Kelkensberg *et al.* (Kelkensberg et al., 2011) Additional experiments analogous to the ones reported in in the first part of this Section have been conducted in other systems such as $O_2$ (Siu et al., 2011).

## 5. Attosecond electron spectroscopy in biorelevant molecules

The current section is dedicated to the investigation of charge transfer (CT) and related phenomena using attosecond spectroscopy. In general, CT describes the spatial redistribution of electronic charge in a molecular system, e.g. the backbone of an extended molecular chain, or between two or more molecules, and is of fundamental importance for photoinduced reactions in biochemistry and biophysics (for a recent review on the topic, s. Ref. (Wörner et al., 2017) and references therein). The CT process is predominantly mediated and influenced by molecular nuclear dynamics. Charge migration (CM), on the other hand, is used to denote the charge transfer driven specifically by electronic coherence and/or electron correlation. As the primary mechanism involves the electronic degrees of freedom, CM takes place much faster, typically on a sub-femtosecond timescale, although the influence of the nuclear motion comes into play provided that the observation time scale is sufficiently long. Whereas CT phenomena have been studied extensively in the past employing various techniques at different temporal resolutions, CM shifted to focus only recently (Calegari et al., 2016), after progress in attosecond technology enabled the generation of attosecond pulses in the EUV regime. Prior to that, CM has been thoroughly investigated from a theoretical perspective, with a particular emphasis on the role of electronic correlations (Breidbach and Cederbaum, 2005, 2003; Hennig et al., 2005; Remacle and Levine, 2006). These efforts were motivated by the experimental work of Weinkauf *et al.* on the fragmentation of small peptides based on natural amino acids (Weinkauf et al., 1995). In these experiments, the initial hole was localized on a given specific site of the molecule, however, analysis of the produced fragments indicated that the fragmentation has occurred at a remote site.

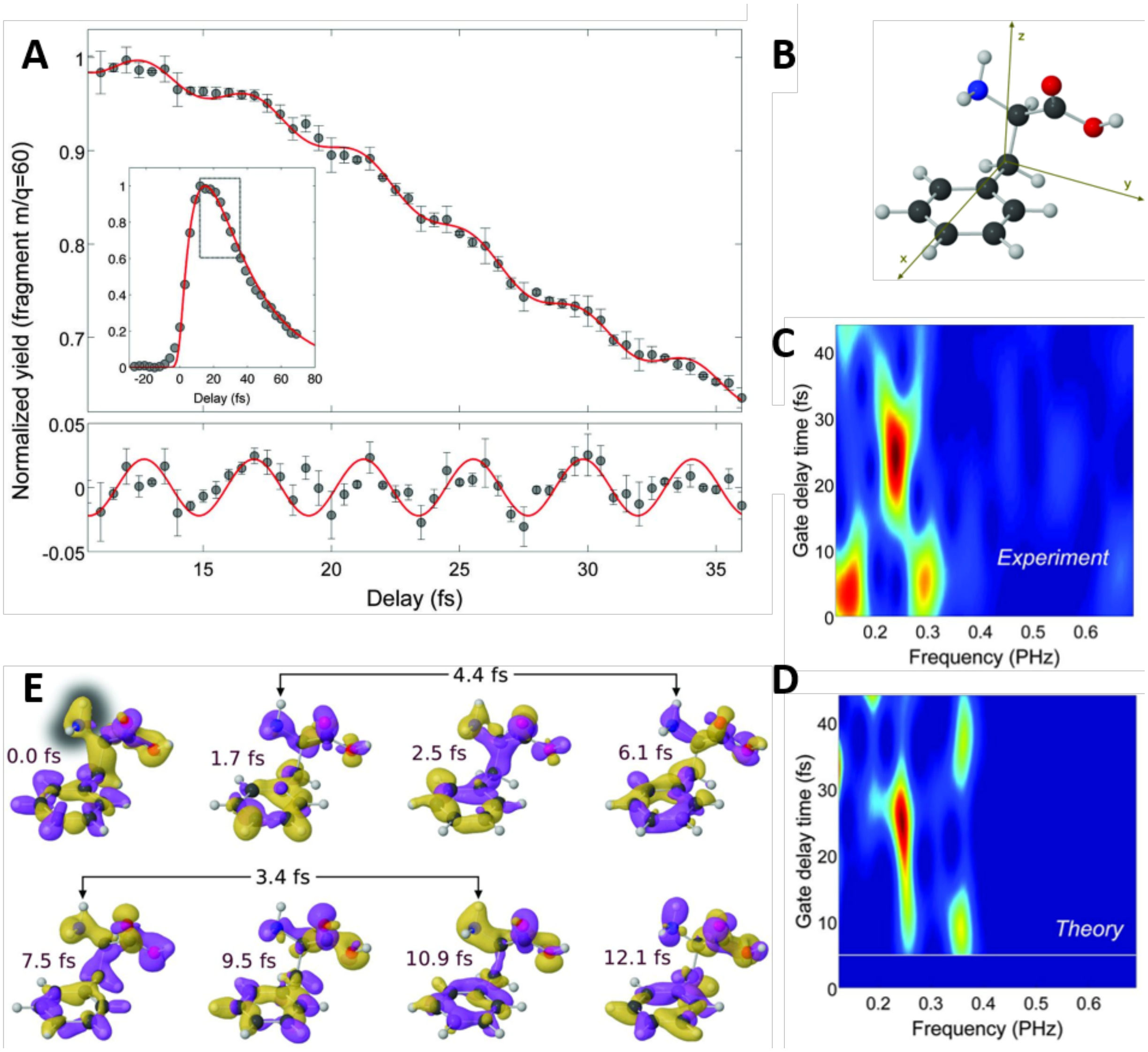

*Figure 5.1: (A) Yield of the doubly-charged immonium ion as a function of the pump-probe delay, measured with a temporal resolution of 3 fs (inset of the top panel) and 0.5 fs (top panel). The bottom panel displays the difference between the experimental data and the exponential fitting curve shown in the inset of the top panel. The error bars indicate the standard error obtained extracted from four measurements. (B) Molecular structure of the most abundant conformer of phenylalanine, obtained from DFT calculations with a B3LYP functional. (C) Sliding-Fourier-transform spectrograms of the data presented in the bottom panel of (A). (D) Same as (C), but extracted from theoretical calculations. (E) Relative variation of the hole density with respect to its time-averaged value as a function of time. The cutoffs of the isosurfaces are set at $+10^{-4}$ (arb.u, yellow) and $-10^{-4}$ (arb.u, purple). The time axis is referenced to the end of the XUV pulse. Figure adapted from (Calegari et al., 2014) with permission.*

The interpretation suggested by the authors involved charge migration of the hole between the two sites of the molecule due to a fast CT mechanism (Weinkauf et al., 1996). However, the poor temporal resolution of the experiments prevented the estimation of the time scale of the process.

Here, the recent implementation of an XUV-pump-NIR probe experiment realized by Calegari *et al*. (Calegari et al., 2014) to investigate ultrafast dynamics in the small amino acid phenylalanine will be discussed. The topic of CM will be revisited again in Section 6.3, where the observation of attosecond charge migration using high-harmonic spectroscopy will be reviewed (P M Kraus et al., 2015).

The experimental approach utilizes an XUV pump-NIR probe scheme, whereby a single isolated, sub-300 as XUV pulse with a bandwidth spanning from 15 to 35 eV is used to trigger ultrafast dynamics via ionization, and the ensuing dynamics is interrogated by a waveform-controlled NIR pulse centered at 1.77 eV with a duration of ≈4 fs. The two pulses are focused into a clean plume of isolated phenylalanine gas-phase molecules (Fig. 5.1 B), generated in-situ via laser-induced evaporation from a thin metal foil. Due to the elevated temperatures of 430 K required for this process, the experiment samples over six equilibrium conformers of phenylalanine. The molecular fragments produced by the probe pulse are collected and analyzed in a linear TOF spectrometer. The analysis focusses on the doubly-charged immonium ion produced via secondary electron ejection by the NIR pulse, whose yield is monitored as a function of the pump-probe delay, with varying step sizes (0.3-3 fs).

The observed dynamics are summarized in Fig. 5.1 A. The long-time evolution, monitored over a temporal window of 100 fs, displays a steep rise time of 10±2 fs, followed by an exponential decay (with a characteristic constant of 25±2 fs). The results recorded with a step size of 0.3 fs, centering on the first 25 fs, are displayed in more detail in the inset resp. the bottom part of Panel A, and reveal an oscillation with a beating period of ≈4.3 fs. More insight can be obtained by performing a sliding-window Fourier-transform (FT) analysis, resulting in the time-frequency map reproduced in Fig. 5.1 C. Two initially present components at 0.14 PHz and 0.3 PHz dominate the response over the first 15 fs. Afterwards, a strong and broad component at 0.24 PHz forms and decays within 35 fs, with a spectral width increasing as the pump-probe delay progresses. The observed fast motion is not compatible with the timescale of typical nuclear dynamics, which, although it cannot be excluded to influence the dynamics at longer delays, evolves on a timescale of 9 fs for the fastest X-H stretching, and takes even longer for the skeleton vibrations. The authors therefore attribute the periodic beatings to CM dynamics mediated by electronic dynamics. A further indication for the participation of electron dynamics is the decrease in the yield of the immonium dication when the bandwidth of the XUV pulse is reduced to 3 eV, which implies that CM entails the participation of higher-lying states of the cation.

Extensive theoretical calculations were conducted to further corroborate the electron-dynamics hypothesis. The electron wave packet, modelled as a coherent superposition of many 1-hole (*1h*) cationic states calculated using time-dependent density-functional theory, was propagated for a delay of 500 as after excitation by the XUV pulse using a density-matrix formalism. The NIR probe was not included in the analysis. The initial wave packet prepared by the large-bandwidth IAP contains contributions from multiple open channels (35 per conformer), and the ionization amplitudes were determined by static-exchange and first-order density functional theories. After averaging over all spatial orientations, the hole density, defined as the difference of the electronic densities of the neutral and the cation was extracted and analyzed in the frequency domain. The temporal modulations of the hole density are found to maximize around the amine group, even though the initially created hole distribution is highly delocalized, preventing an intuitive description in terms of a simple CM from one molecular site to another (cp. Fig. 5.1 E). The resulting frequency spectra, while depending on the choice of the conformer, all contain the three dominant peaks between 0.15 and 0.4 PHz, and correctly reproduce the appearance of the dominant peak at 0.25 PHz and the major features of its temporal evolution (Fig. 5.1 D). These results agree with the experimentally observed beating dynamics in the immonium fragment yield. However, the agreement is only qualitative as the relative intensities and time dependencies of the individual frequency components deviate strongly. The remaining incongruence was attributed to the neglect of the influence of the NIR pulse and the freezing of the nuclear degrees of freedom(Lünnemann et al., 2008), which erases the influence of nuclear motion on longer timescales.

In fact, it remains unclear how the modulation of the electron-hole density at the amino group relates to the experimentally observed yield of the immonium ion. The latter is formed by breaking the C-COOH bond of phenylalanine, rather than the C-NH2 bond that is more likely to be weakened by an increased electron-hole density at the amino group. An alternative theoretical analysis would consist in analyzing the bond order of the C-COOH group as a function of time in a more directed attempt to explain the experimental results. Therefore, additional theoretical work is desirable to relate the experimental and theoretical results.
Following this study, related experimental schemes based on the use of ionizing attosecond XUV pulses in combination with an optical probe were employed to elucidate the response of DNA building blocks to highly-energetic radiation. Examples include the study of ultrafast dynamics in thymidine and thymine (Månsson et al., 2017), attosecond pump-probe experiments in tryptophan (Lara-Astiaso et al., 2018), and the ultrafast hydrogen migration in glycine (Castrovilli et al., 2018).

## 6. High-harmonic spectroscopy

The principle behind high-harmonic spectroscopy (HHS) and its self-probing character have been introduced in Section 2.3. The current section illustrates the capability of HHS to resolve, albeit indirectly, electron and nuclear dynamics with sub-fs resolution. We will cover the following aspects: the resolution of structural rearrangements occurring on a sub-fs-timescale, the laser-induced modification of electronic structure on the generic examples of two small polar polyatomic molecules, and, finally, the reconstruction and laser control of attosecond charge migration.

### *6.1 Observation of sub-fs nuclear dynamics using high-harmonic spectroscopy*

The physical mechanism underlying the sensitivity of HHG to nuclear dynamics can be best illustrated by considering the following expression for the HHG intensity ($I_{\mathrm{HHG}}(N\omega)$), first proposed and validated by M. Lein (Lein, 2005) for the case of the $H_2$ molecule:

$$I_{\mathrm{HHG}}(N\omega) \propto |c(\tau(N\omega)|^2 I_{\mathrm{HHG}}^{\mathrm{FN}}(N\omega), \tag{6.1}$$

where $I_{\mathrm{HHG}}^{\mathrm{FN}}(N\omega)$ denotes the HHG intensity pertaining to the molecule with "frozen" nuclei, and $c(\tau(N\omega)$ is the nuclear autocorrelation function, i.e. the overlap of the nuclear parts $\chi(R,t)$ of the total molecular wavefunction in the ground state and at the instant of recombination, after a continuum excursion of duration $\tau$: $c(\tau) = \int \chi(R,0)\chi(R,\tau)dR$. The unique association of a given electron travel time with the $N^{\mathrm{th}}$-component of the harmonic comb, denoted explicitly through the argument of $\tau$ in Eq. (6.1), implies that the nuclear dynamics can potentially be inferred from a spectrum recorded in a single laser shot.

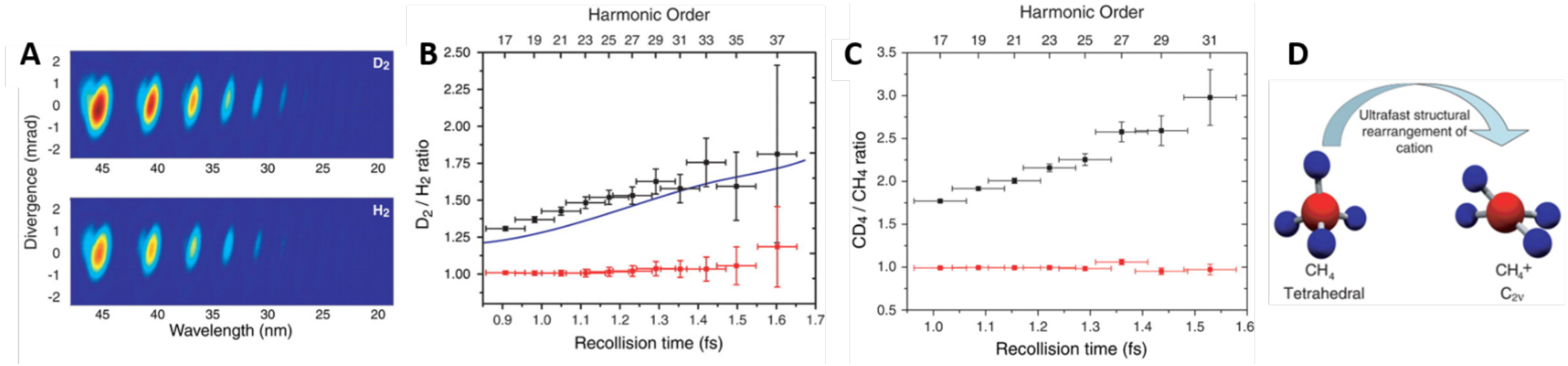

*Figure 6.1: (A) Raw CCD images of the HHG spectra recorded in $D_2$ (top) and $H_2$ (bottom), revealing that at all harmonic orders observed, harmonic emission in $H_2$ is weaker. (B) Ratio of harmonic peak intensities for $D_2$ and $H_2$ (black). The red lines represent the outcome of a control experiment in $H_2$. The blue line corresponds to the ratio obtained from theoretical calculations. (C) Ratio of harmonic signals in $CD_4$ and $CH_4$ (black) as well as the control ratio of two harmonic spectra from $CD_4$ (red). (D) Structures of $CH_4$ and $CH_4^+$ at equilibrium. Upon removal of an electron, a rapid structural rearrangement takes place. Figure adapted from (Baker et al., 2006) with permission.*

In 2006, Baker *et al.* (Baker et al., 2006), used laser pulses centered at 775 nm to generate high-harmonic radiation in a pulsed gas jet of $H_2$ or $D_2$ molecules. Given the fact that the rotational period of the hydrogen molecule is ≈274 fs (and twice that value for $D_2$), the pulses (initially of 30 fs duration) were compressed down to 8 fs in a hollow-core fiber followed by a chirped-mirror assembly in order to preclude the onset of rotational motion from influencing the HHG process. Placing the focus 9 mm before the gas jet isolated the contribution of the short trajectories, which were subsequently detected in a grazing-incidence, angularly-resolving flat-field spectrometer. The experimentally determined ratio of the HHG responses of the two isomers as a function of the transit time in the continuum is reported in Fig. 6.1 B. According to the time-to-frequency mapping discussed in Section 2.3, the experimental time resolution is set by the difference of the recollision times associated with two successive harmonic orders and amounts to 100 as at the wavelength employed (≈800 nm). The HHG efficiency in $D_2$ exceeds the one in $H_2$ (cp. also Fig. 6.1 A) over the entire energy range covered in the experiment, and the $r(D_2{:}H_2)$ ratio exhibits a monotonic increase at longer transit times. This result has been interpreted in the light of the faster nuclear motion in the lighter isomer. The latter interpretation has been further validated with the aid of SFA calculations which incorporate the autocorrelation functions derived from accurate potential energy curves for the two species and additionally include the two-center-interference effect (Lein et al., 2002a, 2002b; Vozzi et al., 2006, 2005), known to affect the HHG emission from the fraction of molecules aligned parallel to the driving field. The good quantitative agreement with the experimental results serves as a confirmation of the validity of the time-to frequency mapping and has been used as a starting point for reconstructing the temporal variation of the mean internuclear distance on a sub-cycle time scale based on the measured data. This approach has been subsequently transferred to the more complex case of the methane molecule ($CH_4$ vs. $CD_4$). The HHG intensity ratio $r(CD_4{:}CH_4)$ depicted in Fig. 6.1 C shows an even more pronounced isotope effect than the $H_2$ / $D_2$ pair despite the presence of heavier nuclei. The faster nuclear dynamics is a direct consequence of the non-adiabatic vibronic couplings (the Jahn-Teller effect) in the methane cation, which are absent in the electronically simple case of $H_2$. Whereas the equilibrium geometry of methane is characterized by a tetrahedral geometry, the cation is unstable with respect to a Jahn-Teller distortion and adopts a $C_{2v}$ structure in equilibrium (Frey and Davidson, 1988; Knight et al., 1984; Vager et al., 1986), with significantly smaller bond angles (<60° vs. 109.4°, cp. Fig. 6.1 D). As the equilibrium structures of the ground state and the ion are highly disparate, this rearrangement has been suggested to take place on an ultrafast time scale. This result, as well as subsequent experiments on the $H_2O/D_2O$ isotopomer pair (Farrell et al., 2011), established that the speed of the nuclear dynamics probed in an HHS experiment is mainly determined by the difference in the potential energy surfaces of the neutral and the ground state. The implications for the Jahn-Teller dynamics in the methane cation have been investigated theoretically in further detail in in Refs. (Mondal and Varandas, 2015, 2014; Patchkovskii, 2009), and, more recently, in Ref. (Patchkovskii and Schuurman, 2017). The frequency of the employed driving field sets an upper limit for the temporal window that can be covered in this type of experiment, which amounts to ≈1.6 fs for an 800-nm-field. Longer driving wavelengths are thus a prerequisite for observing nuclear dynamics of heavier nuclei. In later work, experiments on the $H_2/D_2$-system have been extended to 1.3 μm (Mizutani et al., 2011), with an emphasis on the role of the two-center interference. Other works have exploited HHS-based methods to gain access to the phase of the nuclear wave packet (Haessler et al., 2009a; Kanai et al., 2008).

A further specificity of the HHG-based approach for studying nuclear dynamics is that strong-field ionization, the initiating step of the HHG process, prepares a nuclear wave packet, whose dynamics differ from the ones prepared by single-photon ionization (Kjeldsen and Madsen, 2005; Urbain et al., 2004). A theoretical description based on the Franck-Condon factors and tunneling rates was presented and validated against experimental results on the umbrella mode in ammonia in Ref. (Kraus and Wörner, 2013). These results point out the capability of HHS to probe PES areas far from equilibrium, which are hardly accessible by other methods, thereby interrogating the dynamics with high temporal resolution.

### *6.2 Observation of laser-induced modification of the electronic structure*

The next two selected case studies showcase the sensitivity of HHG to both static electronic structure and electron dynamics and their modification in the presence of a strong electric field. The sensitivity to electronic structure is mediated by the laser-driven recombination step, which gives access to the bound-continuum matrix elements probed via photoionization and photoelectron spectroscopies, but with the additional benefit of the sub-cycle time resolution. HHS has been extensively used to characterize the field-free electronic structure, notable examples include Cooper minima (Bertrand et al., 2012; Higuet et al., 2011; Wong et al., 2013; Wörner et al., 2009), giant resonances (Shiner et al., 2011), shape resonances (Kraus et al., 2014; Ren et al., 2013). However, in all of these studies, the influence of the laser field present in the probe step has tacitly been omitted from consideration. This assumption has been questioned in the work of Kraus *et al*. (P. M. Kraus et al., 2015), where harmonic emission from aligned and/or oriented polar molecules ($CH_3F$ and $CH_3Br$) was shown to encode the modifications of the electronic structure induced by the strong laser electric field. The experimental approach combined high-harmonic generation driven by moderately intense fields (≈$10^{14}$ W/cm$^2$) and techniques for field-free molecular alignment/orientation based on non-resonant impulsive Raman scattering (s. Ref. (Stapelfeldt and Seideman, 2003) and references therein). A moderately strong, temporally stretched fs laser pulse is used for inducing alignment, whereas orientation is achieved by means of a phase-controlled superposition of 800 and 400 nm pulses (two-color field, cp. (De et al., 2009; Frumker et al., 2012; Kraus et al., 2014, 2012)). Imposing orientational order breaks the inversion symmetry of the sample, leading to the emission of even harmonics in addition to the odd ones.

The choice of $CH_3F$ and $CH_3Br$ as generic examples is motivated by their large permanent dipole moments (1.85 D and 1.81 D, respectively) which ensure a substantial degree of coupling with the laser field. The ensembles are probed with an HHG-driving pulse around the first alignment orientational revival time, under otherwise field-free conditions, whereby fine control over the rotational distribution is achieved by varying the pump-probe delay.

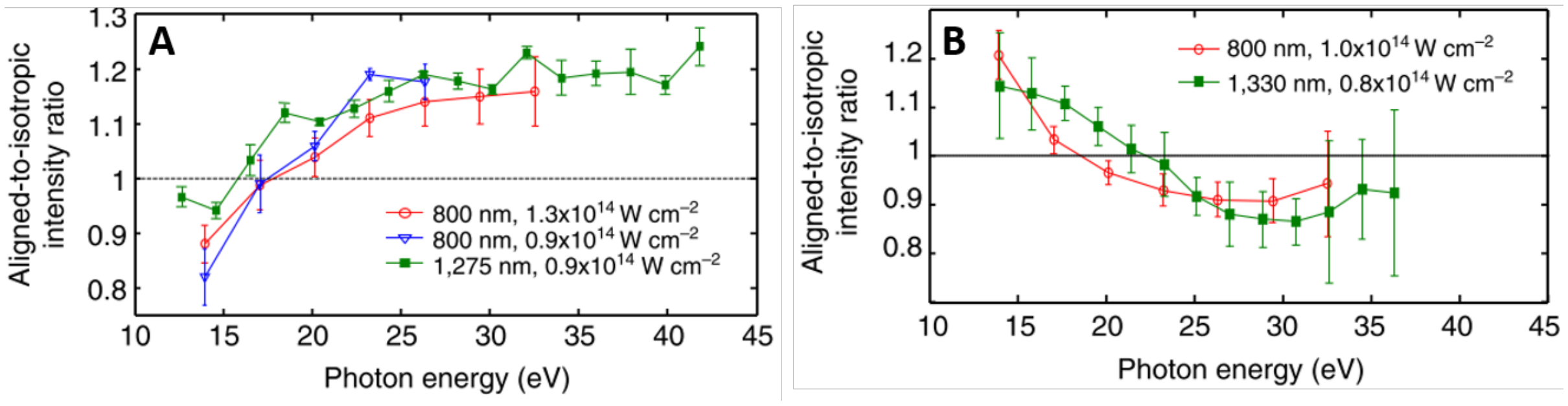

*Figure 6.2: (A) Measured intensity ratio of emission from aligned vs. isotropic molecules at different wavelengths and intensities for $CH_3F$. (B) Same as (A) for $CH_3Br$. Figure adapted from (P. M. Kraus et al., 2015) with permission.*

Quantification of the field-induced electronic-structure modifications is performed with respect to a carefully selected set of robust experimental observables. One of these is the ratio of the emission from aligned vs. isotropically distributed molecules. The photon-energy dependence of this quantity for the two investigated species has been determined at different driving wavelengths (800 nm and 1275 resp. 1330 nm) and intensities (0.8-1.2x$10^{14}$ W/cm$^2$) and is displayed in Fig. 6.2 for $CH_3F$ (panel A) resp. $CH_3Br$ (panel B). The aligned-to-isotropic ratios exhibit a reversal between 16 and 18 eV for $CH_3F$, whereas in the case of methyl bromide the crossing is located between 18 and 21 eV. The insensitivity of the position of the crossing to variations of the laser parameters is a signature of the static electronic structure of the molecules and does not result from an interference effect due to participation of multiple electronic states in the emission process. An additional observable for $CH_3F$ is the even-to-odd ratio, i.e. the ratio of the intensities of the emitted even harmonics relative to the averaged intensities of two adjacent odd ones, determined at the pump-probe delay optimizing the degree of orientation. This observable, plotted in Fig. 6.3 C, exhibits a steep increase until it reaches a maximum at H14, followed by a smooth decline. The theoretical interpretation of these observations required improving the theoretical description of HHG by including the influence of the electric field on each of the steps of the HHG process in a consistent manner. In the case of polar molecules, the treatment of the SFI rates, which is difficult by itself due to the exponential sensitivity to the asymptotic tail of the molecular potential, is additionally complicated due to the presence of a large, angle-dependent Stark-shift caused by the permanent dipoles of the neutral and the cation. This challenge is addressed by employing the recently developed weak-field asymptotic theory, consistently extended to the case of polar molecules (Tolstikhin et al., 2011). In the propagation step, an additional, orientation-dependent phase shift originating from the Stark effect (Dimitrovski et al., 2010; Etches and Madsen, 2010) was introduced, whereas the calculation of the matrix elements governing the recombination step was performed using molecular orbitals that take the distortion due to the static electric field into account (Śpiewanowski et al., 2013). As evident from the different curves shown in Figs. 6.3 A and 3 B, quantitative agreement with the experimental results is achieved only after all field-induced effects are accounted for, whereas the field-free model is qualitatively incorrect and even makes erroneous predictions in the case of $CH_3Br$. This pronounced difference in the level of agreement is interpreted as an evidence for the field-induced modifications of the electronic structure, which take place on the sub-cycle time scale of the HHG process.

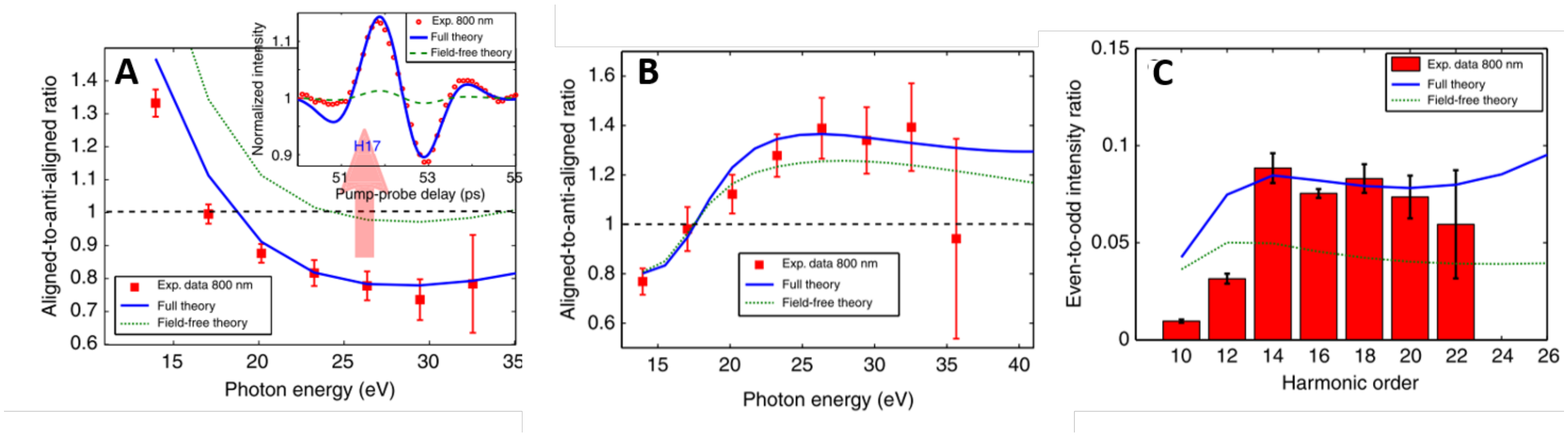

*Figure 6.3: Comparison of the experimental high-harmonic intensity ratios with theoretical results. (A) Ratios of aligned vs anti-aligned $CH_3Br$ molecules compared with the predictions of the full theory and the field-free model (s. text). The inset shows the comparison between the experimentally observed alignment pattern with the theoretical predictions. (B) Same as (A) for $CH_3F$. (C) Comparison of the even-to-odd ratios for $CH_3F$ with the predictions of the two models. Figure adapted from (P. M. Kraus et al., 2015) with permission.*

### *6.3 Measurement and laser control of charge migration in ionized iodoacetylene*

In the studies reviewed so far, HHS has been used to address the static electronic structure of the system under study. The capability of HHS to probe electron dynamics occurring on a sub-cycle time scale has been exploited in the interpretation of intensity minima in $CO_2$ (Rupenyan et al., 2013; O. Smirnova et al., 2009; Olga Smirnova et al., 2009), or the tomographic imaging of orbital wave functions in $N_2^+$ (Haessler et al., 2010). This section illustrates this feature on the example of the reconstruction of the attosecond charge migration (CM) dynamics (P M Kraus et al., 2015) in a spatially oriented polar molecule with a temporal resolution of 100 as from the detailed analysis of the harmonic emission at 800 nm and 1300 nm from impulsively oriented iodoacetylene (HCCI) molecules. The iodoacetylene molecule represents an ideal system for studying CM (cp. energy diagram in Fig. 6.4 A). Its energy structure is characterized by two closely-lying cationic states ($\widetilde{X^+}\ ^2\Pi$ and $\widetilde{A^+}\ ^2\Pi$), separated by 2.2 eV, which can be simultaneously populated by strong-field ionization at the employed intensities, while the population of higher-lying excited cationic states plays a negligible role. The two states are coupled via a strong electric transition dipole moment (≈3.5 D), which is oriented parallel to the internuclear axis in the molecular frame, thus resulting in a strong alignment dependence of the coupling to the external field (Fig. 6.4 B and C). All these characteristics make HCCI a very attractive system for studying field-free as well as field—steered charge migration.
In a manner similar to the experiment presented in the preceding paragraph, a two-color pump pulse is used to induce spatial control (alignment or orientation), and the ensemble is subsequently probed at delays around the first revival time either with an 800 nm or with a 1300 nm pulse. The durations of the electron trajectories at these two driving wavelengths define two observation windows spanning 0.9-1.5 fs resp. 1.3-2.2 fs.

Contrary to the situation encountered in Sec. 6.2, the ratios of the emissions from molecules oriented perpendicular vs. parallel and the spectral intensity ratio of even and odd harmonics for the two employed probe wavelengths as a function of photon energy are sensitive to the wavelength of the driver: for instance, the position of the minimum of the ratio shifts from 23.2 eV at 800 nm to 35.3 eV at 1300 nm, hinting at sub-cycle dynamics taking place between ionization and recombination. An additional observable that was detected in the experiment is the phase of the harmonic radiation as a function of the alignment angle using two-source interferometry between the harmonics emitted from aligned and spatially isotropic molecules (Rupenyan et al., 2012). The variation of the phase as a function of the alignment angle depends not only on the photon energy, but also on the driving wavelength, further consistent with laser-induced dynamics.

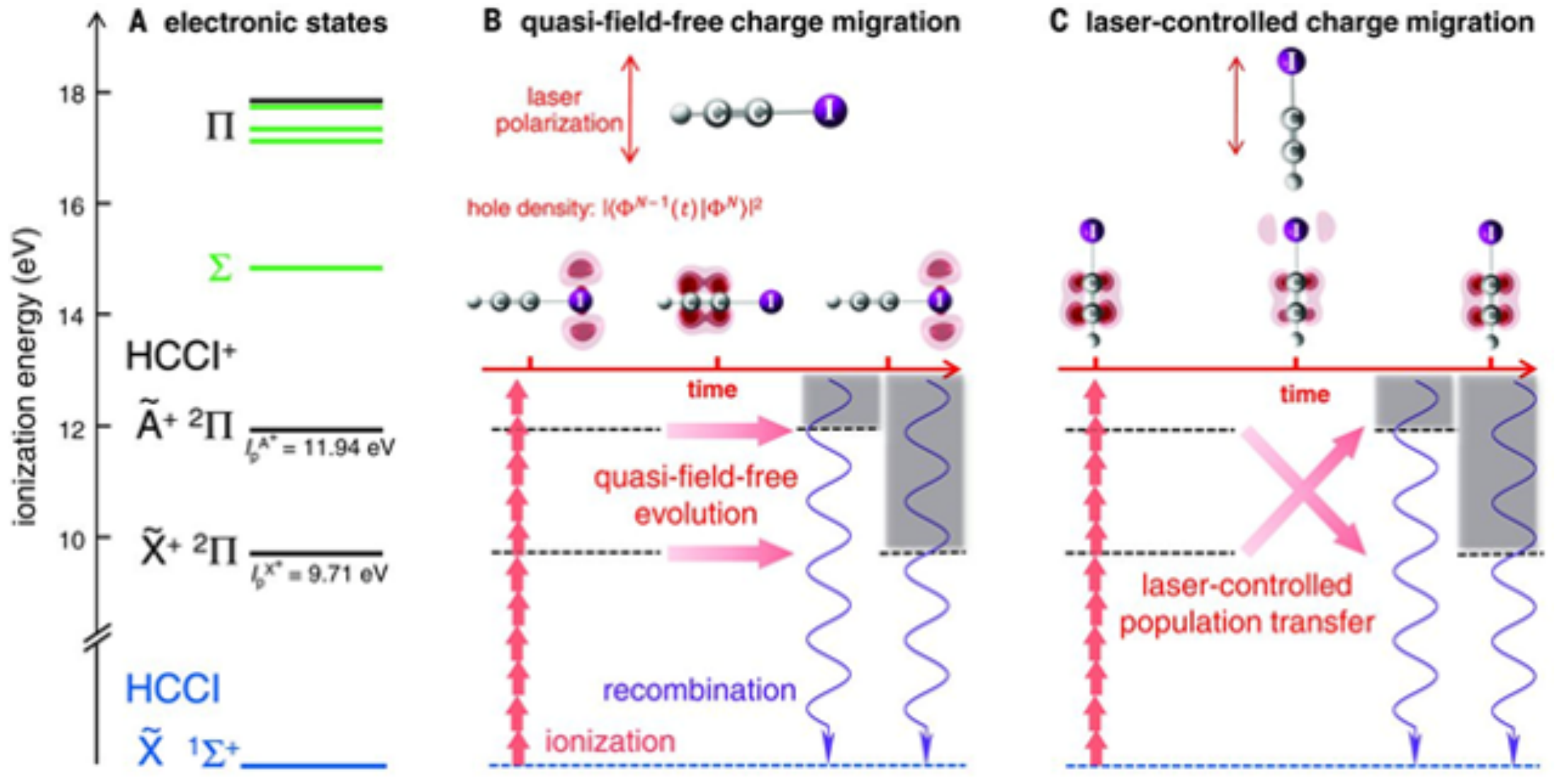

*Figure 6.4: (A) Energy level diagram of the relevant electronic states of HCCI. (B) Schematic illustration of the quasi field-free charge migration and the laser-controlled charge migration. Strong field ionization by the NIR field (red arrows) prepares the electron hole. Its evolution is encoded in the high-harmonic emission (violet) at the recombination instant. For perpendicularly aligned molecules (B), the hole populations are quasi time-independent. For molecules aligned parallelly to the field, the laser field drives a strong population transfer between the $\widetilde{X}^+$ and the $\widetilde{A}^+$ states. Figure adapted from (Wörner et al., 2017) with permission.*

The complete reconstruction of the quantum-mechanical evolution of the coherent superposition of states prepared by SFI until the recombination instant requires knowledge of the relative populations of the two states and their relative phases, as a function of time. These unknowns were retrieved from the experiential observables (parallel-to-perpendicular resp. even-to-od intensity ratios and the measured phase differences) using a Levenberg-Marquardt inversion procedure. The underlying theoretical model incorporates the angular dependence of the SFI step via the WFAT theory, accurate quantum-mechanical scattering states, the effect of the nuclear motion, and the rotational anisotropy. The total emitted dipole is given by a coherent superposition of the emissions from the contributing electronic states at each energy $N\omega$, mapped onto time using the saddle-point trajectories.

The reconstruction for perpendicularly aligned molecules is represented in terms of the hole densities in Fig. 6.4 B. The hole created in the SFI step is initially localized on the iodine atom, then delocalizes over the internuclear axis and localizes at the acetylene end after 930 as. These dynamics correspond to field-free charge migration as the orientation of the transition dipole between the two cationic states $\widetilde{X}^+$ and $\widetilde{A}^+$ precludes coupling to the field in this orientation.

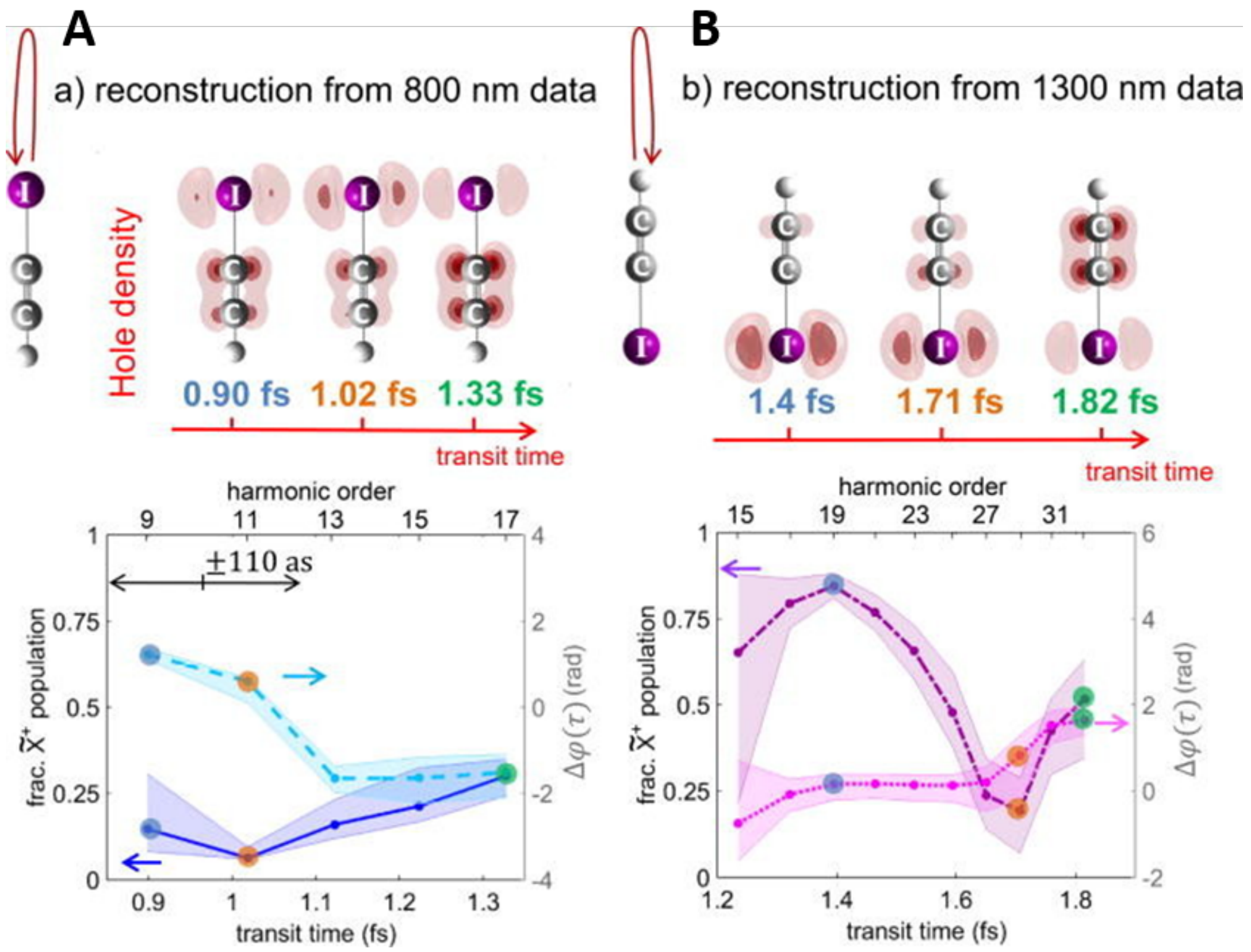

*Figure 6.5: (A) Electron-hole densities for HCCI molecules aligned along the laser polarization axis, reconstructed from high-harmonic emission driven by an 800 nm field. (B) Same as (A) for a 1300 nm driver. The lower panels display the fractional populations of the $\widetilde{X}^+$-state (left vertical axis) as well as the relative phase between the $\widetilde{X}^+$ and the $\widetilde{A}^+$ states (right vertical axis). Figure adapted from Ref.* (Wörner et al., 2017) *with permission.*

The reconstruction for parallelly aligned molecules displayed in Fig. 6.5, on the other hand, shows much richer dynamics. In this case, the electron can tunnel via two possible sites, either via the hydrogen or via the iodine atom, whereby the dynamics is qualitatively similar. Here we focus on the CM using an 800 nm driver, depicted in Fig. 6.5 A. For tunneling via the I-atom, which is the dominant case, the hole is first created on the acetylene fragment, and the relative phase difference between the $\widetilde{X}^+$ and the $\widetilde{A}^+$ states is $\approx\pi$. The reconstructed fractional populations indicate that the $\widetilde{A}^+$ state is strongly depopulated at early transit times, reaching a minimum shortly after $\approx$1 fs and subsequently increasing again. This depopulation is accompanied by a jump in the relative phase. The dynamics in the case of tunneling via the H -atom, on the other hand, reveal a strong depopulation at early delays, and an earlier onset of repopulation. The observed rapid variation of the population of the $\widetilde{X}^+$-state is a clear signature of laser-driven charge migration. The results at 1300 nm, shown only for the case of tunneling via the H-atom, reveal a gradual increase of the $\widetilde{X}^+$ -population until 1.53 fs, followed by a rapid decrease and minimum at 1.73 fs and subsequent rise. These results suggest that both the wavelength as well as the tunneling site offer a possibility to exert control over the CM process.

This study provided evidence that HHS can be used to conduct a spatiotemporal reconstruction of CM triggered by SFI, with a temporal resolution of 100 as. In addition, highlighted the possibility for laser control through the orientational dependence and the driving wavelength.

## 7. Attosecond time delays in molecular photoionization

The recent progress in attosecond chronoscopy, in particular the development of interferometric XUV + IR two-color techniques such as the attosecond streak camera (cp. Section 3.2.1) or RABBIT (cp. Section 3.2.2), has enabled access to the photoionization dynamics in atomic and molecular gases, as well as solids (cp. Refs. (Pazourek et al., 2015) and (Gallmann et al., 2017) for recent reviews). Probing these dynamics in the time-domain provides complementary information to the observables accessible via frequency-domain (i.e. time-integrated) measurements (e.g. partial cross sections or phase shifts) and can thus assist the derivation of a complete quantum-mechanical description of the photoionization process (Jordan et al., 2017). Whereas atomic and solid-phase systems have been a subject to investigation using both streaking and RABBIT techniques (Cavalieri et al., 2007; Gaumnitz et al., 2017; Jain et al., 2018; Jordan et al., 2017; Klünder et al., 2011; Schultze et al., 2010), the relatively paucious amount of experimental data on valence-shell photoionization dynamics in molecules reported to date has been limited to the latter technique. This is due to the presence of multiple close-lying electronic states in molecular cations that fall within the bandwidth of a typical IAP, consequently, application of streaking to molecules is limited to well separated electronic states, such as core-ionized states (Förg et al., 2019). Therefore, in this section we restrict our discussion to recent experimental studies employing APTs phase-locked to multi-cycle IR pulses, where the time-domain observables can be retrieved from the spectral phase information with sufficient energy resolution. The selected examples cover aspects such as the influence of autoionizing Rydberg states (Haessler et al., 2009b) on the phase of side-band oscillations, the manifestation of shape resonances (Huppert et al., 2016) in the measured attosecond photoionization delays, as well as the intricate dependence of the latter on the details of the molecular structure and the anisotropy of the molecular potential (Vos et al., 2018). The final case study addresses the subtle attosecond delays between electrons emitted in the forward vs. backward directions in the two enantiomers of a chiral molecule (Beaulieu et al., 2017). Although this latter study did not make use of the RABBIT technique, we include it here as it illustrates how self-referenced photoelectron interferometry can be used to access dynamics on attosecond time scales without the necessity of employing attosecond pulses / pulse trains, while noting that the interpretation of such experiments is more complicated than in the RABBIT framework.

### *7.1 Phase-resolved near-threshold photoionization of molecular nitrogen*

The experimental characterization of the phase of the two-color two-photon near-threshold photoionization of molecular $N_2$ reported in the study of Haessler *et al.* in 2009 (Haessler et al., 2009b) represents an important preparatory step towards the measurements of photoionization delays in molecules. Notably, this study, as well as the subsequently published theoretical analysis (Caillat et al., 2011), were carried out before the relation between the observables of RABBIT and photoionization delays was established (Klünder et al., 2011). Instead, Haessler et al. explored the manifestations of autoionizing Rydberg states on the phase of side-band oscillations observed in RABBIT measurements. Whereas the experimental publication only reported the measured phases as a function of the side-band order, the theoretical analysis (Caillat et al., 2011) additionally discussed the relation of these phases with the "formation time" of the side-band states, obtained as the derivative of the phases with respect to the photon energy of the dressing (and driving) fields. We note that these formation times are however distinct from the photoionization delays that are discussed in sections 7.2 and 7.3. The variation of the two-color two-photon phases as a function of the photon energy in the neighborhood of a resonance has also been experimentally observed in helium (Swoboda et al., 2010) prior to the publication of Ref. (Klünder et al., 2011). The experiment of Haessler et al. employed an APT derived from a 20 Hz, 50 mJ system delivering 50 fs pulses which are subsequently spatially separated into a pump (HHG, outer part) and a probe (the dressing IR, inner part) arms. The XUV is generated by focusing the outer annular part of the beam in a pulsed argon gas jet, and, after filtering out the residual IR part, is recombined with the probe beam with the aid of a grazing-incidence Au-coated toroidal and focused in an effusive stream of $N_2$ molecules. The relative XUV-IR delay $\tau$ is adjusted with the aid of a piezoelectric transducer, and the generated photoelectrons are collected with a magnetic-bottle time-of flight (TOF) assembly. The accent of the study lies on the spectroscopically rich region just above the ionization potential of $N_2$, and, in particular, the "complex" resonance at around 17.12 eV, which corresponds to an autoionizing state belonging to the Rydberg series converging to the $B^2\Sigma_u^+$ - Hopfield state of the ion (Dehmer et al., 1984). The latter couples most efficiently to the continuum associated with the $X\ ^2\Sigma_u^+$ ground states of $N_2^+$. As the fundamental IR ($\omega$) frequency is centered at 1.565 eV, the resonance falls within the bandwidth of the 11th harmonic (17.21 eV, cp. also Figure 7.1 A).

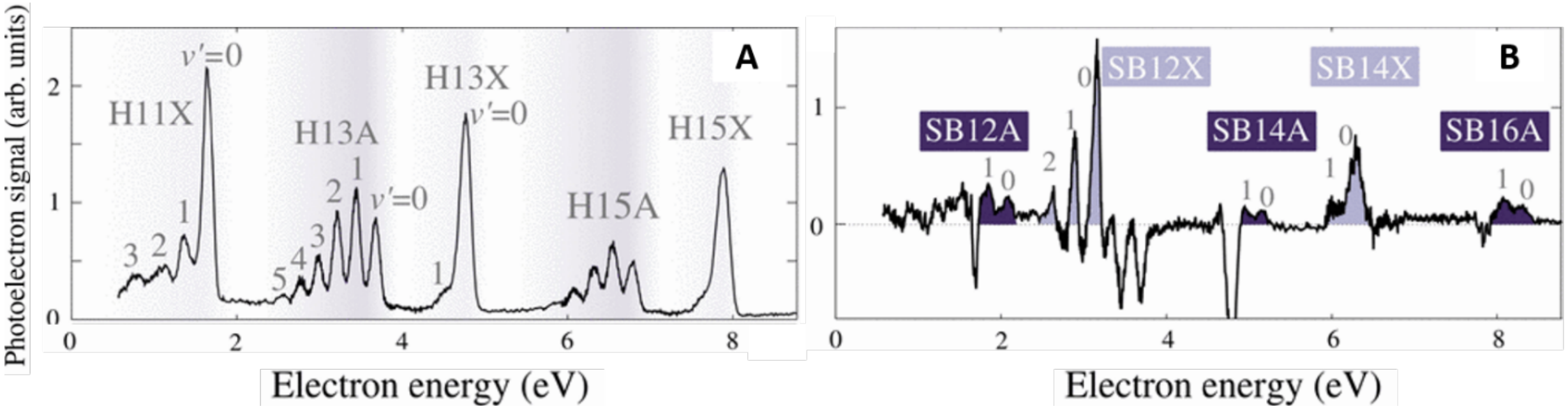

*Figure 7.1: (A) Photoionization spectrum of $N_2$ obtained with the high-harmonic (XUV) comb alone. (B) Difference between XUV-only spectra and spectra obtained in the presence of the perturbing NIR field. Figure adapted from (Haessler et al., 2009b) with permission.*

The XUV photoelectron spectrum in the absence of the dressing IR field is shown in Figure 7.1 A. The spectral bandwidth of the attosecond pulse train results in photoemission from two close-lying photoionization channels associated with the $X^2\Sigma_g^+$- (15.58 eV) and the $A^2\Pi_u$- (16.69 eV) cationic states, which appear as two distinct photoelectron bands with a superimposed vibrational structure. The difference in the shapes of the PE bands reflects the different equilibrium distances in the two states: 1.12 Å vs 1.17 Å. In order to accentuate the differences in the details of the spectra induced by the perturbing IR and mitigate the spectral overlap of the two channels, the two-color spectrum presented in Figure 7.1 B has been background corrected by subtracting the XUV-only response.

The absorption resp. emission of IR photons creates photoelectrons with energies corresponding to the even harmonics $2q$ of the fundamental frequency $\omega$ ("sidebands", label SB), which appear as positive contributions in contrast to the negative areas reflecting the loss of population at the positions of the main peaks. As there are two interfering pathways contributing to each sideband, the intensities of these peaks evolve as a function of the delay between the two colors:

$$SB_{2q} \propto A_{2q} \cos\left(2\omega\tau - \Delta\phi_{\text{mol}} - \Delta\phi_{2q}\right). \quad (7.1)$$

In the experiment, the total phase of each sideband is extracted by means of an FFT analysis of the oscillatory signal. In order to isolate the non-trivial, molecule-specific phases denoted by $\Delta\phi_{\text{mol}}$, the contribution of the phase of the XUV field ($\Delta\phi_{2q}$) to the total phase has been determined by a separate measurement in Ar and subsequently subtracted. This procedure invokes the implicit assumption that the $\Delta\phi_{\text{at}}$ can be reliably calculated from theory (Mairesse et al., 2003), so that the harmonic phases can be directly read off from the phase of the SB oscillations. This analysis is performed for each sideband and each ionization channel and the extracted phases for each vibrational quantum number are plotted as a function of energy (sideband order) in Figure 7.2 B. Whereas most of the measured values for both channels are contained within ±0.2 $\pi$ rad, the phase difference for SB12 in the *X*-state shows a noticeable exception to this trend, assuming a value of -0.35 $\pi$ rad for the *v'*=0 state and reaching up to -0.9 $\pi$ rad for the *v'*=1,2 states. This behavior has been interpreted as a manifestation of the resonance in the vicinity of H11 and can be linked to the phase jump in the complex two-photon photoionization matrix element for the pathways involving absorption of H11. In addition, a remarkable dependence of the magnitude of $\Delta\phi_{\text{mol}}(12)$ on the vibrational level is observed. To aid the rationalization of this behavior, a one-dimensional model is developed. The crucial insight is that the effect of the resonance on the molecular phase is most pronounced when the harmonic energy is situated below the resonance energy. With this knowledge, and given the fact that the H11 is positioned energetically between the *v"*=0 and *v"*=1 levels of the autoionizing B-like state, this behavior can be explained in terms of the Franck-Condon overlap with the vibrational levels of the *X*-cationic state. The latter show that whereas population of the *v'*=0 state of the ion can efficiently take place from both the *v"*=0 (below resonance) and *v"*=1 (above resonance) states of the *B*-like state, the populations of the higher-lying *v'*=1,2 states occurs mainly from the *v"*=1 state (above resonance). Thus, the phase shift $\Delta\phi_{\text{mol}}(12)$ for the *v'*=0-sideband contains contributions from both direct (*v"*=0) as well as resonant (*v"*=1) pathways, which diminishes the magnitude of the total phase shift. In contrast, the higher-lying vibrational states are almost entirely dominated by the resonant contribution and thus exhibit a strong phase shift (≈ $\pi$). The lack of sudden variations of the phases associated with the *A*-channel is attributed to the decreased efficiency of the coupling to the autoionizing state. This work demonstrates the sensitivity of the observables accessible via RABBIT measurements to the location of the one-photon ionization threshold, to the vibrational structure of the PE spectrum, and to the presence of autoionizing resonances. Several of these aspects will be further addressed and explored in more details in the case studies presented in the remaining part of this section.

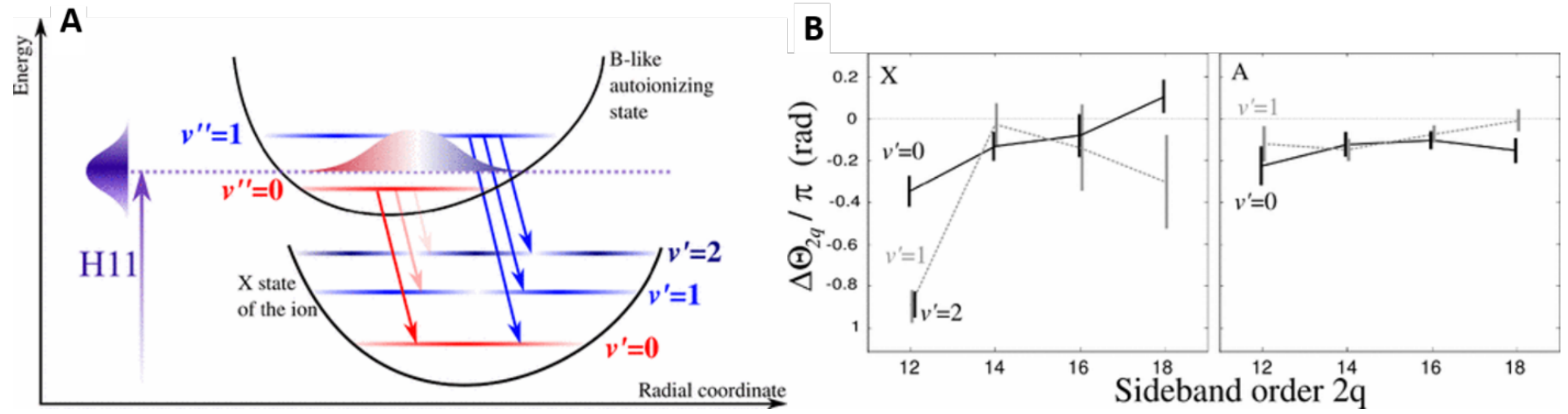

*Figure 7.2: (A) Schematic representation of the photoionization process taking place via vibrational levels of the autoionizing state. (B) Experimental molecular phase difference as a function of the sideband order for the X and A ionization channels. Figure adapted from (Haessler et al., 2009b) with permission.*

## *7.2 Attosecond photoionization delays in the nitrous oxide and water molecules*

The first measurements of time delays in molecular photoionization were reported by Huppert et al (Huppert et al., 2016). While demonstrating the applicability of the RABBIT technique to molecules, the work by Haessler *et al.* did not report time delays for at least two reasons. First, the relation between the observables of RABBIT and time delays in photoionization had not yet been established when the study was published. Second, Haessler et al. investigated the effect of a resonance whose width is much smaller than the energetic separation of two neighboring harmonics. Although such a resonance manifests itself as an additional phase shift in the two-color two-photon matrix elements, the phase shift does not smoothly translate into a time delay because the phase varies too rapidly with energy to validate the finite-difference approximation ((Dahlström et al., 2012; Klünder et al., 2011)) that is used to relate phase shifts to time delays. The study by Haessler et al. simultaneously revealed some of the potential impediments originating from the molecular electronic structure, like the spectral congestion related to the participation of several ionization channels in the valence-shell photoemission process, which is inconvenient given the bandwidth of the employed APTs, or the spectral overlap of the photoelectron spectra associated with different harmonic orders. These limitations have been resolved in the work of Huppert *et al.*(Huppert et al., 2016) by spectrally filtering the attosecond pulse train, thereby removing the problematic spectral overlap that previously limited the reliability of molecular attosecond interferometry. The experiment is performed in an actively-stabilized attosecond beamline (Huppert et al., 2015) and a magnetic TOF spectrometer. High—harmonic generation takes place in an argon-filled gas cell (10 mbar) driven by laser pulses centered at 800 nm (1.5 mJ, 30 fs). The separation and subsequent recombination of the XUV and the dressing IR pulse is performed with the aid of two perforated off-axis parabolic mirrors. The re-focusing of the XUV is done with the aid of a toroidal mirror, and the time delay is adjusted by changing the length of the IR beam path. The complications posed by the spectral overlap are mitigated by introducing several significant experimental advances, most notably via the introduction of thin metal filters (Sn, Ti, Cr) for spectral isolation of harmonics, the single-shot acquisition setup and the insertion of a chopper wheel in the IR beam path which allows for recording high-fidelity (XUV-only-XUV+IR) difference spectrograms. Figure 7.3 A shows a typical photoionization spectrum of $N_2O$ generated with a Sn-filtered APT.

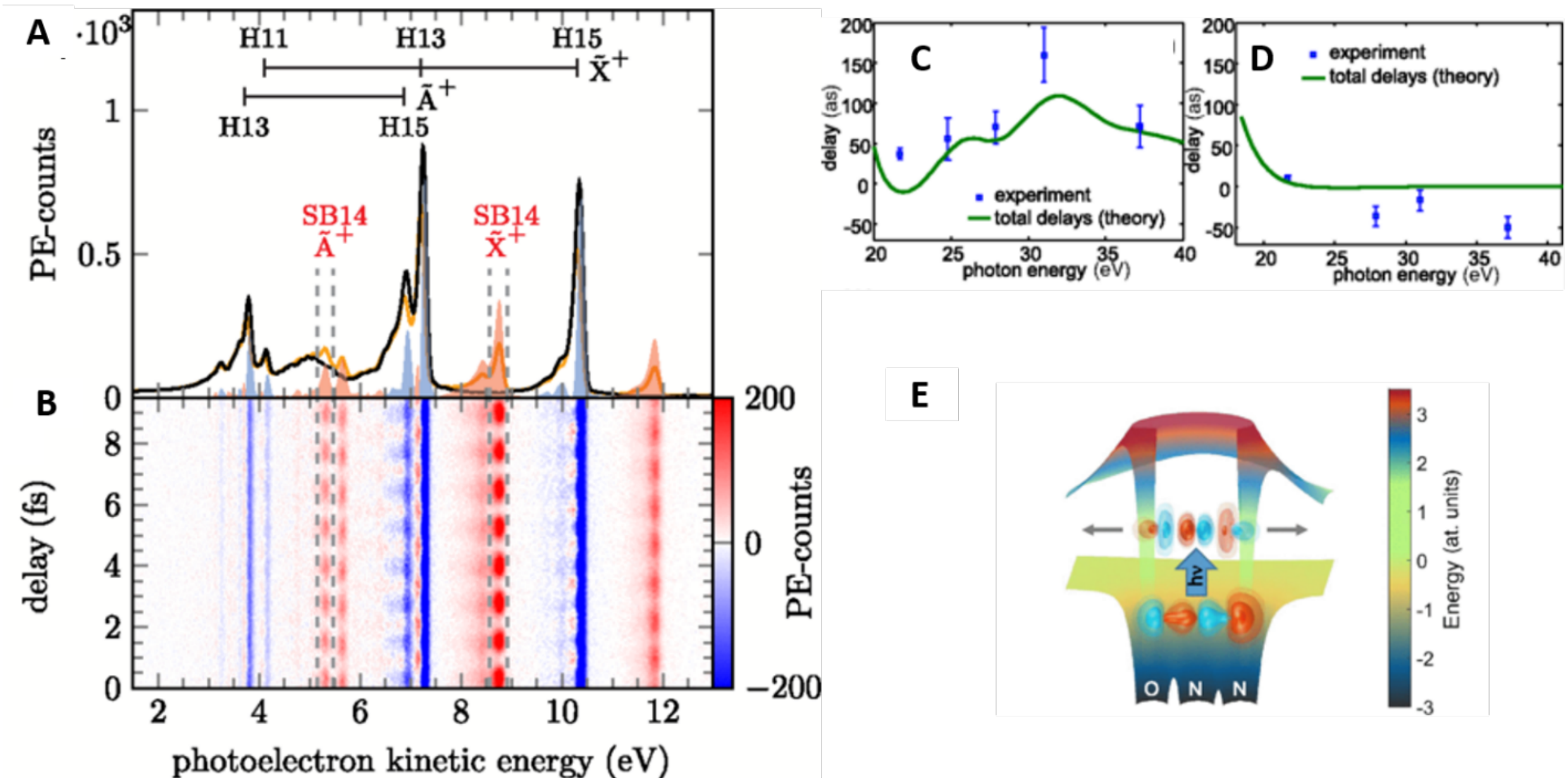

*Figure 7.3: (A) Photoelectron spectrum of $N_2O$ generated by an APT transmitted through an Sn spectral filter without (black) and in the presence (orange) of the perturbing IR field. The red and the blue spectra correspond to difference spectra obtained by subtracting the XUV-only from the XUV+IR spectra and vice versa, respectively. (B) Difference spectrum as a function of the XUV-IR delay and the photoelectron energy. (C) Experimental and theoretical photoionization delays between photoelectrons leaving $N_2O^+$ in its $X^+$ or $A^+$ states. (D) Same as (C), but for $H_2O$. (D) Illustration of the shape resonance of sigma symmetry in the photon energy range of 25-30 eV and associated with the $A^+$ state of $N_2O^+$. The lower surface corresponds to the calculated molecular potential comprising electrostatic and exchange interactions. The upper surface corresponds to the total potential obtained after adding the molecular and the centrifugal potentials for partial wave quantum number $l = 5$. The color-coded isosurfaces indicate the wave functions of the bound orbital and the shape-resonant state. Figure adapted from (Huppert et al., 2015) with permission.*

The study investigates the relative photoionization delays between the two outermost valence shells of two triatomic molecules, nitrous oxide ($N_2O$, linear) and (gaseous) water ($H_2O$, bent). The selection of these two species allows one to survey the effect of a shape resonance on the molecular photoionization delays in a comparative manner. The spectrum of $N_2O$ (cp. Figure 7.3 A) is dominated by photoelectrons corresponding to the first two (bound) cationic states $\tilde{X}^+$ and $\tilde{A}^+$ with ionization energies of 12.89 eV and 16.38 eV, respectively. The difference spectrogram as a function of the two-color delay is shown in Figure 7.3 B, whereby the oscillating positive contributions (red) correspond to the SBs of order 12 and 14 for the $\tilde{X}^+$ state and 14 for the $\tilde{A}^+$ state. Delays of up to 35 ± 6 as between photoionization events originating from the two valence orbitals ($\Delta\tau = \tau(\tilde{A}^+) - \tau(\tilde{X}^+)$) can be extracted after a Fourier analysis of the oscillation of SB14. Varying the metal filter allows one to access different energy regions; and the resulting delays as a function of the sideband order are summarized in Figure 7.3 C. Analogous measurements have been performed in $H_2O$, whereby the PE spectrum contains contributions from the $\tilde{X}^+$ and the $\tilde{A}^+$ states located at 12.62 eV and 14.74 eV, respectively.

A brief inspection of the energy variation of the relative delays between the $\tilde{X}^+$ and the $\tilde{A}^+$ states in the two species (cp. Figure 7.3 C and D) reveals a pronounced enhancement of the magnitude of the attosecond delays in $N_2O$ with respect to the ones determined for $H_2O$, with a maximum value of 160 ± 34 as around 31 eV. Overall, the energy variation of $\Delta\tau$ in $H_2O$ is relatively structureless, decaying monotonically as a function of the SB order.

In order to rationalize these results, the authors resort to a theoretical treatment that takes as its starting point the photoionization matrix elements derived from quantum scattering calculations (Gianturco et al., 1994; Natalense and P. P. Lucchese, 1999). The details of the theoretical work were published separately (Baykusheva and Wörner, 2017). Similar to the atomic case, the molecular photoionization delay $\Delta\tau$ can be decomposed into two contributions, a molecule-specific part ($\tau_{\mathrm{W}}$) and a continuum-continuum ($\tau_{\mathrm{cc}}$, measurement-induced) contribution. Whereas the continuum-continuum ("cc") part is mainly sensitive to the Coulomb part of the asymptotic potential and can thus be approximated by the analytical solution for a hydrogen-like potential, $\tau_{\mathrm{W}}$ is strictly system-specific and encodes the signatures of the molecular potential. The subscript "$W$" stands for "Wigner" time delay, following the established nomenclature (Pazourek et al., 2015) for the system-specific part of the photoionization delays. It does not reduce, however, to the delay associated with the photoionization following the absorption of a single XUV photon. Due to the loss of spherical symmetry in molecular systems, the photoionization matrix elements contain contributions from multiple partial waves of varying angular momenta, which are subsequently coupled by the IR field in the dressing step in a non-trivial manner. The calculated results exhibit an overall good qualitative agreement with the experimental data (cp. Figure 7.3 C and D), correctly predicting the local maximum in $N_2O$ around 31 eV and the smooth variation in the $H_2O$-case. Moreover, the analysis in terms of the quantum-mechanical photoionization matrix elements allows the correlation of the enhancement of the photoionization delay for $N_2O$ with the presence of several shape resonances in the investigated energy range. There are two shape resonances (one of $\sigma$- and one of $\pi$-symmetry) embedded in the energy interval between 21.7 eV and 37.2 eV that previous theoretical studies have linked to continua belonging to the $\tilde{A}^+$ state (Braunstein and McKoy, 1987; Rathbone et al., 2005). Two further shape resonances (of $\sigma$- and $\pi$-symmetry) have been predicted for the $\tilde{X}^+$ continuum (Huppert et al., 2016). Generally, a shape resonance is a single-electron phenomenon that occurs when the combined molecular and centrifugal potential experienced by the electron features a finite barrier, through which the ionizing electron can escape by tunneling, yielding a local enhancement in the photoionization cross section (cp. Figure 7.3 E). In the context of photoionization delays, the SR associated with the $\tilde{A}^+$ state leads to enhancement of the delays around 30 eV. This experiment demonstrated that attosecond interferometry, by providing access to the relative phase differences between different ionization continua, can serve as a probe of the time-domain manifestation of shape resonances.

### *7.3 Stereo-Wigner time delays in molecular photoionization of carbon monoxide*

The photoionization matrix elements, whose phase behavior is intrinsically linked to the attosecond time delays probed in the conventional RABBIT experiments, exhibit a strong anisotropy both in the molecular frame as well as with respect to the mutual orientation of the molecule and the XUV polarization direction. The results presented so far, however, represent averages over all possible molecular orientations. Gaining access to the angular dependence of the delays would provide a direct measurement of the spatial localization of the escaping electron wavepacket (EWP) within the molecular potential. In a simple diatomic heteronuclear molecule like carbon monoxide (CO), one could then introduce a molecular photoionization "stereo-Wigner" time delay (Chacon et al., 2014), i.e. the relative time delay between the two possible escape sites:

$$\tau_{\mathrm{SW}} = \tau_{\mathrm{W}}(C - side) - \tau_{\mathrm{W}}(O - side). \tag{7.2}$$

This possibility, theoretically proposed in Ref. (Chacon et al., 2014) was experimentally realized only very recently by Vos *et al.* (Vos et al., 2018) in a configuration combining the RABBIT excitation scheme with the COLTRIMS detection technique briefly described in Section 3.2.4.

Photoionization dynamics is initiated by an APT (extending from harmonic 15 to 27 and centered at 34 eV) generated in an Ar-gas cell with a 30 fs pulse with a central wavelength of 776 nm. A portion of the generating beam is split prior to the HHG step and serves as a perturbation field, whereby the two-color delay is controlled by a piezo-controlled stage. The two pulses are collinearly focused in a supersonic jet expansion of CO molecules using a toroidal mirror, where dissociative photoionization takes place. The thereby generated electrons and fragment ions are extracted by a combination of (dc) electric and magnetic fields and detected in coincidence, whereby the acceptance angle of 4 $\pi$ for both particles allows for a full 3D reconstruction of the fragment momentum distributions.

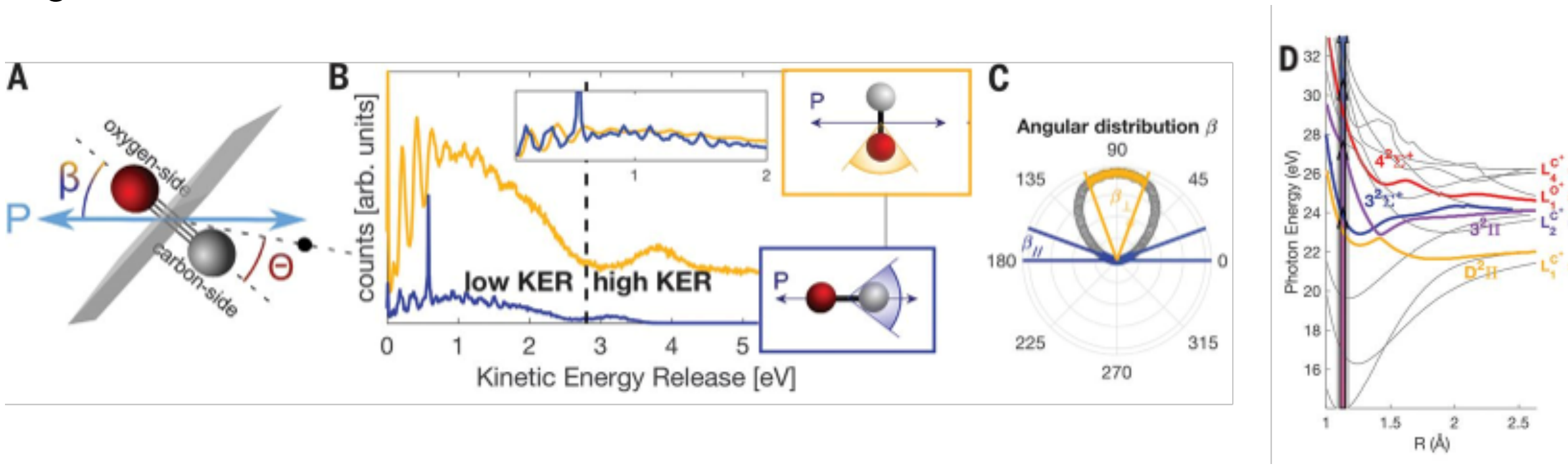

*Figure 7.4: (A) The laboratory frame is defined by the laser polarization axis **P** (blue arrow). The orientation of the molecular axis with respect to the field is denoted by the angle $\beta$, whereas $\theta_e$ corresponds to the emission direction of the electron (black dot) in the molecular frame. (B) Decomposition of the KER spectra in terms of contributions from $\beta_{\parallel}$ (blue) and $\beta_{\perp}$ (yellow), averaged over an emission cone with an opening angle of ± 20° (cp. side panels). The inset indicates the vibrational structure observed in the low-KER (< 2 eV) region. (C) Polar plot of the angular distribution of the recoil angle with respect to the field polarization axis beta. The selection areas corresponding to $\beta_{\perp}$ and $\beta_{\parallel}$ are indicated in orange and blue, respectively. (D) Potential energy curves of the most relevant electronic states of the $CO^+$ molecular ion as a function of the internuclear distance. Figure adapted from (Vos et al., 2018) with permission.*

Out of the three ionization channels lying within the APT bandwidth (one direct channel, $CO + \omega \rightarrow CO^+ + e^-$, and two dissociative channels leading to the production of $C^+$ or $O^+$ ions), only the channel associated with the production of $C^+$-fragments is studied in detail. Due to the substantial degree of spectral congestion (Figure 7.4 D) because of many close-lying states, a detection in state-resolved manner could not be implemented. Therefore, partial electronic state selection in the kinetic-energy release (KER) spectrum is imposed by discriminating between dissociation events where the recoil axis is oriented perpendicular ($\beta_{\perp}$) or parallel ($\beta_{\parallel}$) to the laser polarization (s. Figure 7.4). The angle $\beta$ denotes the relative angle between the internuclear axis and the photoionizing laser field. KER spectra for the two orientations are presented in Figure 7.4 B, whereby the spectrum for perpendicularly oriented molecules is further subdivided into a high-KER region (>2.8 eV) and a low-KER region (< 2.8) eV. With the aid of theoretical calculations of the nuclear dissociation dynamics based on non-adiabatic QM treatment derived from potential energy curves calculated with the MRCI/def2-TZVPP methods, the electronic states contributing to the $\beta_{\perp/\parallel}$-events could be determined. Photoelectron spectra of molecules oriented parallel to laser field are dominated by $^2\Sigma^+$-states ($3^2\Sigma^+$ vs. $4^2\Sigma^+$ at high/low energies, respectively). The spectra of the perpendicularly oriented molecules feature contributions from the $D^2\Pi$- and $3^2\Sigma^+$-states at low KER, and contributions from the $3^2\Pi$-state at high energies. As in the preceding examples, the SB intensity is monitored as a function of the two-color delay and is given by Eq. (7.1), whereby the oscillation phase can be written as: $\Delta\phi_{\text{tot}} = \Delta\phi_{2q} + \Delta\phi_{\text{W}} + \Delta\phi_{\text{cc}}$.
In this case, however, one monitors the relative delay between electrons emitted from different locations of a single molecule. This implies that both the XUV-( $\Delta\phi_{2q}$) and the measurement-induced ($\Delta\phi_{\text{cc}}$) contributions cancel out, yielding a quantity directly proportional to the stereo Wigner time delay (SWTD):

$$\tau_{\text{SW}} = \frac{\Delta\phi_{\text{W}}(C - side) - \Delta\phi_{\text{W}}(O - side)}{2\omega}. \tag{7.3}$$

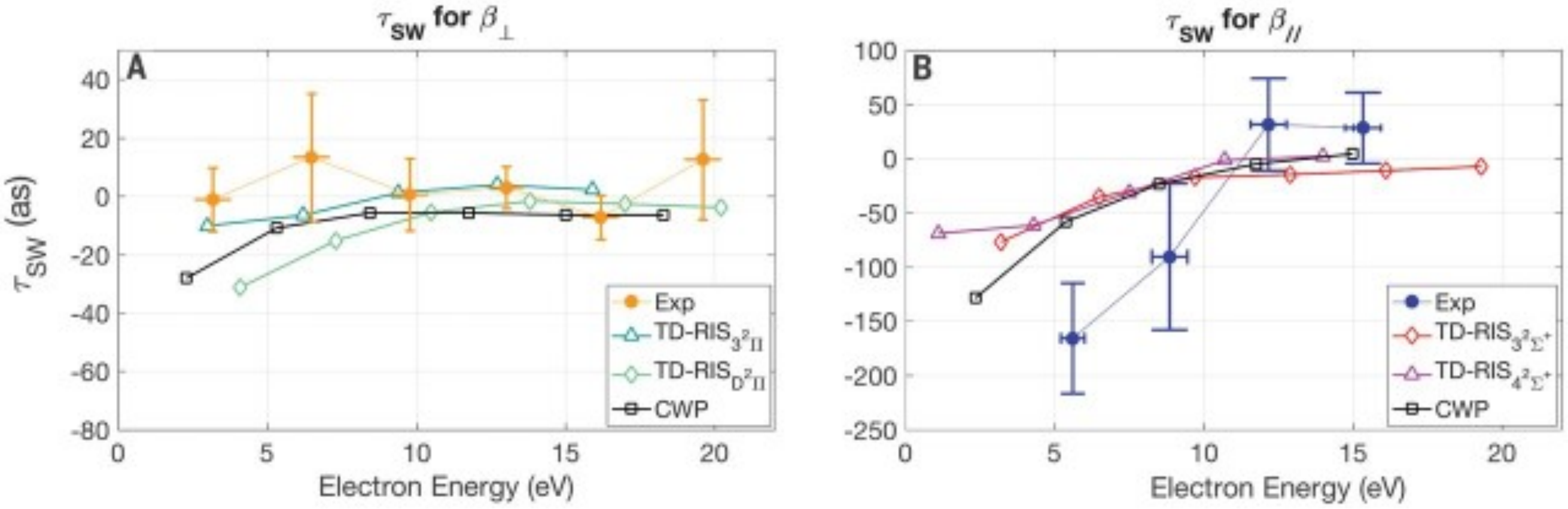

*Figure 7.5: (A) SWTDs for molecules oriented perpendicularly with respect to the laser field, integrated over the entire KER region. The experimental results, shown in orange, suggest that the SWTDs amount to zero within the error bars. This trend is reproduced in the theoretical delays calculated with different models (cp. legend). (B) Same as (A), but for the SWTDs of molecules aligned parallelly to the field. Figure adapted from (Vos et al., 2018) with permission.*

The experimental results for $\beta_{\perp/\parallel}$, obtained after Fourier analysis of the sideband oscillations (averaged over an energy region of 1 eV centered at each SB), are displayed in Figure 7.5. In the case of perpendicularly-oriented molecules, the SWTDs lie within ±35 as and are close to zero resp. slightly positive in the investigated energy interval. This behavior is expected from symmetry considerations and is further confirmed by the results of theoretical calculations based on two different models (s. below). In the case of parallel-oriented CO molecules, the SWTDs exhibit an evolution from strongly negative (-165 as at 5.0 eV) to slightly positive (+30 at 14.4 eV) values. This behavior implies that the SWTDs changes sign as a function of energy, i.e. at low energies, the EWP escapes into the continuum via the C-site, whereas the opposite applies for higher kinetic energies. In order to track the origin of this trend, theoretical calculations based on the following two independent approaches have been considered: (*i*): the TD-RIS method described in detail in Refs.(Spanner and Patchkovskii, 2013, 2009), which accounts for the XUV-initiated dynamics in a fully quantum-mechanical manner, but does not explicitly treat the IR-induced dynamics; and (*ii*): the classical Wigner propagation (CWP) method(Vos et al., 2018) based on propagating the Wigner function of the photoelectron in a classical manner, with the IR-field included. Both calculations capture qualitatively the experimentally observed trend for $\beta_{\perp}$ and $\beta_{\parallel}$, thus revealing the sensitivity of the measured delay to the molecular orientation. There are two potential contributions that could give rise to the observed asymmetry in the SWTDs: either an asymmetry in the initial localization of the EWP, or an asymmetry in the molecular potential experienced by the receding EWP. The contribution of the molecular potential can be ruled out after analysis of the dipole moments of the ionic states associated with each channel, thus identifying the role of the initial localization, i.e. the mean position of the electron along the C-O-bond) of the ionization event as the determining factor for the observed SWTD. This last conclusion is further supported by analyzing the Wigner function of the dipole matrix element of the Dyson orbitals for each of the contributing states. This quantity represents a coordinate and momentum representation of the EWP at the instant of the photoionization event, and its anisotropy directly reflects the asymmetry of the observed SWTD.

### *7.4 Phase-resolved two-color multiphoton ionization of chiral molecules*

A complementary approach to studying photoionization dynamics in the time domain has been introduced in Ref. (Zipp et al., 2014). Instead of employing single-photon ionization by XUV pulses as in the RABBIT technique, this measurement scheme relies on multiphoton ionization by UV pulses centered at 400 nm. Similar to the RABBIT technique, an IR dressing field centered at 800 nm is used to create photoelectron sidebands that can be accessed through two interfering pathways. In both cases, temporal information on the sub-femtosecond time scale is extracted by analyzing the phase of the side-band-intensity oscillations as a function of the delay between the ionizing and dressing pulses. This measurement scheme has been applied to rare gas atoms in Ref. (Zipp et al., 2014) and Ref. (Gong et al., 2017). A combination of a strong UV (400 nm) femtosecond pulse and an 800 nm perturbing field was used in Ref. (Zipp et al., 2014) to extract the intrinsic and measurement-induced phase delays in the ATI of argon atoms in the region 2-15 eV, revealing a strong dependence of the observed phase shifts on the UV field intensity and thus on the ponderomotive potential. In Ref. (Gong et al., 2017), on the other hand, a combination of orthogonally polarized 400 nm + 800 nm fields of equal intensities was used to study the photoemission dynamics of the Freeman resonance via the field-dressed *5p* or *4f* Rydberg states of Ar. While seemingly similar to the RABBIT technique, it is important to emphasize that the multiphoton-ionization scheme does not have any known straightforward relationship to the photoionization dynamics, in pronounced contrast with the RABBIT technique (Dahlström et al., 2012; Klünder et al., 2011). Experimentally, this fact becomes apparent through the strong dependence of the measured delays on the intensity of the ionizing laser pulse, which was reported in Ref. (Zipp et al., 2014) and is absent in the RABBIT scheme. Theoretically, the origin of the time delays measured by the multiphoton technique has been traced to retrapped resonant ionization (Song et al., 2018) and the different numerical values of the time delays compared to those measured by the RABBIT technique have been given (see Fig. S7 therein).

The multiphoton measurement scheme has been applied to the photoionization of chiral molecules by Beaulieu *et al*. (Beaulieu et al., 2017). Specifically, they addressed the topic of the differential response between the two enantiomers of a given molecule when subject to multiphoton ionization. Two-color multiphoton ionization enables access to certain time-domain aspects of the chiral response, which is given in this case by photoemission into the forward or the backward directions (with respect to the polarization of the ionization field).

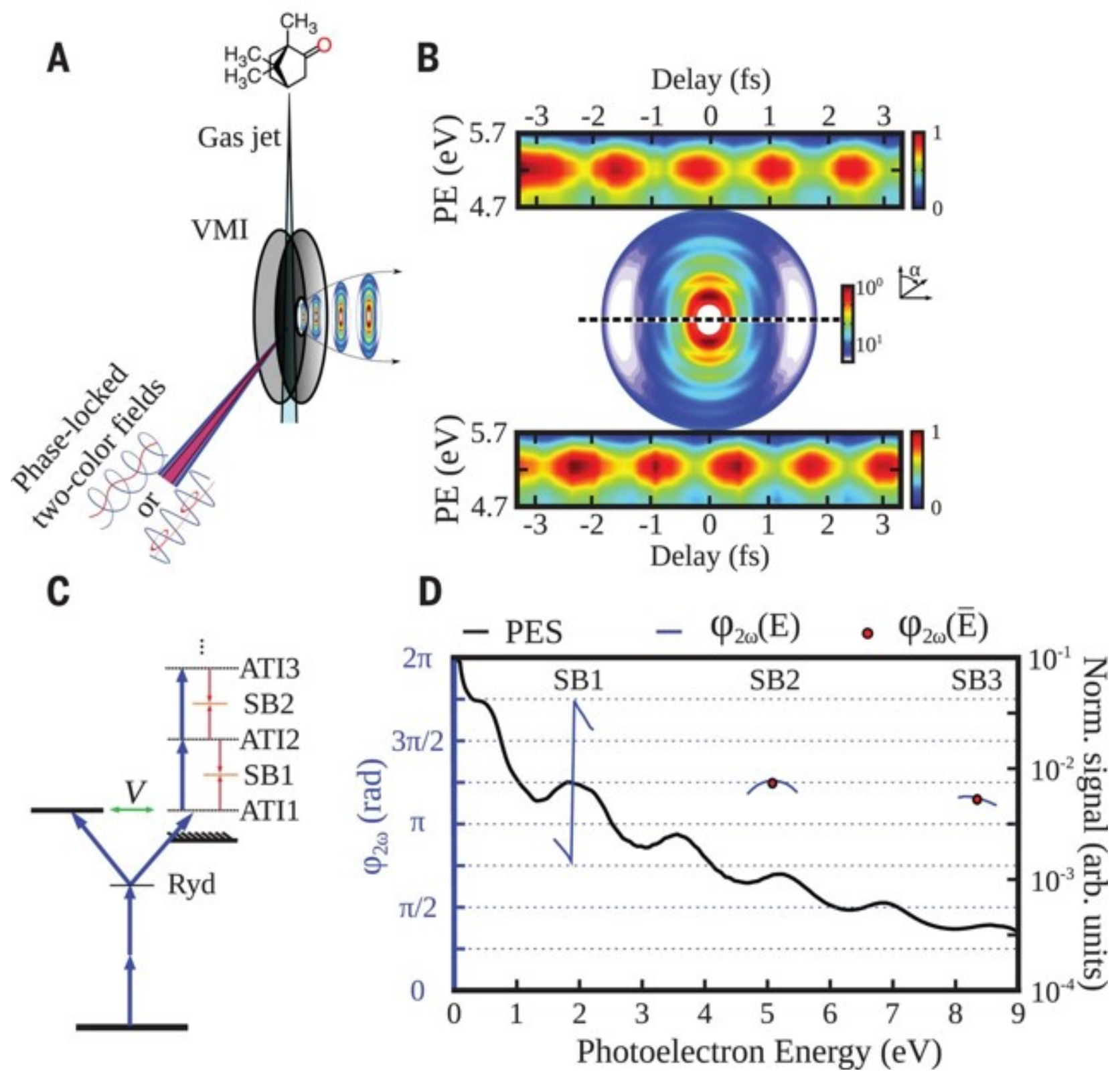

*Figure 7.6: (A) Schematic drawing of the experimental setup, comprising two phase-locked femtosecond laser pulses with either linear or circular polarization which are focused into a jet of enantiopure camphor molecules in the interaction zone of a VMI spectrometer. (B) Typical results for the photoelectron angular distribution as well as the oscillations of the second sideband as a function of the two-color delay. The phase shift (π) between the signals in the upper resp. lower half of the distribution reflect the up-down asymmetry of the ionizing field. (C) Schematic illustration of the physical mechanism underlying photoelectron interferometry. The UV field (blue arrows) induces multiphoton ionization, leading to ATI, whereby the first ionization transition is located close to an autoionizing resonance. The IR field (red arrows) subsequently induces additional transitions, leading to the creation of sidebands between the main ATI peaks. (D) Angle-integrated photoelectron spectrum comprising the main ATI peaks as well as the sidebands of orders 1-3. The oscillation phases (2ω) of each band, integrated over the emission angles, are shown as a function of energy (blue lines) as well as spectrally averaged (red dots, only for the nonresonant SBs 2 and 3). Figure adapted from* (Beaulieu et al., 2017) *with permission.*

The detection scheme of Beaulieu *et al.* is based on photoelectron circular dichroism (PECD) effect (Powis, 2008; Ritchie, 1975), which is defined as the asymmetry in the photoelectron angular distribution (typically recorded in a VMI apparatus, cp. Section 3.2.4) with respect to the light propagation direction and takes place when a chiral species is subject to photoionization in a circularly polarized laser field (s. illustration in Figure 7.6 A).

The interferometric technique employed in this study, relies on the technique introduced in Ref. (Zipp et al., 2014). In the experiment, a UV laser field centered at 400 nm is used to induce ATI in camphor, a bicyclic ketone with an ionization potential of 8.76 eV. The electronic structure of camphor ($C_{10}H_{16}O$) is characterized by Rydberg states around 6.2 eV, and the ionization process takes place via a *2+n*-REMPI pathway, whereby $n$ denotes the order of the ATI band. Superimposing the ionizing UV radiation with a weak IR component at 800 nm leads to the appearance of sidebands between the ATI comb. The two-color field is asymmetric with respect to the PE emission axis, consequently, the electrons ejected in the forward or backward directions are modulated with opposite phases (cp. Figure 7.6 B). This implies that the phases evaluated for the upper and the lower halves of each recorded PAD image are shifted by $\pi$ prior to comparison. The angularly integrated phases of the three sidebands resolved in the experiment are displayed as a function of the photoelectron kinetic energy in Figure 7.6 D. Whereas the second and the third sidebands (SB2 and SB3) exhibit a smooth phase variation across the ATI bandwidth, the first SB (SB1) is characterized by an abrupt discontinuity ($\approx\pi$ in magnitude), which is interpreted as a signature of the autoionizing resonance located at 1.9 eV. Studying the phases of the individual sidebands thus gives access to the chiral photoionization dynamics in two different regimes (resonant vs. non-resonant). In the non-resonant case, the phases reported are averaged over the bandwidth of the corresponding SB, whereas this treatment is not applicable for the first SB. The experiments results in Ref. were analyzed in a framework analogous to Eq. (7.1), although the latter is not applicable to the ATI measurement scheme (see e.g. Ref. (Song et al., 2018)):

$$\tau_{\mathrm{tot}} = \tau_{\mathrm{UV}} + \tau_{\mathrm{W}} + \tau_{\mathrm{cc}}. \tag{7.4}$$

Here $\tau_{\mathrm{UV}}$ corresponds to the instant of the ionization event triggered off by the UV pulse, $\tau_{\mathrm{W}}$ is the delay resulting from the influence of the molecular potential, and $\tau_{\mathrm{cc}}$ is the additional delay induced by the absorption of the weak IR perturbing field, and mainly sensitive to the asymptotic tail of the molecular potential (s. below). In order to eliminate the measurement-induced contribution encoded in $\tau_{\mathrm{UV}}$, one can define and extract a quantity corresponding to the relative photoemission delay $\Delta\tau^{\mathrm{f/b}}$ between electrons emitted in forward (*f*) or backward (*b*) directions, $\Delta\tau^{\mathrm{f/b}} = \Delta\tau^{\mathrm{f}} - \Delta\tau^{\mathrm{b}}$. Disentangling the remaining two contributions associated with $\Delta\tau_{\mathrm{W}}^{\mathrm{f/b}}$ resp. $\Delta\tau_{\mathrm{cc}}^{\mathrm{f/b}}$ is performed by measuring $\Delta\tau^{f/b}$ using two different combinations of linear (LP) vs. circularly-polarized (CP) light in the pump (UV) resp. probe (IR) steps. Choosing the polarization of the UV field to be circular while setting the IR to linear essentially restricts the chiral discrimination to the ionization step, implying that $\Delta\tau^{\mathrm{f/b}} = \Delta\tau_{\mathrm{W}}^{\mathrm{f/b}}$. In contrast, employing a linearly polarized UV pulse erases the asymmetry in the Wigner time delay and allows one to isolate the chiral contribution to the continuum-continuum delay, $\Delta\tau^{\mathrm{f/b}} = \Delta\tau_{\mathrm{cc}}^{\mathrm{f/b}}$. This treatment is based on the tacit assumption that the transitions induced by the UV and the IR fields are strictly independent, in which case the contributions to $\Delta\tau^{\mathrm{f/b}}$ can be treated in an additive manner as implied by Eq. (7.4). They are additionally based on the assumption of a specific time ordering of the interactions, i.e. UV interaction followed by IR interaction. Whereas this approximation is well justified in the RABBIT scheme (see Ref. (Dahlström et al., 2012)), it has no rigorous justification in the multiphoton scheme because of the similar frequencies of the employed laser pulses. Experimentally, $\Delta\tau^{\mathrm{f/b}}$ are measured for each of the two camphor enantiomers and the obtained values are subsequently averaged for the sake of improved accuracy. The values $\Delta\tau^{\mathrm{f/b}}$ switch sign on changing the enantiomer or, equivalently, the helicity of the CPL field, as expected from symmetry reasons.

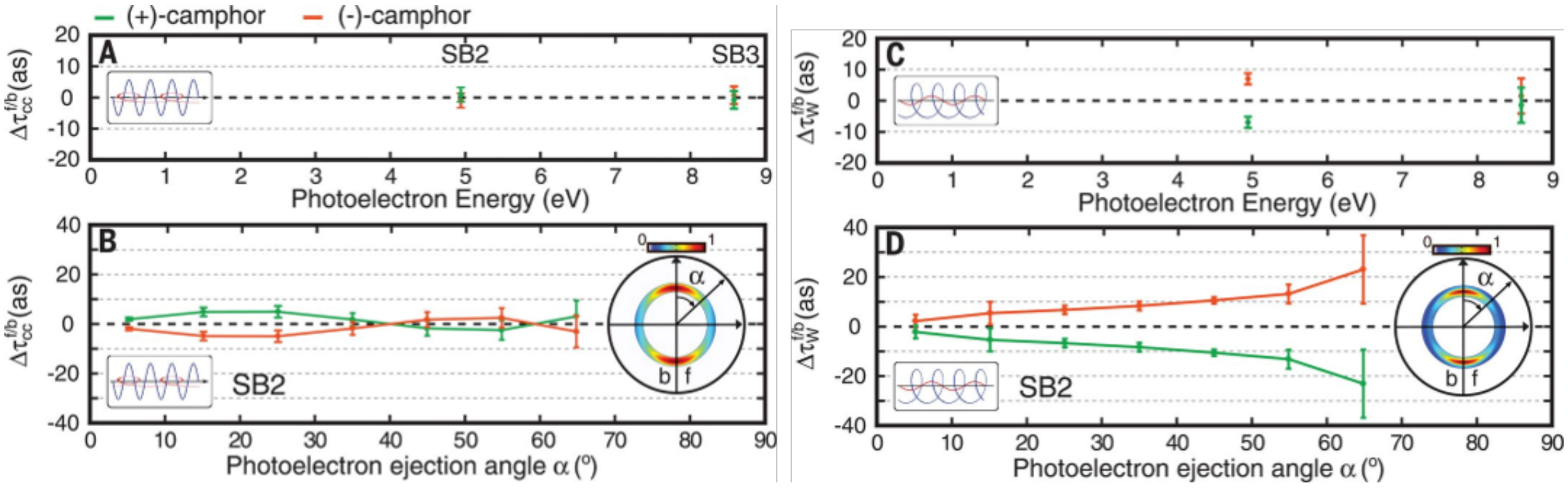

*Figure 7.7: (A) Forward/backward differential delays in nonresonant photoionization of camphor for the experimental scheme employing a linearly-polarized UV field and a left circularly-polarized IR field. (B) Angularly-resolved differential delays corresponding to the results presented in (A) for the second sideband. The inset shows the normalized PAD for the SB2 as well as the definition of the ejection angle alpha. (C), (D): Same for (A), (B) for the experimental configuration employing a left-circularly polarized UV field and a linearly polarized IR. Figure adapted from* (Beaulieu et al., 2017) *with permission.*

In Section 7.2, it was argued that the continuum-continuum contribution is essentially insensitive to the molecular potential as it probes mainly the Coulombic tail of the asymptotic potential. This is also reflected in the current experimental results for angle-averaged $\Delta\tau_{cc}^{f/b}$ for SB2 and SB3 (Figure 7.7 A), which are equal to zero within the experimental accuracy of ±2 as. The chiral signatures can be isolated only after resolving the angular dependence of the photoionization process by integrating the PAD in slices of 10° centered around different ejection angles $\alpha$, measured with respect to the polarization plane of the IR (cp. insets in Figure 7.7 B and D). Even in this case, the chiral signal is weak, and maximal time delays are observed for SB2 for an ejection angle of 25° (5±2 as). For SB3, even in the angular resolved case, no difference between forward and backward emission is detected. This is consistent with the notion that the IR-induced transitions take place away from the core region and $\Delta\tau_{cc}^{f/b}$ is thus less sensitive to the chiral features of the system. For the Wigner-like contribution to the f/b delay, $\Delta\tau_{W}^{f/b}$, (measured by breaking the f/b symmetry in the ionization step by employing a CPL UV pulse), the angle-averaged delays amount to ≈7 as in the case of SB2 (Figure 7.7 C), whereas the photoemission-angle-resolved delays reach up to 24 as for an ejection angle of 60-70° (Figure 7.7 D). Similar to the cc-contribution, the Wigner part $\Delta\tau_{W}^{f/b}$ for SB3 is not sensitive to the chiral character of the PI process. This behavior reflects the diminished sensitivity of electrons with high-kinetic energy to the asymmetric character of the molecular potential.

Finally, we briefly discuss the resonant case, whereby the angle-integrated phase delays for each of the two polarization combinations are presented in Figure 7.8 as a function of the photoelectron energy. The spectral phase exhibits a phase jump with a magnitude of ≈0.75 rad around 2.1 eV when a CPL UV is employed (panel A), with a slight difference between electrons emitted in the forward and the backward directions that is mirrored for the two enantiomers (panel B). Contrary to the non-resonant case, breaking the f/b symmetry with a CPL IR field leads to a very distinct signature in the spectral phase – a phase jump of nearly 0.9 $\pi$, in opposite direction for forward/backward electrons (Figure 7.8 C and D). This result shows that in the vicinity of a resonance, the chiral character of the photoemission process can be probed with the aid of a weak IR-induced cc-transition.

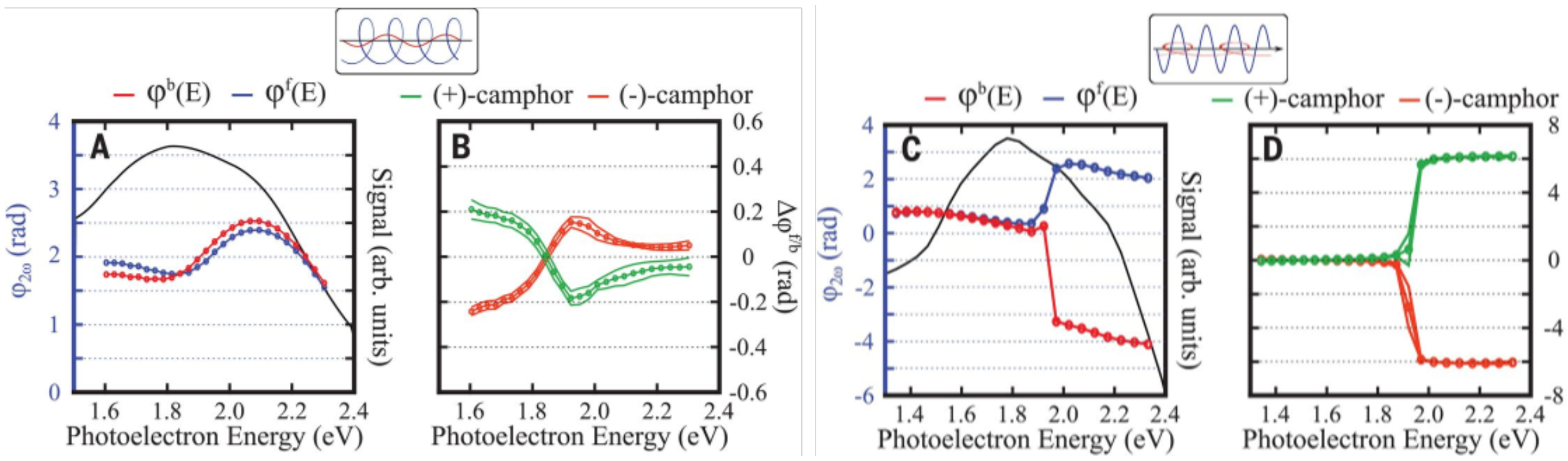

*Figure 7.8: (A) and (C): Spectral amplitudes (black lines) and spectral phases ($\varphi^f(E)$, blue; $\varphi^b(E)$, red) of the resonant sideband SB1 in (1R)-(+)-camphor. In (A), the UV is left-circularly polarized and the IR is linear, whereas in (C) the polarization of the fields is reversed. (B) and (D) Forward/backward spectral phase asymmetry between the two enantiomers of camphor, using a left circularly-polarized UV field and a linearly polarized IR in (B) and a linearly polarized UV/left circularly-polarized IR in (D). Figure adapted from* (Beaulieu et al., 2017) *with permission.*

This work, although restricted to the particular example of camphor, outlines a general approach for tracking the temporal evolution of chiral phenomena with attosecond temporal resolution. Additional theoretical work is required to establish the relation between the measured quantities and the concept of Wigner delays in the photoionization of chiral molecules.

## 8. Attosecond transient absorption spectroscopy

In this section, we review the current advances in attosecond transient absorption spectroscopy (ATAS), a technique which has now become well-established in the realm of femtosecond laser spectroscopy and has been only recently extended to the attosecond domain (for recent reviews, s. (Beck et al., 2015; Gallmann et al., 2013; Kraus et al., 2018; Kraus and Wörner, 2018a; Leone and Neumark, 2016; Ramasesha et al., 2016)). In many respects, ATAS can be viewed as a complementary technique to the HHS method presented in Section 6 as well as the NIR+XUV methods covered in Sections 4-5. HHS possesses a temporal resolution of a fraction of the optical cycle, determined by the duration of an electron's continuum trajectory. However, due to the exponential dependence of the SFI rate on the ionization potential, it is mainly sensitive to valence-state dynamics, and, moreover, relies heavily on the availability of theoretical models for retrieving the electronic structure information from experimental observables. The two-color XUV-pump-NIR-probe techniques (XUV+NIR) presented in Section 4 represent a reliable and versatile approach for directly probing the attosecond dynamics, but also have inherent limitations. The majority of the experiments realized so far rely on the detection of charged particles, providing insight only to the dynamics in the ionized system, typically in a highly-excited state. Moreover, charged particle detection is accompanied by a multitude of drawbacks, such as the susceptibility to space-charge effects, stray magnetic and electric fields, limited detection speed and the presence of background signals due to the strong fields employed. ATAS, on the other hand, relies on photon detection, thus obviating the need for any additional electric or magnetic fields, and can benefit from high detection sensitivities due to the advances in charge-coupled device detector technology.

In addition, high spectral resolution can be achieved, without compromising the bandwidth of the XUV/SXR pulses. The basic experimental scheme does not entail ionization, hence ATAS can serve as a sensitive probe for bound-continuum transitions and resonances.

*8.1 Dynamics of Rydberg and valence states in molecular nitrogen probed by ATAS*

The first example we will cover (Warrick et al., 2016) represents one of the few ATAS studies addressing sub-cycle electron dynamics in molecular systems (Cheng et al., 2014, 2013; Reduzzi et al., 2013). The work of Warrick *et al.* (Warrick et al., 2016) centers on molecular nitrogen ($N_2$) to highlight the possibility of investigating quantum-beat-dynamics between multiple bound and autoionizing electronic states of valence or Rydberg character, a task which has become accessible only with the advent of attosecond laser technology in the view of the wide energy separations between individual electronic states. Prior to this point, coherent wavepacket dynamics were investigated in the vibrational and the rotational domain using narrow-bandwidth femto- or picosecond sources. The electronic wavepacket dynamics in this simple molecule reveals many similarities to the extensively studied case of noble-gas atoms, but also contains novel features unique to diatomic molecules.

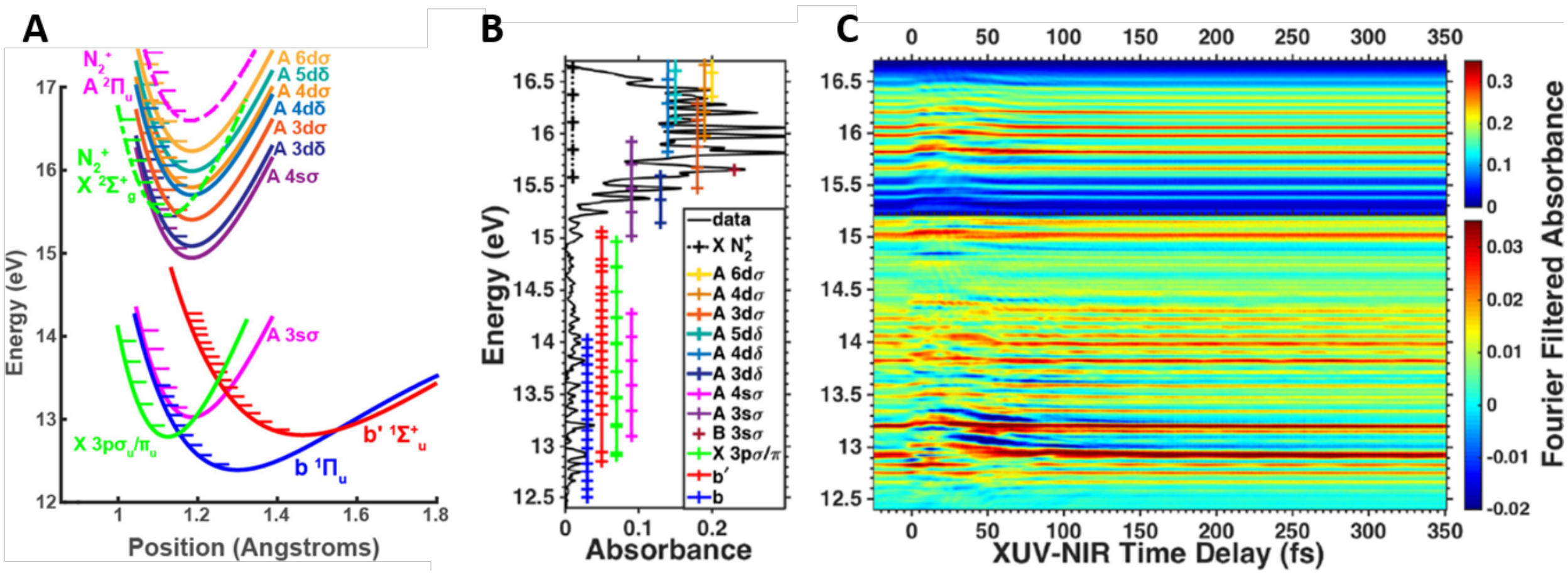

*Figure 8.1: (A) Potential energy curves of molecular nitrogen in the investigated energy region. (B) Energy levels relevant for the static absorption of nitrogen. The diagram lists the positions of the vibrational levels of the valence b $^1\Pi_u$ (blue) and b′ $^1\Sigma_u^+$ (red) states and of the Rydberg series built on the X $^2\Sigma_g^+$ $N_2^+$ (green) and A $^2\Pi_u$ $N_2^+$ cores. (C) Transient absorption spectrum of nitrogen recorded as a function of photon energy (12.5-16.7 eV) and time delay between XUV and IR fields, represented as a Fourier-filtered absorbance. The NIR pulse arrives after the XUV for positive time delays. Figure adapted from (Warrick et al., 2016) with permission.*

The experimental scheme employs 25 fs pulses derived from a 2 mJ, 1kHz CEP-stabilized system which are subsequently spectrally broadened in an HCF filled with Ne-gas and compressed in a chirped-mirror assembly to a duration of 6 fs at 780 nm (0.8 mJ). The more intense part (70 %) is re-shaped in a double-optical-gating scheme (using two quartz plates and a BBO crystal) and used for attosecond pulse generation based on HHG in a static gas cell filled with Xe. An indium foil (200 nm) is used to block the residual IR. The XUV pump and the NIR probe (less-intense portion of the few-cycle pulse) are focused using a toroidal mirror resp. a spherical mirror and recombined collinearly by means of an annular mirror into the interaction region consisting of a 1-mm-long gas cell filled with $N_2$. The temporal resolution of the piezoelectric stage used to delay the NIR is

~100 as. A second indium filter is employed to block the NIR after the target cell, and the transmitted XUV is dispersed and detected using a flat-field spectrometer equipped with an X-ray CCD camera. The experimental data is subject to a Fourier-filtering procedure and presented as absorbance. The static absorption lines in the energy region from 12.5 to 16.7 eV corresponding to the various transitions are indicated in panel B of Fig. 8.1, whereas the transient absorption spectrum covering pump-probe delays up to 350 fs is displayed in Panel C.

The indium filter employed in the pump arm limits the excitation bandwidth of the XUV to 11-17 eV. The electronic structure of $N_2$ in this spectral region (cp. Fig. 8.1 A) is characterized by the presence of two valence states ($b\,^1\Pi_u$ and $b'\,^1\Sigma_u^+$) as well as Rydberg series built on the ground state of the ion (*np* series converging to $X\,^2\Sigma_g^+$ of $N_2^+$) and its first two excited states $A\,^2\Pi_u$ (*ns* and *nd* series) and $B\,^2\Sigma_u^+$ (*3s* series). As evident from Figure 8.1 C, the spectrum at negative delays is dominated by static absorption features associated with the dipole-allowed transitions to the vibrational levels of the valence $b\,^1\Pi_u$ and $b'\,^1\Sigma_u^+$ states (12.5-15 eV), whereby the presence of the first Rydberg states as well as the onset of predissociation cause irregularities in the spectral features. The transitions to the Rydberg series converging to the ionic states start to play a role in the region above the first ionization limit to the $v=0$-state of $X\,^2\Sigma_g^+$ $N_2^+$ located at 15.559 eV. At negative delays, the static features are not affected by the NIR due to its low intensity. In the temporal vicinity of the NIR/XUV-overlap, the features shift to higher energies, and in addition, become weaker or negative (i.e. emissive) at the central transition energy. The temporal duration of the shift affecting states between 15.3 and 16.7 eV is not uniform as a function of energy and shifts to longer values (from 20 to 80 fs) as the transition energy increases. The long-time behavior of some of the transition lines displays oscillatory character, whereby the beating period varies strongly with energy at 12.85 eV, 5-10 fs at 14.5 and 16.5 eV, and 1.3 fs at 13.8 eV. The analysis of these features employs a many-level non-perturbative model that treats each vibrational level or Rydberg level as a separate state. The absorption strength measured in a TAS experiment is directly proportional to the imaginary part of the time-dependent dipole moment in the frequency domain:

$$d_1(t) \propto e^{-\frac{t}{t_1}}\mu_{1g}^2\left[|A_1^1(t-\tau)|\,sin\big(E_1t+\varphi_{A_1^1}\big)+\sum_{n,n\neq 1}\frac{\mu_{ng}}{\mu_{1g}}|A_n^1(t-\tau)|\,sin\big(E_1t+\varphi_{A_n^1}+\Delta E_{n1}\tau\big)\right].\quad(8.1)$$

In the above, $t_1$ denotes the lifetime, $\mu_{ng}$ is the dipole coupling between state *n* and the ground state (*g*), $\tau$ is the NIR-XUV delay, $E_n$ is the energy of the *n*th state, and $\Delta E_{n1} = E_n - E_1$ is the energy difference. The terms $A_1^1$ and $A_n^1$ capture the depletion of state 1 through ionization resp. population transfer to other states *n*. The physical model behind Eq. (8.1) is essentially based on the single-electron approximation discussed in Section 2. Its application to molecular systems can be justified only for the case of Rydberg states, which are essentially decoupled from the molecular core, and does not capture non-adiabatic effects due to coupling with the nuclear motion.

The second term in Eq. (8.1) encodes the population transfer coupling the initial state $E_1$ with another state of energy $E_n$. In the frequency domain, the sinusoidal dependence of this term on the time delay τ and the energy separation $\Delta E_{n1}$ between the two levels translates into a persistent oscillation of the absorption feature. In this framework, the sub-cycle quantum beats with a period of 1.3 fs (half the period of the NIR) can be assigned to interferences between two quantum paths transferring population between two levels spaced by twice the NIR photon energy. This population transfer pathway involves the absorption of two NIR photons and involves the mediation of an XUV dipole-forbidden (dark) state. In molecular $N_2$, the potential candidates for mediating the two-photon coupling pathway are several dark Rydberg states built on the $A$ and $X$ ion-state cores (cp. scheme in Fig. 8.2 A). The most dominant spectral features arising through this pathway are located around 15.8 eV and result from the population transfer to the $A\ 4d\ \delta\ (v = 0)$ -state and the $A\ 3d\ \sigma\ (v = 2)$ state from the $3p\ \sigma/\pi$ – Rydberg states (built on the $X\ {}^2\Sigma_g^+$ $N_2^+$ core at 12.95 eV). Additional examples are the coupling between the $A\ 4d\ \delta\ (v = 2)$ -state (16.4 eV) and the $A\ 3s\ \sigma\ (v = 0)$ at 13.1 eV. Besides the population transfer pathways between Rydberg states of different ion cores and orbital symmetries, which have also been observed in atoms, the current experiment reveals the presence of interferences between states of Rydberg and valence character: the oscillations assigned to the $v = 2 - 4\ b^1\Pi_u$-state levels.

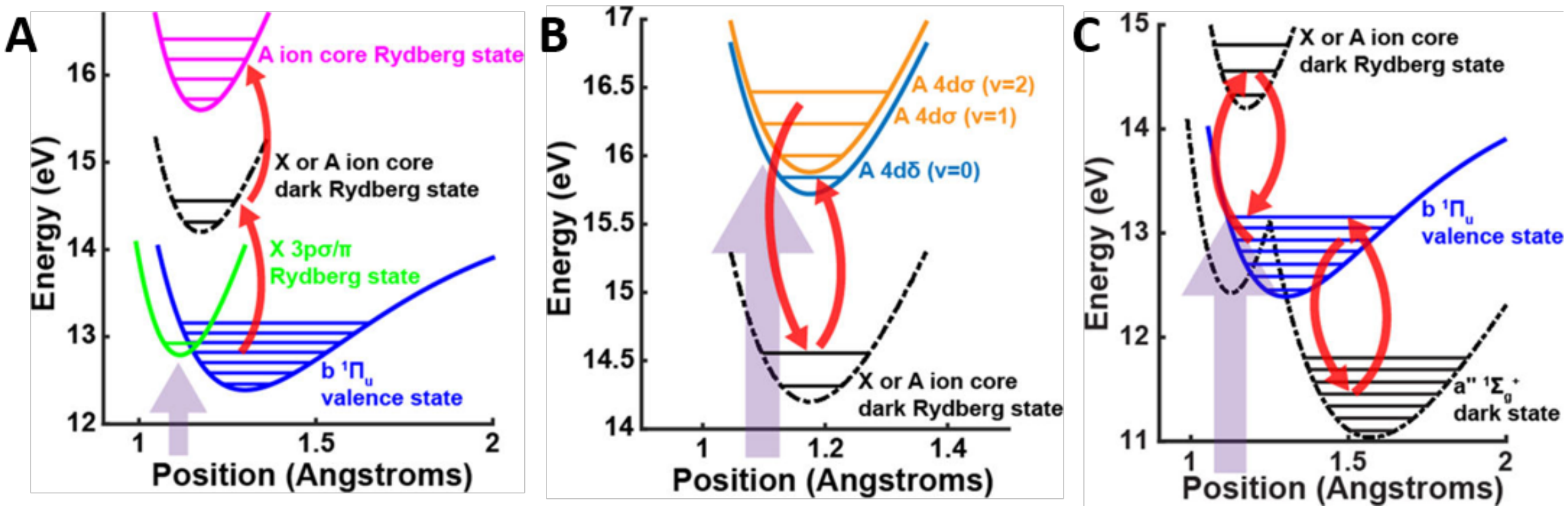

*Figure 8.2: (A) Scheme illustrating the origin of the sub-cycle oscillations caused by interferences between two levels spaced by double the NIR frequency. A dark state built on the $A\ {}^2\Pi_u$ $N_2^+$ core is shown to illustrate the Franck-Condon overlap with the $b\ {}^1\Pi_u$ vibrational levels. (B) Scheme illustrating the origin of the slow oscillations in the $4d\delta\ v = 0$ state at 15.85 eV. The black curve shows a dark state built on the A-core. (C) Scheme illustrating the origin of the slow oscillations in the vibrational levels of the $b\ {}^1\Pi_u$ valence state. The black curves indicate two potential intermediate dark states: a Rydberg state built on the A core at 14 eV as well as the valence $a''$ dark state at 11 eV. In all panels, the population transfer is illustrated by the red arrows, whereas the Franck-Condon overlap with the ground state of nitrogen is indicated with the purple arrow. Figure adapted from (Warrick et al., 2016) with permission.*

The few-fs (5-10 fs) oscillations, on the other hand, originate from the population transfer between two close-lying resonances via a dark state located in energy above or below the resonances via a 'Lambda' or 'V'-like coupling scheme. An essential requirement for the realization of this scheme, which has been also observed in atomic systems, is the presence of two NIR photons of differing energy, which is fulfilled given the substantial bandwidth of the few-cycle NIR (1.4 – 2.2 eV). In molecular nitrogen, the 'V'-coupling scheme is realized by the $A$-Rydberg states above 15 eV that are coupled to the dark $X$ or $A$ ion-core Rydberg states around 14 eV (cp. Fig. 8.2 B). One of the most prominent features in the current experiment are the beatings with frequencies of 11.5 and 7 fs, assigned to the $A\ 4d\ \delta\ (v = 0)$ – state at 15.85 eV. These oscillations result from the population transfer with the $A\ 4d\ \sigma\ (v = 1)$ and $(v = 2)$ levels and represent the first observation of coherent beating between vibrational levels belonging to different electronic potential energy curves, a feature unique to molecular systems. The beatings between vibrational states belonging to the same electronic states give rise to the longer-period oscillations with a vibrational period of 51 fs, matching the vibrational period of the $b$-state. The vibrational beatings thus involve levels of the $b$-state separated by $\Delta v = \pm 1$ that are coupled by the NIR field (cp. scheme in Fig. 8.2 C).

Finally, we briefly comment on the origin of the energy shifts of the absorption features observed in the vicinity of time zero. Following a theoretical analysis developed for atoms, the authors attribute the observed dynamics to the ac-Stark effect induced by the few-cycle NIR pulse. This effect is modeled by the introduction of the laser-imposed phase term $\varphi_{A_n^1}$ in Eq. (8.1). The NIR-mediated coupling between a resonant excited state and other dark states leads to an energy shift of the level that can be incorporated as a phase-shift of the time-dependent dipole moment. The increasing time duration regarding the shifts of the absorption features is explained by the increasing overlap of the energy levels in the high-density region of the $A$-Rydberg states as they approach the ionization potential.

In summary, the observed quantum beating dynamics were attributed to the energy repartitioning between electronic and nuclear degrees of freedom mediate by the perturbative NIR laser field. Whereas the results could to a major extent be interpreted using the theoretical framework built upon the single-electron response in atomic systems, several unique features arising from the nuclear degrees of freedom were identified.

### *8.2 Time-resolved X-ray absorption spectroscopy using a table-top high-harmonic source*

The rest of this section is dedicated to the recent progress in the development of table-top time-resolved X-ray absorption techniques (TR-XAS). Broadband soft-X-ray (SXR) continua with photon energies beyond the XUV range (10-124 eV) represent a powerful diagnostic tool due to their sensitivity to the local chemical structure (via the excitation from a specific core level of a given atom) including spin and oxidation state. Coupled with time-resolved NIR or optical ultrafast excitation schemes, this fingerprinting-capability has the potential to track the dynamics of oxidation and spin states with ultrafast temporal resolution and element specificity. Until recently, time-resolved studies employing XAS were limited to large-scale synchrotron facilities or free-electron lasers and were mostly confined to the study of sub-ps dynamics in the condensed phase. The advent of HHG-based light sources capable of producing sub-50 fs SXR pulses with full spatial and temporal coherence permitted the realization of time-resolved x-ray absorption measurements in a table-top manner.

We will illustrate this recent progress on the example of two pioneering studies (Attar et al., 2017; Pertot et al., 2017) reporting the realization of TR-XAS experiments with femtosecond temporal resolution employing SXR supercontinua extending beyond 160 eV and up to 350 eV, i.e. partially covering the chemically and biologically highly relevant energy range of 282 eV to 533 eV where water becomes transparent ("water window") and where the K-edge of C is located (282 eV). Prior to these two studies, XAS experiments were limited to ≈100 eV, covering only the element-specific edges in the XUV. These experiments enabled the tracking of the dynamics following strong-field ionization in a variety of systems such as Xe atoms (Loh et al., 2007), Kr (Goulielmakis et al., 2010) and thin-film silicon (Schultze et al., 2014). Studies centering on molecular systems have so far addressed the SFI-induced vibrational wavepacket dynamics in $Br_2$ (Hosler and Leone, 2013), the SFI-induced dissociation of dibromomethane (Chatterley et al., 2016a) and ferrocene (Chatterley et al., 2016b), the evolution of transition state during the SFI-induced dissociation of $CH_3I$ (Attar et al., 2015), and the ring-opening reaction of selenophene (Lackner et al., 2016) following ionization.

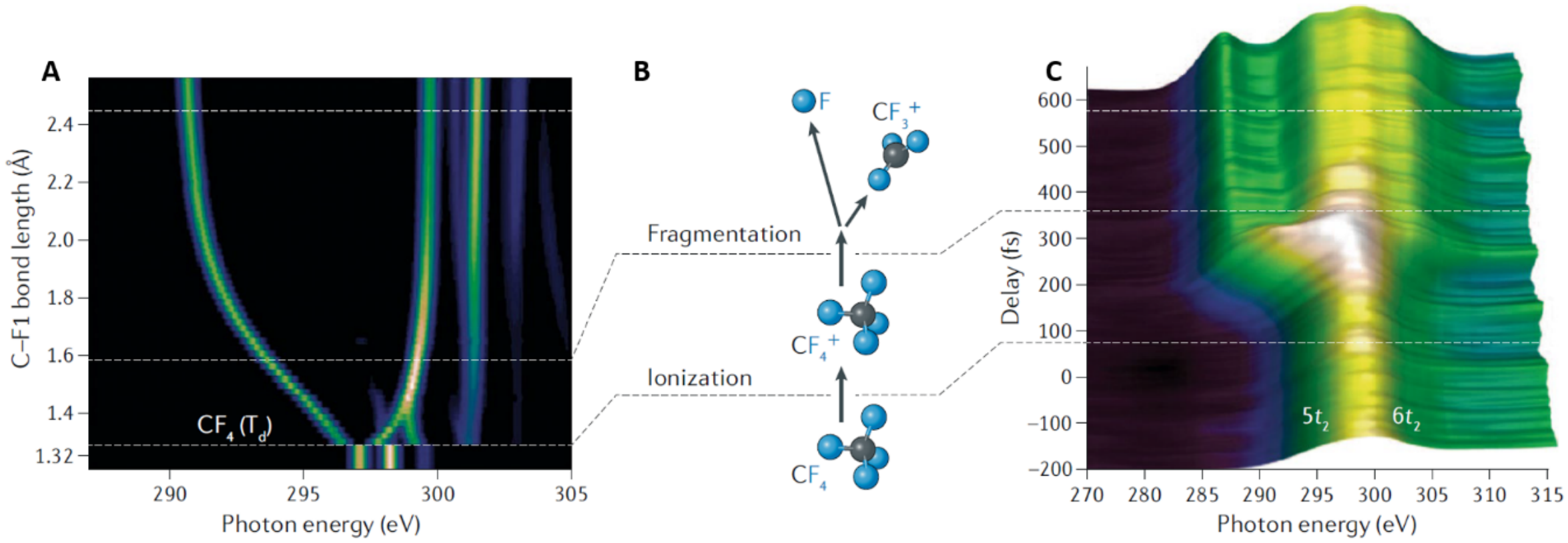

*Figure 8.3: (A) Calculated X-ray absorption spectra as a function of photon energy and C-F bond distance for the light-induced dissociation $CF_4^+$-> $CF_3^+$ + F. (B) Schematic illustration of the light-induced dissociation of $CF_3^+$ after strong-field ionization of $CF_4$. (C) Experimental transient absorption spectrum at the carbon K-edge as a function of photon energy and NIR-XUV time delay. Figure adapted from (Kraus et al., 2018) with permission.*

In the experiment of Pertot *et al.*(Pertot et al., 2017), a high-power laser source based on a cryogenically-cooled Ti:Sa system is employed to generate 40 fs pulses (800 nm) of exceptional power (reaching 18 mJ at 1 kHz). The major part of the energy is down-converted in an OPA to yield MIR pulses centered at 1800 nm with an energy of 2.5 mJ, which are subsequently used to drive HHG in Ne, thereby resulting in an SXR supercontinuum extending up to 350 eV. A portion of the NIR beam (800 nm) is used to strong-field-ionize tetrafluoromethane ($CF_4$) molecules. The $CF_4^+$ cation is unstable with respect to a Jahn-Teller distortion and has an energy minimum at a lower symmetry ($C_{2v}$) compared to the tetrahedral neutral species. This symmetry lowering lifts the degeneracy of the initially triply degenerate $5t_2$-orbital of the neutral (cp. calculated spectra in Fig. 8.3 A). In addition, the cation is unstable and undergoes a dissociation reaction to yield the trigonal-planar ($D_{3h}$) $CF_3^+$ and F (Fig. 8.3 B). The geometry changes accompanying the C-F bond breakage and the structural rearrangement from $C_{2v}$ to the $D_{3h}$-structure in $CF_3^+$ are tracked at each NIR-pump-SXR-probe delay by monitoring the X-ray absorption lines at the C-edge. The static absorption spectrum of the tetrahedral $CF_4$ is characterized by two intense lines (cp. Fig. 8.3 C) which can be assigned to transitions to the $5t_2$ and $6t_2$ orbitals. Symmetry arguments dictate that the transition of the $C\ 1s \rightarrow t_2$-type associated with the initially tetrahedral carbon atom will split into two $C\ 1a \rightarrow a_2''$ -lines and one $1s \rightarrow e'$-line as the environment around C progressively changes to trigonal-planar. The experimentally recorded transient-absorption spectra shown in Fig. 8.3 C reflect this situation (cp. labels in Panel C). With progressing time, the spectrum splits into multiple bands, with one band shifting down to 288 eV (by ≈10 eV), two further bands shifting up/down by 1 eV, and a fourth line appearing at 302 eV as a shoulder structure in the absorption spectrum. Thus, the TAS spectrum reflects the changes in the geometry as the C-F bond breakage takes its course. Further, the intensity evolution of the $5e'$- and the $6e'$-transitions (cp. Fig. 8.3 A) is interpreted as an evidence of the mixing of Rydberg and valence character of these orbitals in the context of the fluorine "cage effect" (Dehmer et al., 1979). The $5e'$ orbital, initially a valence-type, well-localized orbital in the tetrahedral neutral species, develops a partial Rydberg character, whereas the 6e' orbital (of Rydberg character, localized outside of the region defined by the four F atoms in $CF_4$) develops a partial valence character. The difference in the overlap with the highly localized core orbital accounts for the evolution of the intensity distribution as the reaction progresses. This study first demonstrated the feasibility of table-top-based TR-XAS experiments and highlighted the potential of this technique in elucidating the dynamics of chemical reactions.

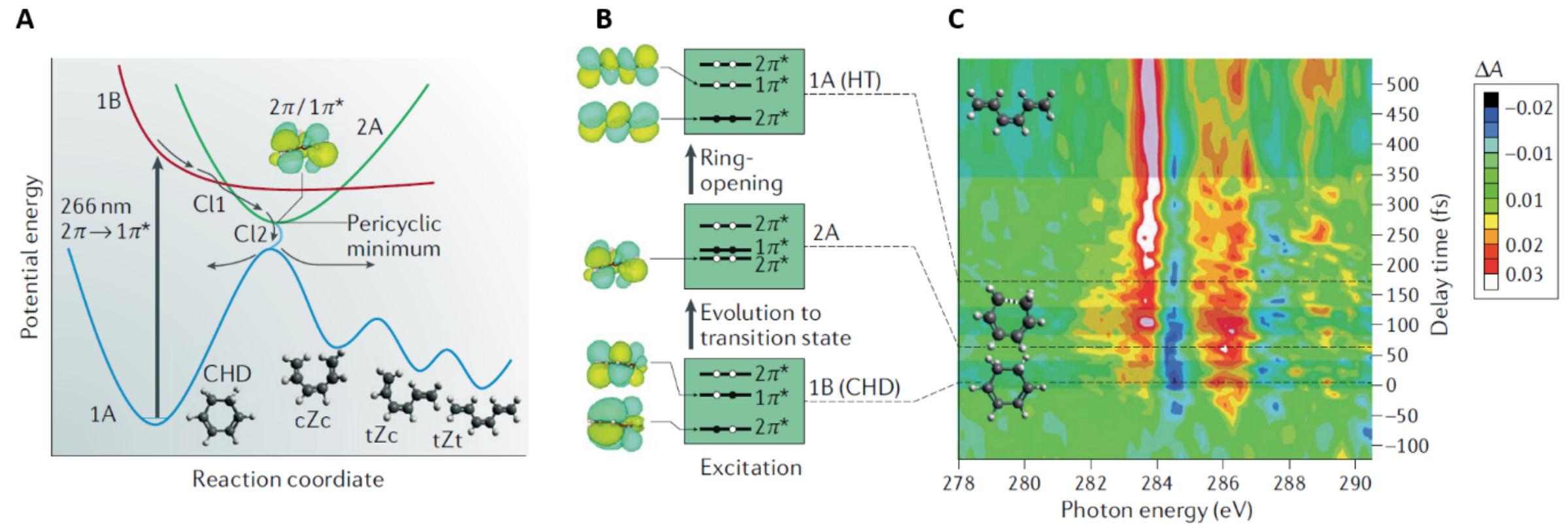

*Figure 8.4: (A) Schematic view of the potential energy surfaces relevant for the electrocyclic ring-opening of cyclohexadiene, which leads to the formation of three stereoisomers of 1,3,5-hexatriene (HT): s- cis,Z,s- cis (cZc); s- trans,Z,s- cis (tZc,) and s- trans,Z,s- trans (tZt). (B) Orbital diagram and electronic configurations of the relevant transition states. (C) Experimental transient absorption spectrum at the carbon K-edge as a function of photon energy and NIR-XUV time delay. The black arrows highlight the connection of the transient intensity modulations in the absorption spectrum with the populations of various transition states (panel (B)) during the reaction. Figure adapted from (Kraus et al., 2018) with permission.*

In a subsequent study, Attar *et al.*(Attar et al., 2017) used a 266 nm, 100 fs pulse to induce the isomerization reaction of ring-closed cyclohexadiene (CHD) to the open-chain 1,3,5-hexatriene (HT). Mechanistically, the UV photoexcitation promotes the system to the 1B excited state via a strong symmetry-allowed ($\pi \rightarrow \pi^*$) transition. The ring-opening reaction proceeds near the conical intersection (CI) of the 1B state with the dark 2A excited state and reaches a "pericyclic minimum" in the transition state of 2A symmetry (cp. scheme in Fig. 8.4 A). The latter is characterized by an electron configuration with a doubly-occupied antibonding $\pi^*$ - orbital (s. orbital scheme in Fig. 8.4 B). After reaching a second conical intersection between the excited 2A and the ground 1A states, the wavepacket can bifurcate either to the open-chain hexatriene or back towards the cyclic isomer. The reaction progress has been tracked by monitoring the time-resolved XAS-signal near the C K-edge. With the aid of advanced TD-DFT calculations, the reaction constants pertaining to the major steps in the mechanistic pathway were determined: the intermediate transition-state minimum is reached within 60±20 fs, whereas the decay to the open photoproduct occurs in further 110±60 fs.

Although the highlighted examples involving molecular systems are still limited to the femtosecond domain, extension to the attosecond range is in near sight given the recent progress in the generation and characterization of IAP-based SXR continua in the water window (Cousin et al., 2017; Silva et al., 2015).

## References

Abraham, H., Lemoine, J., 1899. Disparition instantanée du phénomène de Kerr. Compt. Rend.(Paris) 129, 206–208.

Ammosov, M. V., Delone, N.B., Krainov, V.P., 1986. Tunnel ionization of complex atoms and of atomic ions in an alternating electromagnetic field. Sov. Phys. JETP 64, 1191–1194. https://doi.org/10.1117/12.938695

Andriukaitis, G., Balčiūnas, T., Ališauskas, S., Pugžlys, A., Baltuška, A., Popmintchev, T., Chen, M.-C., Murnane, M.M., Kapteyn, H.C., 2011. 90 GW peak power few-cycle mid-infrared pulses from an optical parametric amplifier. Opt. Lett. 36, 2755–2757. https://doi.org/10.1364/OL.36.002755

Attar, A.R., Bhattacherjee, A., Leone, S.R., 2015. Direct Observation of the Transition-State Region in the Photodissociation of $CH_3I$ by Femtosecond Extreme Ultraviolet Transient Absorption Spectroscopy. J. Phys. Chem. Lett. 6, 5072–5077. https://doi.org/10.1021/acs.jpclett.5b02489

Attar, A.R., Bhattacherjee, A., Pemmaraju, C.D., Schnorr, K., Closser, K.D., Prendergast, D., Leone, S.R., 2017. Femtosecond x-ray spectroscopy of an electrocyclic ring-opening reaction. Science . 356, 54–59. https://doi.org/10.1126/science.aaj2198

Baker, S., Robinson, J.S., Lein, M., Chirila, C.C., Bandulet, H.C., Comtois, D., Villeneuve, D., Kieffer, J.C., Tisch, J.W.G., Marangos, J.P., 2006. Probing proton dynamics in molecules on an attosecond timescale. Science . 312, 424–427. https://doi.org/10.1109/CLEO.2007.4453253

Baltuška, A., Fuji, T., Kobayashi, T., 2002. Controlling the Carrier-Envelope Phase of Ultrashort Light Pulses with Optical Parametric Amplifiers. Phys. Rev. Lett. 88, 133901. https://doi.org/10.1103/PhysRevLett.88.133901

Bandrauk, A.D., Fillion-Gourdeau, F., Lorin, E., 2013. Atoms and molecules in intense laser fields: Gauge invariance of theory and models. J. Phys. B At. Mol. Opt. Phys. 46. https://doi.org/10.1088/0953-4075/46/15/153001

Baykusheva, D., Wörner, H.J., 2017. Theory of attosecond delays in molecular photoionization. J. Chem. Phys. 146, 124306. https://doi.org/10.1063/1.4977933

Beaulieu, S., Comby, A., Clergerie, A., Caillat, J., Descamps, D., Dudovich, N., Fabre, B., Géneaux, R., Légaré, F., Petit, S., Pons, B., Porat, G., Ruchon, T., Taïeb, R., Blanchet, V., Mairesse, Y., 2017. Attosecond-resolved photoionization of chiral molecules. Science . 358, 1288–1294. https://doi.org/10.1126/science.aao5624

Beck, A.R., Neumark, D.M., Leone, S.R., 2015. Probing ultrafast dynamics with attosecond transient absorption. Chem. Phys. Lett. 624, 119–130. https://doi.org/10.1016/j.cplett.2014.12.048

Bertrand, J.B., Wörner, H.J., Hockett, P., Villeneuve, D.M., Corkum, P.B., 2012. Revealing the Cooper minimum of $N_2$ by Molecular Frame High-Harmonic Spectroscopy. Phys. Rev. Lett. 109, 143001. https://doi.org/10.1103/PhysRevLett.109.143001

Bohman, S., Suda, A., Kanai, T., Yamaguchi, S., Midorikawa, K., 2010. Generation of 5.0 fs, 5.0 mJ pulses at 1 kHz using hollow-fiber pulse compression. Opt. Lett. 35, 1887–1889. https://doi.org/10.1364/OL.35.001887

Bouhal, A., Salières, P., Breger, P., Agostini, P., Hamoniaux, G., Mysyrowicz, A., Antonetti, A., Constantinescu, R., Muller, H.G., 1998. Temporal dependence of high-order harmonics in the presence of strong ionization. Phys. Rev. A 58, 389–399.

https://doi.org/10.1103/PhysRevA.58.389
Braunstein, M., McKoy, V., 1987. Shape resonances in the photoionization of $N_2O$. J. Chem. Phys. 87, 224–228. https://doi.org/10.1063/1.453620
Breidbach, J., Cederbaum, L.S., 2005. Universal attosecond response to the removal of an electron. Phys. Rev. Lett. 94, 1–4. https://doi.org/10.1103/PhysRevLett.94.033901
Breidbach, J., Cederbaum, L.S., 2003. Migration of holes: Formalism, mechanisms, and illustrative applications. J. Chem. Phys. 118, 3983–3996. https://doi.org/10.1063/1.1540618
Butkus, R., Danielius, R., Dubietis, A., Piskarskas, A., Stabinis, A., 2004. Progress in chirped pulse optical parametric amplifiers. Appl. Phys. B 79, 693–700. https://doi.org/10.1007/s00340-004-1614-3
Caillat, J., Maquet, A., Haessler, S., Fabre, B., Ruchon, T., Salières, P., Mairesse, Y., Taïeb, R., 2011. Attosecond resolved electron release in two-color near-threshold photoionization of $N_2$. Phys. Rev. Lett. 106, 1–4. https://doi.org/10.1103/PhysRevLett.106.093002
Calegari, F., Ayuso, D., Trabattoni, A., Belshaw, L., De Camillis, S., Anumula, S., Frassetto, F., Poletto, L., Palacios, A., Decleva, P., Greenwood, J.B., Martín, F., Nisoli, M., 2014. Ultrafast electron dynamics in phenylalanine initiated by attosecond pulses. Science . 346, 336–339. https://doi.org/10.1126/science.1254061
Calegari, F., Sansone, G., Nisoli, M., 2015. Optical Technologies for Extreme-Ultraviolet and Soft X-ray Coherent Sources, 1st ed. Springer-Verlag Berlin Heidelberg. https://doi.org/10.1007/978-3-662-47443-3
Calegari, F., Trabattoni, A., Palacios, A., Ayuso, D., Castrovilli, M.C., Greenwood, J.B., Decleva, P., Martín, F., Nisoli, M., 2016. Charge migration induced by attosecond pulses in bio-relevant molecules. J. Phys. B At. Mol. Opt. Phys. 49, 142001. https://doi.org/10.1088/0953-4075/49/14/142001
Canova, L., Chen, X., Trisorio, A., Jullien, A., Assion, A., Tempea, G., Forget, N., Oksenhendler, T., Lopez-Martens, R., 2009. Carrier-envelope phase stabilization and control using a transmission grating compressor and an AOPDF. Opt. Lett. 34, 1333–1335. https://doi.org/10.1364/OL.34.001333
Castrovilli, M.C., Trabattoni, A., Bolognesi, P., O'Keeffe, P., Avaldi, L., Nisoli, M., Calegari, F., Cireasa, R., 2018. Ultrafast Hydrogen Migration in Photoionized Glycine. J. Phys. Chem. Lett. 9, 6012–6016. https://doi.org/10.1021/acs.jpclett.8b02089
Cavalieri, A.L., Müller, N., Uphues, T., Yakovlev, V.S., Baltuska, A., Horvath, B., Schmidt, B., Blümel, L., Holzwarth, R., Hendel, S., Drescher, M., Kleineberg, U., Echenique, P.M., Kienberger, R., Krausz, F., Heinzmann, U., 2007. Attosecond spectroscopy in condensed matter. Nature 449, 1029–1032. https://doi.org/10.1038/nature06229
Cederbaum, L.S., Domcke, W., Schirmer, J., Niessen, W. von, 1986. Correlation Effects in the Ionization of Molecules : Breakdown of the Molecular Orbital Picture. Adv. Chem. Phys. LXV, 115. https://doi.org/10.1002/9780470142899.ch3
Cerullo, G., De Silvestri, S., 2003. Ultrafast optical parametric amplifiers. Rev. Sci. Instrum. 74, 1–18. https://doi.org/10.1063/1.1523642
Chacon, A., Lein, M., Ruiz, C., 2014. Asymmetry of Wigner's time delay in a small molecule. Phys. Rev. A - At. Mol. Opt. Phys. 89, 1–7. https://doi.org/10.1103/PhysRevA.89.053427
Chalus, O., Thai, A., Bates, P.K., Biegert, J., 2010. Six-cycle mid-infrared source with 3.8 μJ at 100 kHz. Opt. Lett. 35, 3204–3206. https://doi.org/10.1364/OL.35.003204
Chang, Z., 2007. Controlling attosecond pulse generation with a double optical gating. Phys. Rev. A 76, 51403. https://doi.org/10.1103/PhysRevA.76.051403

Chang, Z., Rundquist, A., Wang, H., Murnane, M.M., Kapteyn, H.C., 1997. Generation of Coherent Soft X Rays at 2.7 nm Using High Harmonics. Phys. Rev. Lett. 79, 2967–2970. https://doi.org/10.1103/PhysRevLett.79.2967

Chatterley, A.S., Lackner, F., Neumark, D.M., Leone, S.R., Gessner, O., 2016a. Tracking dissociation dynamics of strong-field ionized 1,2-dibromoethane with femtosecond XUV transient absorption spectroscopy. Phys. Chem. Chem. Phys. 18, 14644–14653. https://doi.org/10.1039/c6cp02598f

Chatterley, A.S., Lackner, F., Pemmaraju, C.D., Neumark, D.M., Leone, S.R., Gessner, O., 2016b. Dissociation Dynamics and Electronic Structures of Highly Excited Ferrocenium Ions Studied by Femtosecond XUV Absorption Spectroscopy. J. Phys. Chem. A 120, 9509–9518. https://doi.org/10.1021/acs.jpca.6b09724

Cheng, Y., Chini, M., Wang, X., Wu, Y., Chang, Z., 2014. Attosecond Transient Absorption in Molecular Hydrogen, in: CLEO: 2014. Optical Society of America, p. FM2B.3. https://doi.org/10.1364/CLEO_QELS.2014.FM2B.3

Cheng, Y., Chini, M., Wang, X., Wu, Y., Chang, Z., 2013. Probing Hydrogen and Deuterium Molecular Dynamics Using Attosecond Transient Absorption, in: Frontiers in Optics 2013. Optical Society of America, p. LW2H.2. https://doi.org/10.1364/LS.2013.LW2H.2

Chu, Y., Gan, Z., Liang, X., Yu, L., Lu, X., Wang, C., Wang, X., Xu, L., Lu, H., Yin, D., Leng, Y., Li, R., Xu, Z., 2015. High-energy large-aperture Ti:sapphire amplifier for 5 PW laser pulses. Opt. Lett. 40, 5011–5014. https://doi.org/10.1364/OL.40.005011

Ciriolo, G.A., Negro, M., Devetta, M., Cinquanta, E., Faccialà, D., Pusala, A., De Silvestri, S., Stagira, S., Vozzi, C., 2017. Optical Parametric Amplification Techniques for the Generation of High-Energy Few-Optical-Cycles IR Pulses for Strong Field Applications. Appl. Sci. 7. https://doi.org/10.3390/app7030265

Cohen, E.R., Cvitas, T., Frey, J.G., Holmstrom, B., Kuchitsu, K., Marquardt, R., Mills, I., Pavese, F., Quack, M., Stohner, J., Strauss, H.L., Takami, M., Thor, A.J., 2008. Quantities, Units and Symbols in Physical Chemistry, 3rd ed. IUPAC and RSC Publishing, Cambridge.

Corkum, P.B., 1993. Plasma perspective on strong field multiphoton ionization. Phys. Rev. Lett. 71, 1994–1997.

Corkum, P.B., Burnett, N.H., Ivanov, M.Y., 1994. Subfemtosecond pulses. Opt. Lett. 19, 1870–1872. https://doi.org/10.1364/OL.19.001870

Cousin, S.L., Di Palo, N., Buades, B., Teichmann, S.M., Reduzzi, M., Devetta, M., Kheifets, A., Sansone, G., Biegert, J., 2017. Attosecond streaking in the water window: A new regime of attosecond pulse characterization. Phys. Rev. X 7, 041030. https://doi.org/10.1103/PhysRevX.7.041030

Dahlström, J.M., L'Huillier, A., Maquet, A., 2012. Introduction to attosecond delays in photoionization. J. Phys. B At. Mol. Opt. Phys. 45, 183001. https://doi.org/10.1088/0953-4075/45/18/183001

Danson, C., Hillier, D., Hopps, N., Neely, D., 2015. Petawatt class lasers worldwide. High Power Laser Sci. Eng. 3, e3. https://doi.org/DOI: 10.1017/hpl.2014.52

De, S., Znakovskaya, I., Ray, D., Anis, F., Johnson, N.G., Bocharova, I.A., Magrakvelidze, M., Esry, B.D., Cocke, C.L., Litvinyuk, I. V, Kling, M.F., 2009. Field-Free Orientation of CO Molecules by Femtosecond Two-Color Laser Fields. Phys. Rev. Lett. 103, 153002. https://doi.org/10.1103/PhysRevLett.103.153002

Dehmer, J.L., Dill, D., Wallace, S., 1979. Shape-resonance-enhanced nuclear-motion effects in molecular photoionization. Phys. Rev. Lett. 43, 1005–1008.

https://doi.org/10.1103/PhysRevLett.43.1005
Dehmer, P.M., Miller, P.J., Chupka, W.A., 1984. Photoionization of $N_2$ X $^1\Sigma_g{}^+$, v″=0 and 1 near threshold. Preionization of the Worley–Jenkins Rydberg series. J. Chem. Phys. 80, 1030–1038. https://doi.org/10.1063/1.446829
Deng, Y., Schwarz, A., Fattahi, H., Ueffing, M., Gu, X., Ossiander, M., Metzger, T., Pervak, V., Ishizuki, H., Taira, T., Kobayashi, T., Marcus, G., Krausz, F., Kienberger, R., Karpowicz, N., 2012. Carrier-envelope-phase-stable, 1.2 mJ, 1.5 cycle laser pulses at 2.1 μm. Opt. Lett. 37, 4973–4975. https://doi.org/10.1364/OL.37.004973
Dimitrovski, D., Martiny, C.P.J., Madsen, L.B., 2010. Strong-field ionization of polar molecules: Stark-shift-corrected strong-field approximation. Phys. Rev. A - At. Mol. Opt. Phys. 82. https://doi.org/10.1103/PhysRevA.82.053404
Dörner, R., Mergel, V., Jagutzki, O., Spielberger, L., Ullrich, J., Moshammer, R., Schmidt-Böcking, H., 2000. Cold Target Recoil Ion Momentum Spectroscopy: a 'momentum microscope' to view atomic collision dynamics. Phys. Rep. 330, 95–192. https://doi.org/https://doi.org/10.1016/S0370-1573(99)00109-X
Drescher, M., Hentschel, M., Kienberger, R., Uiberacker, M., Yakovlev, V., Scrinzi, a, Westerwalbesloh, T., Kleineberg, U., Heinzmann, U., Krausz, F., 2002. Time-resolved atomic inner-shell spectroscopy. Nature 419, 803–807. https://doi.org/10.1038/nature01143
Dubietis, A., Butkus, R., Piskarskas, A.P., 2006. Trends in chirped pulse optical parametric amplification. IEEE J. Sel. Top. Quantum Electron. 12, 163–172. https://doi.org/10.1109/JSTQE.2006.871962
Dubietis, A., Jonušauskas, G., Piskarskas, A., 1992. Powerful femtosecond pulse generation by chirped and stretched pulse parametric amplification in BBO crystal. Opt. Commun. 88, 437–440. https://doi.org/https://doi.org/10.1016/0030-4018(92)90070-8
Dudley, J.M., Genty, G., Coen, S., 2006. Supercontinuum generation in photonic crystal fiber. Rev. Mod. Phys. 78, 1135–1184. https://doi.org/10.1103/RevModPhys.78.1135
Dutin, C.F., Dubrouil, A., Petit, S., Mével, E., Constant, E., Descamps, D., 2010. Post-compression of high-energy femtosecond pulses using gas ionization. Opt. Lett. 35, 253–255. https://doi.org/10.1364/OL.35.000253
Eigen, M., 1954. Methods for investigation of ionic reactions in aqueous solutions with half-times as short as $10^{-9}$ sec. Application to neutralization and hydrolysis reactions. Discuss. Faraday Soc. 17, 194–205. https://doi.org/10.1039/DF9541700194
Etches, A., Madsen, L.B., 2010. Extending the strong-field approximation of high-order harmonic generation to polar molecules: gating mechanisms and extension of the harmonic cutoff. J. Phys. B At. Mol. Opt. Phys. 43, 155602. https://doi.org/10.1088/0953-4075/43/15/155602
Farkas, G., Tóth, C., 1992. Proposal for attosecond light pulse generation using laser induced multiple-harmonic conversion processes in rare gases. Phys. Lett. A 168, 447–450. https://doi.org/https://doi.org/10.1016/0375-9601(92)90534-S
Farrell, J.P., Petretti, S., Förster, J., McFarland, B.K., Spector, L.S., Vanne, Y. V., Decleva, P., Bucksbaum, P.H., Saenz, A., Gühr, M., 2011. Strong field ionization to multiple electronic states in water. Phys. Rev. Lett. 107, 1–4. https://doi.org/10.1103/PhysRevLett.107.083001
Feng, X., Gilbertson, S., Mashiko, H., Wang, H., Khan, S.D., Chini, M., Wu, Y., Zhao, K., Chang, Z., 2009. Generation of Isolated Attosecond Pulses with 20 to 28 Femtosecond Lasers. Phys. Rev. Lett. 103, 183901. https://doi.org/10.1103/PhysRevLett.103.183901
Ferrari, F., Calegari, F., Lucchini, M., Vozzi, C., Stagira, S., Sansone, G., Nisoli, M., 2010.

High-energy isolated attosecond pulses generated by above-saturation few-cycle fields. Nat. Photonics 4, 875.
Förg, B., Biswas, S., Schötz, J., Schweinberger, W., Ortmann, L., Zimmerman, T., Pi, L., Baykusheva, D., Masood, H.A., Liontos, I., Kamal, A.M., Kling, N.G., Alharbi, M., Krausz, F., Azzee, A.M., Wörner, H.J., Landsman, A.S., Kling, M.F., 2019. Probing molecular environment through photoemission delays (under rev.
Fork, R.L., Cruz, C.H.B., Becker, P.C., Shank, C. V, 1987. Compression of optical pulses to six femtoseconds by using cubic phase compensation. Opt. Lett. 12, 483–485. https://doi.org/10.1364/OL.12.000483
Frey, R.F., Davidson, E.R., 1988. Potential energy surfaces of $CH_4^+$. J. Chem. Phys. 88, 1775–1785. https://doi.org/10.1063/1.454101
Frumker, E., Kajumba, N., Bertrand, J.B., Wörner, H.J., Hebeisen, C.T., Hockett, P., Spanner, M., Patchkovskii, S., Paulus, G.G., Villeneuve, D.M., Naumov, A., Corkum, P.B., 2012. Probing polar molecules with high harmonic spectroscopy. Phys. Rev. Lett. 109, 1–5. https://doi.org/10.1103/PhysRevLett.109.233904
Fu, Y., Midorikawa, K., Takahashi, E.J., 2018. Towards a petawatt-class few-cycle infrared laser system via dual-chirped optical parametric amplification. Sci. Rep. 8, 7692. https://doi.org/10.1038/s41598-018-25783-0
Fu, Y., Takahashi, E.J., Midorikawa, K., 2015. High-energy infrared femtosecond pulses generated by dual-chirped optical parametric amplification. Opt. Lett. 40, 5082–5085. https://doi.org/10.1364/OL.40.005082
Fuji, T., Rauschenberger, J., Apolonski, A., Yakovlev, V.S., Tempea, G., Udem, T., Gohle, C., Hänsch, T.W., Lehnert, W., Scherer, M., Krausz, F., 2005. Monolithic carrier-envelope phase-stabilization scheme. Opt. Lett. 30, 332–334. https://doi.org/10.1364/OL.30.000332
Gallmann, L., Cirelli, C., Keller, U., 2012. Attosecond Science: Recent Highlights and Future Trends. Annu. Rev. Phys. Chem. 63, 447–469. https://doi.org/10.1146/annurev-physchem-032511-143702
Gallmann, L., Herrmann, J., Locher, R., Sabbar, M., Ludwig, A., Lucchini, M., Keller, U., 2013. Resolving intra-atomic electron dynamics with attosecond transient absorption spectroscopy. Mol. Phys. 111, 2243–2250. https://doi.org/10.1080/00268976.2013.799298
Gallmann, L., Jordan, I., Wörner, H.J., Castiglioni, L., Hengsberger, M., Osterwalder, J., Arrell, C.A., Chergui, M., Liberatore, E., Rothlisberger, U., Keller, U., 2017. Photoemission and photoionization time delays and rates. Struct. Dyn. 4, 061502. https://doi.org/10.1063/1.4997175
Gan, Z., Yu, L., Li, S., Wang, C., Liang, X., Liu, Y., Li, W., Guo, Z., Fan, Z., Yuan, X., Xu, L., Liu, Z., Xu, Y., Lu, J., Lu, H., Yin, D., Leng, Y., Li, R., Xu, Z., 2017. 200 J high efficiency Ti:sapphire chirped pulse amplifier pumped by temporal dual-pulse. Opt. Express 25, 5169–5178. https://doi.org/10.1364/OE.25.005169
Gaumnitz, T., Jain, A., Pertot, Y., Huppert, M., Jordan, I., Ardana-Lamas, F., Wörner, H.J., 2017. Streaking of 43-attosecond soft-X-ray pulses generated by a passively CEP-stable mid-infrared driver. Opt. Express 25, 27506–27518. https://doi.org/10.1364/OE.25.027506
Gianturco, F.A., Lucchese, R.R., Sanna, N., 1994. Calculation of low-energy elastic cross sections for electron-$CF_4$ scattering. J. Chem. Phys. 100, 6464–6471. https://doi.org/http://dx.doi.org/10.1063/1.467237
Gong, X., Lin, C., He, F., Song, Q., Lin, K., Ji, Q., Zhang, W., Ma, J., Lu, P., Liu, Y., Zeng, H., Yang, W., Wu, J., 2017. Energy-Resolved Ultrashort Delays of Photoelectron Emission

Clocked by Orthogonal Two-Color Laser Fields. Phys. Rev. Lett. 118, 143203. https://doi.org/10.1103/PhysRevLett.118.143203
Goulielmakis, E., Loh, Z.-H., Wirth, A., Santra, R., Rohringer, N., Yakovlev, V.S., Zherebtsov, S., Pfeifer, T., Azzeer, A.M., Kling, M.F., Leone, S.R., Krausz, F., 2010. Real-time observation of valence electron motion. Nature 466, 739–743. https://doi.org/10.1038/nature09212
Goulielmakis, E., Schultze, M., Hofstetter, M., Yakovlev, V.S., Gagnon, J., Uiberacker, M., Aquila, A.L., Gullikson, E.M., Attwood, D.T., Kienberger, R., Krausz, F., Kleineberg, U., 2008. Single-Cycle Nonlinear Optics. Science . 320, 1614 LP – 1617. https://doi.org/10.1126/science.1157846
Gruson, V., Ernotte, G., Lassonde, P., Laramée, A., Bionta, M.R., Chaker, M., Di Mauro, L., Corkum, P.B., Ibrahim, H., Schmidt, B.E., Legaré, F., 2017. 2.5 TW, two-cycle IR laser pulses via frequency domain optical parametric amplification. Opt. Express 25, 27706–27714. https://doi.org/10.1364/OE.25.027706
Guénot, D., Kroon, D., M, E.B., Larsen, E.W., Kotur, M., Miranda, M., Fordell, T., Johnsson, P., Mauritsson, J., Gisselbrecht, M., Varjù, K., Arnold, C.L., Carette, T., Kheifets, A.S., Lindroth, E., L'Huillier, A., Dahlström, J.M., 2014. Measurements of relative photoemission time delays in noble gas atoms. J. Phys. B At. Mol. Opt. Phys. 47, 245602. https://doi.org/10.1088/0953-4075/47/24/245602
Haessler, S., Boutu, W., Stankiewicz, M., Frasinski, L.J., Weber, S., Caillat, J., Taeb, R., Maquet, A., Breger, P., Monchicourt, P., Carré, B., Salières, P., 2009a. Attosecond chirp-encoded dynamics of light nuclei. J. Phys. B At. Mol. Opt. Phys. 42. https://doi.org/10.1088/0953-4075/42/13/134002
Haessler, S., Caillat, J., Boutu, W., Giovanetti-Teixeira, C., Ruchon, T., Auguste, T., Diveki, Z., Breger, P., Maquet, A., Carré, B., Taïeb, R., Salières, P., 2010. Attosecond imaging of molecular electronic wavepackets. Nat. Phys. 6, 200–206. https://doi.org/10.1038/nphys1511
Haessler, S., Fabre, B., Higuet, J., Caillat, J., Ruchon, T., Breger, P., Carré, B., Constant, E., Maquet, A., Mével, E., Salières, P., Taïeb, R., Mairesse, Y., 2009b. Phase-resolved attosecond near-threshold photoionization of molecular nitrogen. Phys. Rev. A - At. Mol. Opt. Phys. 80, 011404. https://doi.org/10.1103/PhysRevA.80.011404
Hassan, M.T., Luu, T.T., Moulet, A., Raskazovskaya, O., Zhokhov, P., Garg, M., Karpowicz, N., Zheltikov, A.M., Pervak, V., Krausz, F., Goulielmakis, E., 2016. Optical attosecond pulses and tracking the nonlinear response of bound electrons. Nature 530, 66.
Hauri, C.P., Kornelis, W., Helbing, F.W., Heinrich, A., Couairon, A., Mysyrowicz, A., Biegert, J., Keller, U., 2004. Generation of intense, carrier-envelope phase-locked few-cycle laser pulses through filamentation. Appl. Phys. B 79, 673–677. https://doi.org/10.1007/s00340-004-1650-z
Hennig, H., Breidbach, J., Cederbaum, L.S., 2005. Charge transfer driven by electron correlation: A non-Dyson propagator approach. J. Chem. Phys. 122, 134104. https://doi.org/10.1063/1.1869473
Hentschel, M., Kienberger, R., Spielmann, C., Reider, G.A., Milosevic, N., Brabec, T., Corkum, P., Heinzmann, U., Drescher, M., Krausz, F., 2001. Attosecond metrology. Nature 414, 509.
Heuser, S., Jiménez Galán, Á., Cirelli, C., Marante, C., Sabbar, M., Boge, R., Lucchini, M., Gallmann, L., Ivanov, I., Kheifets, A.S., Dahlström, J.M., Lindroth, E., Argenti, L., Martín, F., Keller, U., 2016. Angular dependence of photoemission time delay in helium. Phys. Rev.

A 94, 63409. https://doi.org/10.1103/PhysRevA.94.063409
Heyl, C.M., Arnold, C.L., Couairon, A., L'Huillier, A., 2016. Introduction to macroscopic power scaling principles for high-order harmonic generation. J. Phys. B At. Mol. Opt. Phys. 50, 13001. https://doi.org/10.1088/1361-6455/50/1/013001
Higuet, J., Ruf, H., Thiré, N., Cireasa, R., Constant, E., Cormier, E., Descamps, D., Mével, E., Petit, S., Pons, B., Mairesse, Y., Fabre, B., 2011. High-order harmonic spectroscopy of the Cooper minimum in argon: Experimental and theoretical study. Phys. Rev. A - At. Mol. Opt. Phys. 83, 1–12. https://doi.org/10.1103/PhysRevA.83.053401
Hosler, E.R., Leone, S.R., 2013. Characterization of vibrational wave packets by core-level high-harmonic transient absorption spectroscopy. Phys. Rev. A - At. Mol. Opt. Phys. 88, 023420. https://doi.org/10.1103/PhysRevA.88.023420
Huppert, M., Jordan, I., Baykusheva, D., von Conta, A., Wörner, H.J., 2016. Attosecond Delays in Molecular Photoionization. Phys. Rev. Lett. 117, 093001. https://doi.org/10.1103/PhysRevLett.117.093001
Huppert, M., Jordan, I., Wörner, H.J., 2015. Attosecond beamline with actively stabilized and spatially separated beam paths. Rev. Sci. Instrum. 86, 123106. https://doi.org/10.1063/1.4937623
Ishii, N., Kaneshima, K., Kitano, K., Kanai, T., Watanabe, S., Itatani, J., 2014. Carrier-envelope phase-dependent high harmonic generation in the water window using few-cycle infrared pulses. Nat. Commun. 5, 3331.
Itatani, J., Quéré, F., Yudin, G.L., Ivanov, M.Y., Krausz, F., Corkum, P.B., 2002. Attosecond Streak Camera. Phys. Rev. Lett. 88, 173903. https://doi.org/10.1103/PhysRevLett.88.173903
Ivanov, M.Y., Spanner, M., Smirnova, O., 2005. Anatomy of strong field ionization. J. Mod. Op. 52, 165–184. https://doi.org/10.1080/0950034042000275360
Jain, A., Gaumnitz, T., Bray, A., Kheifets, A., Wörner, H.J., 2018. Photoionization delays in xenon using single-shot referencing in the collinear back-focusing geometry. Opt. Lett. 43, 4510–4513. https://doi.org/10.1364/OL.43.004510
Jordan, I., Huppert, M., Pabst, S., Kheifets, A.S., Baykusheva, D., Wörner, H.J., 2017. Spin-orbit delays in photoemission. Phys. Rev. A 95, 13404. https://doi.org/10.1103/PhysRevA.95.013404
Jullien, A., Pfeifer, T., Abel, M.J., Nagel, P.M., Bell, M.J., Neumark, D.M., Leone, S.R., 2008. Ionization phase-match gating for wavelength-tunable isolated attosecond pulse generation. Appl. Phys. B 93, 433. https://doi.org/10.1007/s00340-008-3187-z
Kakehata, M., Takada, H., Kobayashi, Y., Torizuka, K., Fujihira, Y., Homma, T., Takahashi, H., 2001. Single-shot measurement of carrier-envelope phase changes by spectral interferometry. Opt. Lett. 26, 1436–1438. https://doi.org/10.1364/OL.26.001436
Kanai, T., Takahashi, E.J., Nabekawa, Y., Midorikawa, K., 2008. Observing the attosecond dynamics of nuclear wavepackets in molecules by using high harmonic generation in mixed gases. New J. Phys. 10, 025036. https://doi.org/10.1088/1367-2630/10/2/025036
Keldysh, L. V., 1965. Ionization in the field of a strong electromagnetic wave. JETP 20, 1307.
Kelkensberg, F., Siu, W., Pérez-Torres, J.F., Morales, F., Gademann, G., Rouzée, A., Johnsson, P., Lucchini, M., Calegari, F., Sanz-Vicario, J.L., Martín, F., Vrakking, M.J.J., 2011. Attosecond control in photoionization of hydrogen molecules. Phys. Rev. Lett. 107, 043002. https://doi.org/10.1103/PhysRevLett.107.043002
Keller, U., 'tHooft, G.W., Knox, W.H., Cunningham, J.E., 1991. Femtosecond pulses from a continuously self-starting passively mode-locked Ti:sapphire laser. Opt. Lett. 16, 1022–1024. https://doi.org/10.1364/OL.16.001022
Kim, K.T., Zhang, C., Ruchon, T., Hergott, J.-F., Auguste, T., Villeneuve, D.M., Corkum, P.B., Quéré, F., 2013. Photonic streaking of attosecond pulse trains. Nat. Photonics 7, 651.
Kjeldsen, T.K., Madsen, L.B., 2005. Vibrational excitation of diatomic molecular ions in strong field ionization of diatomic molecules. Phys. Rev. Lett. 95, 1–4. https://doi.org/10.1103/PhysRevLett.95.073004
Kling, M.F., Verhoef, A.J., Khan, J.I., Schultze, M., Ni, Y., Uiberacker, M., Drescher, M., Krausz, F., Vrakking, M.J.J., Siedschlag, C., Verhoef, A.J., Khan, J.I., Schultze, M., Uphues, T., Ni, Y., Uiberacker, M., Drescher, M., Krausz, F., Vrakking, M.J.J., 2006. Control of Electron Localization in Molecular Dissociation. Science . 312, 246–248. https://doi.org/10.1126/science.1126259
Kling, M.F., Vrakking, M.J.J.J., 2008. Attosecond Electron Dynamics. Annu. Rev. Phys. Chem. 59, 463–492. https://doi.org/10.1146/annurev.physchem.59.032607.093532
Klünder, K., Dahlström, J.M., Gisselbrecht, M., Fordell, T., Swoboda, M., Guénot, D., Johnsson, P., Caillat, J., Mauritsson, J., Maquet, A., Taïeb, R., L'Huillier, A., 2011. Probing Single-Photon Ionization on the Attosecond Time Scale. Phys. Rev. Lett. 106, 143002. https://doi.org/10.1103/PhysRevLett.106.143002
Knight, L.B., Steadman, J., Feller, D., Davidson, E.R., 1984. Experimental evidence for a $C_{2v}$ ($^2B_1$) ground-state structure of the methane cation radical: ESR and ab initio CI investigations of methane cation radicals ($CH_4^+$ and $CD_2H_2^+$) in neon matrixes at 4 K. J. Am. Chem. Soc. 106, 3700–3701. https://doi.org/10.1021/ja00324a066
Koke, S., Grebing, C., Frei, H., Anderson, A., Assion, A., Steinmeyer, G., 2010. Direct frequency comb synthesis with arbitrary offset and shot-noise-limited phase noise. Nat. Photonics 4, 462.
Kraus, P.M., Baykusheva, D., Wörner, H.J., 2014. Two-pulse field-free orientation reveals anisotropy of molecular shape resonance. Phys. Rev. Lett. 113, 023001. https://doi.org/10.1103/PhysRevLett.113.023001
Kraus, P M, Mignolet, B., Baykusheva, D., Rupenyan, A., Horný, L., Penka, E.F., Grassi, G., Tolstikhin, O.I., Schneider, J., Jensen, F., Madsen, L.B., Bandrauk, A.D., Remacle, F., Wörner, H.J., 2015. Measurement and laser control of attosecond charge migration in ionized iodoacetylene. Science . 350, 790.
Kraus, P.M., Rupenyan, A., Wörner, H.J., 2012. High-harmonic spectroscopy of oriented ocs molecules: Emission of even and odd harmonics. Phys. Rev. Lett. 109, 233903. https://doi.org/10.1103/PhysRevLett.109.233903
Kraus, P. M., Tolstikhin, O.I., Baykusheva, D., Rupenyan, A., Schneider, J., Bisgaard, C.Z., Morishita, T., Jensen, F., Madsen, L.B., Wörner, H.J., 2015. Observation of laser-induced electronic structure in oriented polyatomic molecules. Nat. Commun. 6, 7039. https://doi.org/10.1038/ncomms8039
Kraus, P.M., Wörner, H.J., 2018a. Perspectives of Attosecond Spectroscopy for the Understanding of Fundamental Electron Correlations. Angew. Chemie Int. Ed. 57, 5228–5247. https://doi.org/10.1002/anie.201702759
Kraus, P.M., Wörner, H.J., 2018b. Perspektiven für das Verständnis fundamentaler Elektronenkorrelationen durch Attosekundenspektroskopie. Angew. Chemie 130, 5324–5344. https://doi.org/10.1002/ange.201702759
Kraus, P.M., Wörner, H.J., 2013. Attosecond nuclear dynamics in the ammonia cation: Relation

between high-harmonic and photoelectron spectroscopies. ChemPhysChem 14, 1445–1450. https://doi.org/10.1002/cphc.201201022
Kraus, P.M., Zürch, M., Cushing, S.K., Neumark, D.M., Leone, S.R., 2018. The ultrafast X-ray spectroscopic revolution in chemical dynamics. Nat. Rev. Chem. 2, 82–94. https://doi.org/10.1038/s41570-018-0008-8
Krausz, F., 2016. The birth of attosecond physics and its coming of age. Phys. Scr. 91. https://doi.org/10.1088/0031-8949/91/6/063011
Krausz, F., Ivanov, M., 2009. Attosecond physics. Rev. Mod. Phys. 81, 163–234. https://doi.org/10.1103/RevModPhys.81.163
Krehl, P., Engemann, S., 1995. August Toepler - The first who visualized shock waves. Shock Waves 5, 1–18. https://doi.org/10.1007/BF02425031
Kreß, M., Löffler, T., Thomson, M.D., Dörner, R., Gimpel, H., Zrost, K., Ergler, T., Moshammer, R., Morgner, U., Ullrich, J., Roskos, H.G., 2006. Determination of the carrier-envelope phase of few-cycle laser pulses with terahertz-emission spectroscopy. Nat. Phys. 2, 327.
Kulander, K.C., Schafer, K.J., Krause, J.L., 1993. Dynamics of Short-Pulse Excitation, Ionization and Harmonic Conversion - Super-Intense Laser-Atom Physics, in: Piraux, B., L'Huillier, A., Rzążewski, K. (Eds.), . Springer US, Boston, MA, pp. 95–110. https://doi.org/10.1007/978-1-4615-7963-2_10
Lackner, F., Chatterley, A.S., Pemmaraju, C.D., Closser, K.D., Prendergast, D., Neumark, D.M., Leone, S.R., Gessner, O., 2016. Direct observation of ring-opening dynamics in strong-field ionized selenophene using femtosecond inner-shell absorption spectroscopy. J. Chem. Phys. 145, 234313. https://doi.org/10.1063/1.4972258
Lara-Astiaso, M., Galli, M., Trabattoni, A., Palacios, A., Ayuso, D., Frassetto, F., Poletto, L., De Camillis, S., Greenwood, J., Decleva, P., Tavernelli, I., Calegari, F., Nisoli, M., Martín, F., 2018. Attosecond Pump–Probe Spectroscopy of Charge Dynamics in Tryptophan. J. Phys. Chem. Lett. 9, 4570–4577. https://doi.org/10.1021/acs.jpclett.8b01786
Lein, M., 2005. Attosecond probing of vibrational dynamics with high-harmonic generation. Phys. Rev. Lett. 94, 1–4. https://doi.org/10.1103/PhysRevLett.94.053004
Lein, M., Hay, N., Velotta, R., Marangos, J.P., Knight, P.L., 2002a. Role of the Intramolecular Phase in High-Harmonic Generation. Phys. Rev. Lett. 88, 183903. https://doi.org/10.1103/PhysRevLett.88.183903
Lein, M., Hay, N., Velotta, R., Marangos, J.P., Knight, P.L., 2002b. Interference effects in high-order harmonic generation with molecules. Phys. Rev. A 66, 23805. https://doi.org/10.1103/PhysRevA.66.023805
Leone, S.R., Neumark, D.M., 2016. Attosecond science in atomic, molecular, and condensed matter physics. Faraday Discuss. 194, 15–39. https://doi.org/10.1039/c6fd00174b
Lepetit, L., Chériaux, G., Joffre, M., 1995. Linear techniques of phase measurement by femtosecond spectral interferometry for applications in spectroscopy. J. Opt. Soc. Am. B 12, 2467–2474. https://doi.org/10.1364/JOSAB.12.002467
Lépine, F., Ivanov, M.Y., Vrakking, M.J.J., 2014. Attosecond molecular dynamics: Fact or fiction? Nat. Photonics 8, 195–204. https://doi.org/10.1038/nphoton.2014.25
Lewenstein, M., Balcou, P., Ivanov, M.Y., L'Huillier, A., Corkum, P.B., 1994. Theory of high-harmonic generation by low-frequency laser fields. Phys. Rev. A 49, 2117–2132.
Li, J., Ren, X., Yin, Y., Zhao, K., Chew, A., Cheng, Y., Cunningham, E., Wang, Y., Hu, S., Wu, Y., Chini, M., Chang, Z., 2017. 53-attosecond X-ray pulses reach the carbon K-edge. Nat.

Commun. 8, 186. https://doi.org/10.1038/s41467-017-00321-0
Loh, Z.H., Khalil, M., Correa, R.E., Santra, R., Buth, C., Leone, S.R., 2007. Quantum state-resolved probing of strong-field-ionized Xenon atoms using femtosecond high-order harmonic transient absorption spectroscopy. Phys. Rev. Lett. 98, 143601. https://doi.org/10.1103/PhysRevLett.98.143601
Louisy, M., Arnold, C.L., Miranda, M., Larsen, E.W., Bengtsson, S.N., Kroon, D., Kotur, M., Guénot, D., Rading, L., Rudawski, P., Brizuela, F., Campi, F., Kim, B., Jarnac, A., Houard, A., Mauritsson, J., Johnsson, P., L'Huillier, A., Heyl, C.M., 2015. Gating attosecond pulses in a noncollinear geometry. Optica 2, 563–566. https://doi.org/10.1364/OPTICA.2.000563
Ludwig, A., Maurer, J., Mayer, B.W., Phillips, C.R., Gallmann, L., Keller, U., 2014. Breakdown of the dipole approximation in strong-field ionization. Phys. Rev. Lett. 113, 1–5. https://doi.org/10.1103/PhysRevLett.113.243001
Lünnemann, S., Kuleff, A.I., Cederbaum, L.S., 2008. Charge migration following ionization in systems with chromophore-donor and amine-acceptor sites. J. Chem. Phys. 129, 104305. https://doi.org/10.1063/1.2970088
Maine, P., Strickland, D., Bado, P., Pessot, M., Mourou, G., 1988. Generation of ultrahigh peak power pulses by chirped pulse amplification. IEEE J. Quantum Electron. 24, 398–403. https://doi.org/10.1109/3.137
Mairesse, Y., de Bohan, A., Frasinski, L.J., Merdji, H., Dinu, L.C., Monchicourt, P., Breger, P., Kovačev, M., Taïeb, R., Carré, B., Muller, H.G., Agostini, P., Salières, P., 2003. Attosecond Synchronization of High-Harmonic Soft X-rays. Science . 302, 1540 LP – 1543. https://doi.org/10.1126/science.1090277
Månsson, E.P., De Camillis, S., Castrovilli, M.C., Galli, M., Nisoli, M., Calegari, F., Greenwood, J.B., 2017. Ultrafast dynamics in the DNA building blocks thymidine and thymine initiated by ionizing radiation. Phys. Chem. Chem. Phys. 19, 19815–19821. https://doi.org/10.1039/C7CP02803B
McPherson, A., Gibson, G., Jara, H., Johann, U., Luk, T.S., McIntyre, I.A., Boyer, K., Rhodes, C.K., 1987. Studies of multiphoton production of vacuum-ultraviolet radiation in the rare gases. J. Opt. Soc. Am. B 4, 595–601. https://doi.org/10.1364/JOSAB.4.000595
Mehendale, M., Mitchell, S.A., Likforman, J.-P., Villeneuve, D.M., Corkum, P.B., 2000. Method for single-shot measurement of the carrier envelope phase of a few-cycle laser pulse. Opt. Lett. 25, 1672–1674. https://doi.org/10.1364/OL.25.001672
Mizutani, H., Minemoto, S., Oguchi, Y., Sakai, H., 2011. Effect of nuclear motion observed in high-order harmonic generation from $D_2$/$H_2$ molecules with intense multi-cycle 1300 nm and 800 nm pulses. J. Phys. B At. Mol. Opt. Phys. 44, 081002. https://doi.org/10.1088/0953-4075/44/8/081002
Mondal, T., Varandas, A.J.C., 2015. Structural evolution of the methane cation in subfemtosecond photodynamics. J. Chem. Phys. 143, 14304. https://doi.org/10.1063/1.4922906
Mondal, T., Varandas, A.J.C., 2014. On Extracting Subfemtosecond Data from Femtosecond Quantum Dynamics Calculations: The Methane Cation. J. Chem. Theory Comput. 10, 3606–3616. https://doi.org/10.1021/ct500388k
Moulton, P.F., 1986. Spectroscopic and laser characteristics of Ti:$Al_2O_3$. J. Opt. Soc. Am. B 3, 125–133. https://doi.org/10.1364/JOSAB.3.000125
Natalense, A., P. P. Lucchese, R.R., 1999. Cross section and asymmetry parameter calculation for sulfur 1s photoionization of $SF_6$. J. Chem. Phys. 111, 5344–5348.

https://doi.org/http://dx.doi.org/10.1063/1.479794
Neidel, C., Klei, J., Yang, C.H., Rouzée, A., Vrakking, M.J.J., Klünder, K., Miranda, M., Arnold, C.L., Fordell, T., L'Huillier, A., Gisselbrecht, M., Johnsson, P., Dinh, M.P., Suraud, E., Reinhard, P.G., Despré, V., Marques, M.A.L., Lépine, F., 2013. Probing time-dependent molecular dipoles on the attosecond time scale. Phys. Rev. Lett. 111, 033001. https://doi.org/10.1103/PhysRevLett.111.033001
Neppl, S., Ernstorfer, R., Bothschafter, E.M., Cavalieri, A.L., Menzel, D., Barth, J. V, Krausz, F., Kienberger, R., Feulner, P., 2012. Attosecond Time-Resolved Photoemission from Core and Valence States of Magnesium. Phys. Rev. Lett. 109, 87401. https://doi.org/10.1103/PhysRevLett.109.087401
Nisoli, M., De Silvestri, S., Svelto, O., 1996. Generation of high energy 10 fs pulses by a new pulse compression technique. Appl. Phys. Lett. 68, 2793–2795. https://doi.org/10.1063/1.116609
Nisoli, M., Decleva, P., Calegari, F., Palacios, A., Martín, F., 2017. Attosecond Electron Dynamics in Molecules. Chem. Rev. 117, 10760–10825. https://doi.org/10.1021/acs.chemrev.6b00453
Norrish, R.G.W., Porter, G., 1949. Chemical Reactions Produced by Very High Light Intensities. Nature 164, 658.
Okell, W.A., Witting, T., Fabris, D., Arrell, C.A., Hengster, J., Ibrahimkutty, S., Seiler, A., Barthelmess, M., Stankov, S., Lei, D.Y., Sonnefraud, Y., Rahmani, M., Uphues, T., Maier, S.A., Marangos, J.P., Tisch, J.W.G., 2015. Temporal broadening of attosecond photoelectron wavepackets from solid surfaces. Optica 2, 383–387. https://doi.org/10.1364/OPTICA.2.000383
Ossiander, M., Riemensberger, J., Neppl, S., Mittermair, M., Schäffer, M., Duensing, A., Wagner, M.S., Heider, R., Wurzer, M., Gerl, M., Schnitzenbaumer, M., Barth, J. V, Libisch, F., Lemell, C., Burgdörfer, J., Feulner, P., Kienberger, R., 2018. Absolute timing of the photoelectric effect. Nature 561, 374–377. https://doi.org/10.1038/s41586-018-0503-6
Ossiander, M., Siegrist, F., Shirvanyan, V., Pazourek, R., Sommer, A., Latka, T., Guggenmos, A., Nagele, S., Feist, J., Burgdörfer, J., Kienberger, R., Schultze, M., 2016. Attosecond Correlation Dynamics. Nat. Phys. 1, 10.1038/nphys3941. https://doi.org/10.1038/nphys3941
Patchkovskii, S., 2009. Nuclear dynamics in polyatomic molecules and high-order harmonic generation. Phys. Rev. Lett. 102, 8–11. https://doi.org/10.1103/PhysRevLett.102.253602
Patchkovskii, S., Schuurman, M.S., 2017. Full-dimensional treatment of short-time vibronic dynamics in a molecular high-order-harmonic-generation process in methane. Phys. Rev. A 96, 053405. https://doi.org/10.1103/PhysRevA.96.053405
Paul, P.M., Toma, E.S., Breger, P., Mullot, G., Augé, F., Balcou, P., Muller, H.G., Agostini, P., 2001. Observation of a Train of Attosecond Pulses from High Harmonic Generation. Science . 292, 1689 LP – 1692. https://doi.org/10.1126/science.1059413
Paulus, G.G., Grasbon, F., Walther, H., Villoresi, P., Nisoli, M., Stagira, S., Priori, E., De Silvestri, S., 2001. Absolute-phase phenomena in photoionization with few-cycle laser pulses. Nature 414, 182–184. https://doi.org/10.1038/35102520
Pazourek, R., Nagele, S., Burgdörfer, J., 2015. Attosecond chronoscopy of photoemission. Rev. Mod. Phys. 87. https://doi.org/10.1103/RevModPhys.87.765
Perelomov, A.M., Popov, V.S., Terent'ev, M. V., 1966. Ionization of atoms in an alternating electrical field. Sov. Phys. JETP 23, 924.
Pertot, Y., Schmidt, C., Matthews, M., Chauvet, A., Huppert, M., Svoboda, V., von Conta, A.,

Tehlar, A., Baykusheva, D., Wolf, J.-P., Wörner, H.J., 2017. Time-resolved X-ray absorption spectroscopy with a water window high-harmonic source. Science . 355, 264 LP – 267.
Pfeifer, T., Jullien, A., Abel, M.J., Nagel, P.M., Gallmann, L., Neumark, D.M., Leone, S.R., 2007. Generating coherent broadband continuum soft-x-ray radiation by attosecond ionization gating. Opt. Express 15, 17120–17128. https://doi.org/10.1364/OE.15.017120
Popmintchev, T., Chen, M.-C., Bahabad, A., Gerrity, M., Sidorenko, P., Cohen, O., Christov, I.P., Murnane, M.M., Kapteyn, H.C., 2009. Phase matching of high harmonic generation in the soft and hard X-ray regions of the spectrum. Proc. Natl. Acad. Sci. U. S. A. 106, 10516–10521. https://doi.org/10.1073/pnas.0903748106
Popruzhenko, S. V., 2014. Keldysh theory of strong field ionization: history, applications, difficulties and perspectives. J. Phys. B At. Mol. Opt. Phys. 47, 204001. https://doi.org/10.1088/0953-4075/47/20/204001
Powis, I., 2008. Photoelectron Circular Dichroism in Gas Phase Chiral Molecules. Adv. Chem. Phys. 138, 267–329.
Ramasesha, K., Leone, S.R., Neumark, D.M., 2016. Real-Time Probing of Electron Dynamics Using Attosecond Time-Resolved Spectroscopy. Annu. Rev. Phys. Chem. 67, 41–63. https://doi.org/10.1146/annurev-physchem-040215-112025
Rathbone, G.J., Poliakoff, E.D., Bozek, J.D., Toffoli, D., Lucchese, R.R., 2005. Photoelectron trapping in $N_2O$ 7σ→kσ resonant ionization. J. Chem. Phys. 123, 14307. https://doi.org/10.1063/1.1946738
Reduzzi, M., Feng, C., Chu, W.-C., Dubrouil, A., Calegari, F., Nisoli, M., Frassetto, F., Poletto, L., Lin, C.-D., Sansone, G., 2013. Attosecond Absorption Spectroscopy in Molecules, in: CLEO: 2013. Optical Society of America, p. QF2C.1. https://doi.org/10.1364/CLEO_QELS.2013.QF2C.1
Reiss, H.R., 2008a. Limits on tunneling theories of strong-field ionization. Phys. Rev. Lett. 101, 1–4. https://doi.org/10.1103/PhysRevLett.101.043002
Reiss, H.R., 2008b. Erratum: Limits on Tunneling Theories of Strong-Field Ionization [Phys. Rev. Lett. 101 , 043002 (2008)]. Phys. Rev. Lett. 101, 159901. https://doi.org/10.1103/PhysRevLett.101.159901
Reiss, H.R., 1980. Gauges for intense-field electrodynamics. Phys. Rev. A 22, 770–772. https://doi.org/10.1103/PhysRevA.22.770
Remacle, F., Levine, R.D., 2006. An electronic time scale in chemistry. Proc. Natl. Acad. Sci. 103, 6793–6798. https://doi.org/10.1073/pnas.0601855103
Ren, X., Makhija, V., Le, A.-T., Troß, J., Mondal, S., Jin, C., Kumarappan, V., Trallero-Herrero, C., 2013. Measuring the angle-dependent photoionization cross section of nitrogen using high-harmonic generation. Phys. Rev. A 88, 43421. https://doi.org/10.1103/PhysRevA.88.043421
Ritchie, B., 1975. Theory of the angular distribution of photoelectrons ejected from optically active molecules and molecular negative ions. Phys. Rev. A 13, 1411–1415.
Rothhardt, J., Demmler, S., Hädrich, S., Limpert, J., Tünnermann, A., 2012. Octave-spanning OPCPA system delivering CEP-stable few-cycle pulses and 22 W of average power at 1 MHz repetition rate. Opt. Express 20, 10870–10878. https://doi.org/10.1364/OE.20.010870
Rupenyan, A., Bertrand, J.B., Villeneuve, D.M., Wörner, H.J., 2012. All-Optical Measurement of High-Harmonic Amplitudes and Phases in Aligned Molecules. Phys. Rev. Lett. 108, 33903. https://doi.org/10.1103/PhysRevLett.108.033903

Rupenyan, A., Kraus, P.M., Schneider, J., Wörner, H.J., 2013. High-harmonic spectroscopy of isoelectronic molecules: Wavelength scaling of electronic-structure and multielectron effects. Phys. Rev. A - At. Mol. Opt. Phys. 87, 033409. https://doi.org/10.1103/PhysRevA.87.033409

Sabbar, M., Heuser, S., Boge, R., Lucchini, M., Carette, T., Lindroth, E., Gallmann, L., Cirelli, C., Keller, U., 2015. Resonance Effects in Photoemission Time Delays. Phys. Rev. Lett. 115, 1–5. https://doi.org/10.1103/PhysRevLett.115.133001

Salières, P., Carré, B., Le Déroff, L., Grasbon, F., Paulus, G.G., Walther, H., Kopold, R., Becker, W., Milošević, D.B., Sanpera, A., Lewenstein, M., 2001. Feynman's Path-Integral Approach for Intense-Laser-Atom Interactions. Science . 292, 902 LP – 905.

Sansone, G., Benedetti, E., Calegari, F., Vozzi, C., Avaldi, L., Flammini, R., Poletto, L., Villoresi, P., Altucci, C., Velotta, R., Stagira, S., De Silvestri, S., Nisoli, M., 2006. Isolated Single-Cycle Attosecond Pulses. Science . 314, 443 LP – 446. https://doi.org/10.1126/science.1132838

Sansone, G., Kelkensberg, F., Pérez-Torres, J.F., Morales, F., Kling, M.F., Siu, W., Ghafur, O., Johnsson, P., Swoboda, M., Benedetti, E., Ferrari, F., Lépine, F., Sanz-Vicario, J.L., Zherebtsov, S., Znakovskaya, I., L'huillier, A., Ivanov, M.Y., Nisoli, M., Martín, F., Vrakking, M.J.J., 2010. Electron localization following attosecond molecular photoionization. Nature 465, 763–766. https://doi.org/10.1038/nature09084

Sayler, A.M., Rathje, T., Müller, W., Rühle, K., Kienberger, R., Paulus, G.G., 2011. Precise, real-time, every-single-shot, carrier-envelope phase measurement of ultrashort laser pulses. Opt. Lett. 36, 1–3. https://doi.org/10.1364/OL.36.000001

Schafer, K.J., Yang, B., DiMauro, L.F., Kulander, K.C., 1993. Above threshold ionization beyond the high harmonic cutoff. Phys. Rev. Lett. 70, 1599–1602. https://doi.org/10.1103/PhysRevLett.70.1599

Schmidt, B.E., Béjot, P., Giguère, M., Shiner, A.D., Trallero-Herrero, C., Bisson, É., Kasparian, J., Wolf, J.-P., Villeneuve, D.M., Kieffer, J.-C., Corkum, P.B., Légaré, F., 2010. Compression of 1.8 μm laser pulses to sub two optical cycles with bulk material. Appl. Phys. Lett. 96, 121109. https://doi.org/10.1063/1.3359458

Schmidt, B.E., Shiner, A.D., Lassonde, P., Kieffer, J.-C., Corkum, P.B., Villeneuve, D.M., Légaré, F., 2011. CEP stable 1.6 cycle laser pulses at 1.8 μm. Opt. Express 19, 6858–6864. https://doi.org/10.1364/OE.19.006858

Schmidt, B.E., Thiré, N., Boivin, M., Laramée, A., Poitras, F., Lebrun, G., Ozaki, T., Ibrahim, H., Légaré, F., 2014. Frequency domain optical parametric amplification. Nat. Commun. 5, 3643.

Schultze, M., Fieß, M., Karpowicz, N., Gagnon, J., Korbman, M., Hofstetter, M., Neppl, S., Cavalieri, A.L., Komninos, Y., Mercouris, T., Nicolaides, C.A., Pazourek, R., Nagele, S., Feist, J., Burgdörfer, J., Azzeer, A.M., Ernstorfer, R., Kienberger, R., Kleineberg, U., Goulielmakis, E., Krausz, F., Yakovlev, V.S., 2010. Delay in Photoemission. Science . 328, 1658–1662. https://doi.org/10.1126/science.1189401

Schultze, M., Ramasesha, K., Pemmaraju, C.D., Sato, S.A., Whitmore, D., Gandman, A., Prell, J.S., Borja, L.J., Prendergast, D., Yabana, K., Neumark, D.M., Leone, S.R., 2014. Attosecond band-gap dynamics in silicon. Science . 346, 1348–1353.

Seres, E., Seres, J., Spielmann, C., 2006. X-ray absorption spectroscopy in the keV range with laser generated high harmonic radiation. Appl. Phys. Lett. 89, 181919. https://doi.org/10.1063/1.2364126

Shafir, D., Soifer, H., Bruner, B.D., Dagan, M., Mairesse, Y., Patchkovskii, S., Ivanov, M.Y., Smirnova, O., Dudovich, N., 2012. Resolving the time when an electron exits a tunnelling barrier. Nature 485, 343–6. https://doi.org/10.1038/nature11025

Shiner, A.D., Schmidt, B.E., Trallero-Herrero, C., Wörner, H.J., Patchkovskii, S., Corkum, P.B., Kieffer, J.-C., Légaré, F., Villeneuve, D.M., 2011. Probing collective multi-electron dynamics in xenon with high-harmonic spectroscopy. Nat. Phys. 7, 464–467. https://doi.org/10.1038/nphys1940

Silva, F., Teichmann, S.M., Cousin, S.L., Hemmer, M., Biegert, J., 2015. Spatiotemporal isolation of attosecond soft X-ray pulses in the water window. Nat. Commun. 6, 1–6. https://doi.org/10.1038/ncomms7611

Siu, W., Kelkensberg, F., Gademann, G., Rouzée, A., Johnsson, P., Dowek, D., Lucchini, M., Calegari, F., De Giovannini, U., Rubio, A., Lucchese, R.R., Kono, H., Lépine, F., Vrakking, M.J.J., 2011. Attosecond control of dissociative ionization of $O_2$ molecules. Phys. Rev. A - At. Mol. Opt. Phys. 84, 063412. https://doi.org/10.1103/PhysRevA.84.063412

Skantzakis, E., Tzallas, P., Kruse, J., Kalpouzos, C., Charalambidis, D., 2009. Coherent continuum extreme ultraviolet radiation in the sub-100-nJ range generated by a high-power many-cycle laser field. Opt. Lett. 34, 1732–1734. https://doi.org/10.1364/OL.34.001732

Smirnova, Olga, Mairesse, Y., Patchkovskii, S., Dudovich, N., Villeneuve, D., Corkum, P., Ivanov, M.Y., 2009. High harmonic interferometry of multi-electron dynamics in molecules. Nature 460, 972–977. https://doi.org/10.1038/nature08253

Smirnova, O., Patchkovskii, S., Mairesse, Y., Dudovich, N., Ivanov, M.Y., 2009. Strong-field control and spectroscopy of attosecond electron-hole dynamics in molecules. Proc. Natl. Acad. Sci. 106, 16556–16561. https://doi.org/10.1073/pnas.0907434106

Smirnova, O., Spanner, M., Ivanov, M., 2007. Anatomy of strong field ionization II: to dress or not to dress? J. Mod. Opt. 54, 1019–1038. https://doi.org/10.1080/09500340701234656

Sola, I.J., Mével, E., Elouga, L., Constant, E., Strelkov, V., Poletto, L., Villoresi, P., Benedetti, E., Caumes, J.-P., Stagira, S., Vozzi, C., Sansone, G., Nisoli, M., 2006. Controlling attosecond electron dynamics by phase-stabilized polarization gating. Nat. Phys. 2, 319.

Song, X., Shi, G., Zhang, G., Xu, J., Lin, C., Chen, J., Yang, W., 2018. Attosecond Time Delay of Retrapped Resonant Ionization. Phys. Rev. Lett. 121, 103201. https://doi.org/10.1103/PhysRevLett.121.103201

Spanner, M., Patchkovskii, S., 2013. Molecular strong field ionization and high harmonic generation: A selection of computational illustrations. Chem. Phys. 414, 10–19. https://doi.org/https://doi.org/10.1016/j.chemphys.2011.12.016

Spanner, M., Patchkovskii, S., 2009. One-electron ionization of multielectron systems in strong nonresonant laser fields. Phys. Rev. A 80, 63411. https://doi.org/10.1103/PhysRevA.80.063411

Spence, D.E., Kean, P.N., Sibbett, W., 1991. 60-fsec pulse generation from a self-mode-locked Ti:sapphire laser. Opt. Lett. 16, 42–44. https://doi.org/10.1364/OL.16.000042

Śpiewanowski, M.D., Etches, A., Madsen, L.B., 2013. High-order-harmonic generation from field-distorted orbitals. Phys. Rev. A - At. Mol. Opt. Phys. 87, 043424. https://doi.org/10.1103/PhysRevA.87.043424

Stapelfeldt, H., Seideman, T., 2003. Colloquium: Aligning molecules with strong laser pulses. Rev. Mod. Phys. 75, 543–557. https://doi.org/10.1103/RevModPhys.75.543

Strickland, D., Mourou, G., 1985. Compression of amplified chirped optical pulses. Opt. Commun. 56, 219–221. https://doi.org/https://doi.org/10.1016/0030-4018(85)90120-8

Suda, A., Hatayama, M., Nagasaka, K., Midorikawa, K., 2005. Generation of sub-10-fs, 5-mJ-optical pulses using a hollow fiber with a pressure gradient. Appl. Phys. Lett. 86, 111116. https://doi.org/10.1063/1.1883706

Sutter, D.H., Steinmeyer, G., Gallmann, L., Matuschek, N., Morier-Genoud, F., Keller, U., Scheuer, V., Angelow, G., Tschudi, T., 1999. Semiconductor saturable-absorber mirror–assisted Kerr-lens mode-locked Ti:sapphire laser producing pulses in the two-cycle regime. Opt. Lett. 24, 631–633. https://doi.org/10.1364/OL.24.000631

Swoboda, M., Fordell, T., Klünder, K., Dahlström, J.M., Miranda, M., Buth, C., Schafer, K.J., Mauritsson, J., L'Huillier, A., Gisselbrecht, M., 2010. Phase Measurement of Resonant Two-Photon Ionization in Helium. Phys. Rev. Lett. 104, 103003. https://doi.org/10.1103/PhysRevLett.104.103003

Takahashi, E.J., Kanai, T., Ishikawa, K.L., Nabekawa, Y., Midorikawa, K., 2008. Coherent Water Window X Ray by Phase-Matched High-Order Harmonic Generation in Neutral Media. Phys. Rev. Lett. 101, 253901. https://doi.org/10.1103/PhysRevLett.101.253901

Tcherbakoff, O., Mével, E., Descamps, D., Plumridge, J., Constant, E., 2003. Time-gated high-order harmonic generation. Phys. Rev. A 68, 43804. https://doi.org/10.1103/PhysRevA.68.043804

Telle, H.R., Steinmeyer, G., Dunlop, A.E., Stenger, J., Sutter, D.H., Keller, U., 1999. Carrier-envelope offset phase control: A novel concept for absolute optical frequency measurement and ultrashort pulse generation. Appl. Phys. B 69, 327–332. https://doi.org/10.1007/s003400050813

Timmers, H., Sabbar, M., Hellwagner, J., Kobayashi, Y., Neumark, D.M., Leone, S.R., 2016. Polarization-assisted amplitude gating as a route to tunable, high-contrast attosecond pulses. Optica 3, 707–710. https://doi.org/10.1364/OPTICA.3.000707

Tolstikhin, O.I., Morishita, T., Madsen, L.B., 2011. Theory of tunneling ionization of molecules: Weak-field asymptotics including dipole effects. Phys. Rev. A - At. Mol. Opt. Phys. 84, 1–5. https://doi.org/10.1103/PhysRevA.84.053423

Tong, X.M., Zhao, Z.X., Lin, C.D., 2002. Theory of molecular tunneling ionization. Phys. Rev. A 66, 33402.

Tzallas, P., Skantzakis, E., Kalpouzos, C., Benis, E.P., Tsakiris, G.D., Charalambidis, D., 2007. Generation of intense continuum extreme-ultraviolet radiation by many-cycle laser fields. Nat. Phys. 3, 846.

Uiberacker, M., Uphues, T., Schultze, M., Verhoef, A.J., Yakovlev, V., Kling, M.F., Rauschenberger, J., Kabachnik, N.M., Schröder, H., Lezius, M., Kompa, K.L., Muller, H.-G., Vrakking, M.J.J., Hendel, S., Kleineberg, U., Heinzmann, U., Drescher, M., Krausz, F., 2007. Attosecond real-time observation of electron tunnelling in atoms. Nature 446, 627.

Ullrich, J., Moshammer, R., Dorn, A., Dörner, R., Schmidt, L.P.H., Schmidt-Böcking, H., 2003. Recoil-ion and electron momentum spectroscopy: reaction-microscopes. Reports Prog. Phys. 66, 1463–1545. https://doi.org/10.1088/0034-4885/66/9/203

Ullrich, J., Moshammer, R., Dörner, R., Jagutzki, O., Mergel, V., Schmidt-Böcking, H., Spielberger, L., 1997. Recoil-ion momentum spectroscopy. J. Phys. B At. Mol. Opt. Phys. 30, 2917–2974. https://doi.org/10.1088/0953-4075/30/13/006

Urbain, X., Fabre, B., Staicu-Casagrande, E.M., De Ruette, N., Andrianarijaona, V.M., Jureta, J., Posthumus, J.H., Saenz, A., Baldit, E., Cornaggia, C., 2004. Intense-laser-field ionization of molecular hydrogen in the tunneling regime and its effect on the vibrational excitation of $H_2^+$. Phys. Rev. Lett. 92, 163004–1. https://doi.org/10.1103/PhysRevLett.92.163004

Vager, Z., Kanter, E.P., Both, G., Cooney, P.J., Faibis, A., Koenig, W., Zabransky, B.J., Zajfman, D., 1986. Direct Determination of the Stereochemical Structure of $CH_4^+$. Phys. Rev. Lett. 57, 2793–2795. https://doi.org/10.1103/PhysRevLett.57.2793
Vaupel, A., Bodnar, N., Webb, B., Shah, L., Richardson, M.C., 2013. Concepts, performance review, and prospects of table-top, few-cycle optical parametric chirped-pulse amplification. Opt. Eng. 53, 1–12.
Vincenti, H., Quéré, F., 2012. Attosecond Lighthouses: How To Use Spatiotemporally Coupled Light Fields To Generate Isolated Attosecond Pulses. Phys. Rev. Lett. 108, 113904. https://doi.org/10.1103/PhysRevLett.108.113904
Vos, J., Cattaneo, L., Patchkovskii, S., Zimmermann, T., Cirelli, C., Lucchini, M., Kheifets, A., Landsman, A.S., Keller, U., 2018. Orientation-dependent stereo Wigner time delay and electron localization in a small molecule. Science . 360, 1326–1330. https://doi.org/10.1126/science.aao4731
Vozzi, C., Calegari, F., Benedetti, E., Berlasso, R., Sansone, G., Stagira, S., Nisoli, M., Altucci, C., Velotta, R., Torres, R., Heesel, E., Kajumba, N., Marangos, J.P., 2006. Probing two-centre interference in molecular high harmonic generation. J. Phys. B At. Mol. Opt. Phys. 39, S457–S466. https://doi.org/10.1088/0953-4075/39/13/s19
Vozzi, C., Calegari, F., Benedetti, E., Caumes, J.P., Sansone, G., Stagira, S., Nisoli, M., Torres, R., Heesel, E., Kajumba, N., Marangos, J.P., Altucci, C., Velotta, R., 2005. Controlling two-center interference in molecular high harmonic generation. Phys. Rev. Lett. 95, 1–4. https://doi.org/10.1103/PhysRevLett.95.153902
Vrakking, M.J.J., 2014. Attosecond imaging. Phys. Chem. Chem. Phys. 16, 2775–2789.
Vrakking, M.J.J., Lepine, F., 2019. Attosecond Molecular Dynamics, Theoretical and Computational Chemistry Series. The Royal Society of Chemistry. https://doi.org/10.1039/9781788012669
Warrick, E.R., Cao, W., Neumark, D.M., Leone, S.R., 2016. Probing the Dynamics of Rydberg and Valence States of Molecular Nitrogen with Attosecond Transient Absorption Spectroscopy. J. Phys. Chem. A 120, 3165–3174. https://doi.org/10.1021/acs.jpca.5b11570
Weinkauf, R., Schanen, P., Metsala, A., Schlag, E.W., Bürgle, M., Kessler, H., 1996. Highly Efficient Charge Transfer in Peptide Cations in the Gas Phase: Threshold Effects and Mechanism. J. Phys. Chem. 100, 18567–18585. https://doi.org/10.1021/jp960926m
Weinkauf, R., Schanen, P., Yang, D., Soukara, S., Schlag, E.W., 1995. Elementary Processes in Peptides: Electron Mobility and Dissociation in Peptide Cations in the Gas Phase. J. Phys. Chem. 99, 11255–11265. https://doi.org/10.1021/j100028a029
Wheeler, J.A., Borot, A., Monchocé, S., Vincenti, H., Ricci, A., Malvache, A., Lopez-Martens, R., Quéré, F., 2012. Attosecond lighthouses from plasma mirrors. Nat. Photonics 6, 829.
Wirth, A., Hassan, M.T., Grguraš, I., Gagnon, J., Moulet, A., Luu, T.T., Pabst, S., Santra, R., Alahmed, Z.A., Azzeer, A.M., Yakovlev, V.S., Pervak, V., Krausz, F., Goulielmakis, E., 2011. Synthesized Light Transients. Science . 334, 195 LP – 200. https://doi.org/10.1126/science.1210268
Witte, S., Eikema, K.S.E., 2012. Ultrafast Optical Parametric Chirped-Pulse Amplification. IEEE J. Sel. Top. Quantum Electron. 18, 296–307. https://doi.org/10.1109/JSTQE.2011.2118370
Wolter, B., Pullen, M.G., Baudisch, M., Sclafani, M., Hemmer, M., Senftleben, A., Schröter, C.D., Ullrich, J., Moshammer, R., Biegert, J., 2015. Strong-Field Physics with Mid-IR Fields. Phys. Rev. X 5, 21034. https://doi.org/10.1103/PhysRevX.5.021034
Wong, M.C.H., Le, A.-T., Alharbi, A.F., Boguslavskiy, A.E., Lucchese, R.R., Brichta, J.-P., Lin,

C.D., Bhardwaj, V.R., 2013. High Harmonic Spectroscopy of the Cooper Minimum in Molecules. Phys. Rev. Lett. 110, 33006. https://doi.org/10.1103/PhysRevLett.110.033006
Wörner, H.J., Arrell, C.A., Banerji, N., Cannizzo, A., Chergui, M., Das, A.K., Hamm, P., Keller, U., Kraus, P.M., Liberatore, E., Lopez-Tarifa, P., Lucchini, M., Meuwly, M., Milne, C., Moser, J.-E., Rothlisberger, U., Smolentsev, G., Teuscher, J., van Bokhoven, J.A., Wenger, O., 2017. Charge migration and charge transfer in molecular systems. Struct. Dyn. 4, 61508. https://doi.org/10.1063/1.4996505
Wörner, H.J., Corkum, P.B., 2011. Attosecond Spectroscopy, in: Merkt, F., Quack, M. (Eds.), Handbook of High-resolution Spectroscopy, Major Reference Works. John Wiley & Sons, Ltd. https://doi.org/doi:10.1002/9780470749593.hrs085
Wörner, H.J., Niikura, H., Bertrand, J.B., Corkum, P.B., Villeneuve, D.M., 2009. Observation of electronic structure minima in high-harmonic generation. Phys. Rev. Lett. 102, 1–4. https://doi.org/10.1103/PhysRevLett.102.103901
Xu, L., Spielmann, C., Poppe, A., Brabec, T., Krausz, F., Hänsch, T.W., 1996. Route to phase control of ultrashort light pulses. Opt. Lett. 21, 2008–2010. https://doi.org/10.1364/OL.21.002008
Yudin, G., Ivanov, M., 2001. Nonadiabatic tunnel ionization: Looking inside a laser cycle. Phys. Rev. A 64, 4. https://doi.org/10.1103/PhysRevA.64.013409
Zewail, A.H., 2000. Femtochemistry:  Atomic-Scale Dynamics of the Chemical Bond. J. Phys. Chem. A 104, 5660–5694. https://doi.org/10.1021/jp001460h
Zhang, Q., Takahashi, E.J., Mücke, O.D., Lu, P., Midorikawa, K., 2011. Dual-chirped optical parametric amplification for generating few hundred mJ infrared pulses. Opt. Express 19, 7190–7212. https://doi.org/10.1364/OE.19.007190
Zipp, L.J., Natan, A., Bucksbaum, P.H., 2014. Probing electron delays in above-threshold ionization. Optica 1, 361. https://doi.org/10.1364/OPTICA.1.000361